\documentclass{aa}

\usepackage{amsmath}
\usepackage{amssymb}
\usepackage{graphicx}
\usepackage{color}
\usepackage{comment}
\usepackage{ulem}
\usepackage{xspace}

\usepackage{dcolumn}
\usepackage{txfonts}
\usepackage{epstopdf} 
\usepackage{siunitx}
\usepackage{caption}
\usepackage{booktabs}

\usepackage{hyperref}

\usepackage{array}
\newcolumntype{L}[1]{>{\raggedright\let\newline\\\arraybackslash\hspace{0pt}}m{#1}}
\newcolumntype{C}[1]{>{\centering\let\newline\\\arraybackslash\hspace{0pt}}m{#1}}
\newcolumntype{R}[1]{>{\raggedleft\let\newline\\\arraybackslash\hspace{0pt}}m{#1}}

\newcommand{\al}{f}
\newcommand{\ac}{h}
\newcommand{\alPixScale}{p_{\tau}}
\newcommand{\acPixScale}{p_{\mu}}
\newcommand{\gsw}{g}
\newcommand{\paperI}{Paper~I\xspace}
\newcommand{\gaia}{\textit{Gaia}\xspace}
\newcommand{\nueff}{$\nu_{\text{eff}}$\xspace}

\begin{document} 

\title{\gaia Data Release 4}

\subtitle{Modelling of drift-scan related effects in \gaia's point
spread function}

\titlerunning{\gaia DR4 -- Modelling of drift-scan related effects in
\gaia's PSF}
\authorrunning{Rowell et al.}

\author{
N.~Rowell\inst{\ref{inst:uoe}}\fnmsep\thanks{Corresponding author: N.
Rowell\newline
e-mail: \href{mailto:nr@roe.ac.uk}{\tt nr@roe.ac.uk}}
\and M.~Davidson\inst{\ref{inst:uoe}}
\and N.~C.~Hambly\inst{\ref{inst:uoe}}
\and L.~Lindegren\inst{\ref{inst:lund}}
\and J.~Casta{\~n}eda\inst{\ref{inst:dapcom}}
\and C.~Fabricius\inst{\ref{inst:ieec}\ref{inst:iccub}\ref{inst:fqa}}
\and J.~Hern{\'a}ndez\inst{\ref{inst:esac}}
\and D.~W.~Evans\inst{\ref{inst:ioa}}
}

\institute{
Institute for Astronomy, School of Physics and Astronomy, University of Edinburgh, 
Royal Observatory, Blackford Hill, Edinburgh, EH9~3HJ, 
United Kingdom
\label{inst:uoe} 
\and
Lund Observatory, Division of Astrophysics, Department of Physics, Lund
University, Box 43, SE-22100 Lund, Sweden \label{inst:lund} 
\and
DAPCOM for Institut de Ci\`{e}ncies del Cosmos (ICCUB), Universitat de Barcelona
(IEEC-UB), Mart\'{i} Franqu\`{e}s 1, E-08028 Barcelona, Spain
\label{inst:dapcom} 
\and
Institut d$^\prime$Estudis Espacials de Catalunya (IEEC), Esteve Terradas 1,
08860 Castelldefels, Spain \label{inst:ieec} 
\and
Institut de Ci\`encies del Cosmos (ICCUB), Universitat de Barcelona (UB),
Mart\'i Franqu\`es 1, E-08028 Barcelona, Spain 
\label{inst:iccub} 
\and
Departament de F\'{i}sica Qu\`{a}ntica i Astrof\'{i}sica (FQA), Universitat de Barcelona (UB), Mart\'{\i} i Franqu\`{e}s 1, 08028 Barcelona, Spain
\label{inst:fqa} 
\and
ESA, European Space Astronomy Centre, Camino Bajo del Castillo s/n, 28691
 Villanueva de la Ca{\~n}ada, Spain
\label{inst:esac} 
\and
Institute of Astronomy, University of Cambridge, Madingley Road, Cambridge
 CB3~0HA, UK \label{inst:ioa} 
}

\date{}

\abstract{
An accurate model of the point spread function (PSF) is required in order to
estimate positions and brightnesses of stars in digitized images. The PSF of the
\gaia space telescope is unusual due to the use of drift-scan mode and
time-delayed integration (TDI), in which the satellite spins and
precesses while images are captured. This induces several systematic and
periodic distortions in the PSF that are unique to \gaia.}
{We identify several effects that distort \gaia's PSF. These include systematic
variations in the stellar image drift rate with respect to the charge transfer
rate, and spatial variations in the CCD response that are, contrary to
expectations, not marginalised by the use of TDI mode. These must be
incorporated into the PSF model in order to reduce systematic errors in \gaia's
data products.}
{We have developed a semi-analytic model of the PSF, in which the blurring
effects of along- and across-scan stellar image motion are modelled
analytically, and dependences of the PSF shape on source colour and position
within the CCD are calibrated empirically.
Constraints on the PSF origin are introduced in order to break a degeneracy
with the geometric instrument calibration.}
{Our PSF model successfully reproduces several drift-scan related effects and
leads to significant improvements in the modelling of observations, particularly
around the 11--13 magnitude range in $G$. This will contribute to reductions in
the astrometric and photometric uncertainties in the derived data products.}
{Our PSF model represents a significant advance over earlier models applied to
\gaia data. It was deployed in the \gaia cyclic data processing systems and used
in the production of the forthcoming Data Release 4.
The linear part of \gaia's PSF is now well understood.
Future development work will focus on optimised configuration of the model, and
the handling of several nonlinear effects that depend on the signal level,
including charge transfer inefficiency and the brighter-fatter effect.
This work will provide a useful reference for users of \gaia data and for 
other missions that use the same observing principles, in particular the
proposed GaiaNIR mission.}


\maketitle
\nolinenumbers
\section{Introduction \label{sect:intro}}
The European Space Agency's \gaia mission aims to investigate the
composition, formation and evolution of the Milky Way galaxy, primarily by
mapping the precise three-dimensional positions and motions of a large number of
its constituent stars \citep{2016A&A...595A...1G}.
The raw observations on which this catalogue is based are collected using a
dedicated pair of space telescopes mounted on a single observing platform, the
\gaia satellite, which operated almost continuously from 2014 to 2025
from its orbit around the second Lagrange point in the Sun-Earth system.
The conversion of the raw data to science-ready catalogues suitable for public
release is the task of the Gaia Data Processing and Analysis Consortium (DPAC).
Several data releases have already been made since the beginning of the mission
as the quantity of raw data accumulates
\citep{DR1-DPACP-8, DR2-DPACP-36, EDR3-DPACP-130, DR3-DPACP-185},
with each successive release based on a complete reprocessing of the available
data using the latest software. This allows gradual improvements in the
instrument modelling and calibration pipelines to contribute to reductions in
the systematic errors, boosting the accuracy of the resulting catalogues beyond
that expected purely from the increased number of observations.
The fourth data release (DR4) is based on the first 66 months of observations,
and is expected to be published in December 2026.

A key component of the data processing, particularly for the astrometry and
$G$-band photometry, is the modelling and calibration of the point spread
function (PSF). In the context of the
\gaia instrument modelling, the PSF model predicts the distribution of
photoelectrons among the digitised samples for a particular observation, and it
is used to estimate the instantaneous positions and fluxes of all sources in the
\gaia data stream \citep[see e.g.][]{DR1-DPACP-7}.
These in turn are used to determine the astrometric and ($G$-band) photometric
properties for all sources, as well as various auxiliary calibrations such as
the satellite attitude and focal plane geometry \citep{Lindegren2012}.
The PSF model must incorporate many physical effects including the telescope
optics at the time of observation, image pixelisation, detector properties such
as spatial response variations, source properties such as colour,
and potentially several nonlinear effects that depend on the signal level.
In addition to these somewhat conventional PSF dependences, there are additional
major systematic effects uniquely present in \gaia's PSF due to the
unusual way in which the images of stars are acquired.
The \gaia telescopes are operated in drift-scan mode, in which the
satellite spins about an axis perpendicular to 
both telescopes at a constant rate of one revolution every six hours.
The telescopes are swept continuously across the sky, with the images from
each being combined onto a single shared focal plane. The images of stars take
around a minute to drift through the field of view,
crossing each of 12--15 charge-coupled devices (CCDs) in turn that
are used to acquire different types of observations of the star. The CCDs are operated in
Time-Delayed Integration (TDI) mode, with charge moved along the pixel columns
at a rate that is matched to the average motion of the stars, allowing the
signal to accumulate.
In principle, this forms a continuous, ribbon-like image of the sky.
However, only a small fraction of the resulting charge is actually read out: the
data is highly windowed in order to optimise the telemetry budget, with windows
positioned to coincide with on-board detections of
sources\footnote{Additional windows (referred to as `virtual
objects') are assigned according to a fixed pattern, for use in several
auxiliary calibrations. See \citet{2016A&A...595A...1G} section 3.3.5 for
more details of the windowing.}

This neat picture hides a lot of complexity. As \gaia spins it also precesses,
such that the images of stars do not travel perfectly along the CCD columns but
drift in the orthogonal direction at a varying rate. The motion in the
along-scan direction also deviates systematically from the fixed charge transfer
rate due to several effects. Therefore, the resulting integrated images are
significantly smeared out. This then induces a sensitivity to along-scan spatial
variations in the CCD response that was somewhat unexpected, and which further
modulates the resulting observation.
These effects have not been properly recognised and accounted for in previous
\gaia data processing, and must be incorporated into the PSF modelling.

The PSF model actually implemented in the \gaia data processing has
advanced considerably over the course of the data releases. This has been driven
by a combination of two factors. First, experience of working with the
observations has deepened our knowledge of how the in-flight instrument behaves
and our understanding of the data that it produces. Second, each successive data
processing cycle brings with it progressive improvements in the various
auxiliary instrument calibrations and source astrometry on which the PSF
modelling relies, allowing more subtle systematic effects to be revealed. The
model used in the production of Early Data Release 3 \citep[EDR3;
see][]{EDR3-DPACP-130} represented a major step forward, and is
described in detail in \citet{EDR3-DPACP-73}, henceforth referred to as \paperI.

In \paperI Sect.~6.1 we reported some major systematic errors in the PSF
model; these were known to be caused by the incomplete modelling of certain
drift-scan related effects, but were not fully understood at the time.
In the present paper we describe these effects in detail, and explain how the
systematics present in our earlier work have been overcome in the recent data
processing by the development of a new PSF model. This model has been
implemented and deployed in the \gaia data processing, and was used in the
production of the forthcoming DR4.
The purpose of this paper is to present the PSF model itself; the performance
and calibration results in the DR4 processing will be published as part of the
official documentation at the time of the data release.
This paper is organised as follows.
In Sect.~\ref{sec:terms} we introduce some terminology that will be used
throughout the paper.
In Sect.~\ref{sec:driftscan} we discuss the drift-scan mode employed by
\gaia and the resulting effects on the PSF.
In Sect.~\ref{sec:dr3plsf} we briefly recap the PSF model from \paperI.
In Sect.~\ref{sec:dr4psf} we derive a new PSF model that incorporates all
known drift-scan related effects in a consistent and efficient manner.
In Sect.~\ref{sec:results} we present some results to demonstrate the features
of the new model, the reduction in systematic errors in the reconstruction of
\gaia observations, and improvements in the estimated source locations.
Quantitative improvements in the DR4 derived data products, such as the source
astrometry and $G$-band photometry, are by necessity deferred to other
publications.

In Sect.~\ref{sec:disc} we discuss some limitations of the new PSF model,
such as the choice of parameterisation for the empirically calibrated
dependences. We present a brief analysis of the brighter-fatter effect and
charge transfer inefficiency in \gaia data, neither of which are currently modelled.
We also set some expectations for DR4 and plans for the future,
including possible implications for the proposed GaiaNIR mission.
In Sect.~\ref{sec:conc} we draw some conclusions.
In the Appendices we present some implementation details and describe two new
auxiliary instrument calibrations that are required by the PSF model.
\section{Terminology \label{sec:terms}}
This paper will make use of certain \gaia-specific terminology to describe the
instruments, observation strategy, data collection and PSF modelling. We closely
follow the terms defined in section 2 of \paperI, to which the reader is
referred. Briefly, these describe:
\begin{itemize}
  \item the two telescopes, referred to as field of view 1 and 2 (FOV1 and
  FOV2)\footnote{Sometimes also referred to as the `preceding' and
  `following' field of view (PFoV and FFoV) in other \gaia
  publications.},
  \item the fundamental along-scan (AL) and across-scan (AC) directions in the
  focal plane,
  \item the layout of the CCDs, their designation by row and strip, and
  assignment to the Sky Mapper (SM) and Astrometric Field (AF) instruments
  (see also Fig.~\ref{fig:deviceTypes}),
  \item the windowing, sampling and marginalisation of the data,
  \item the 2D point spread function (PSF) and 1D line spread function (LSF),
  collectively referred to as the PLSF,
  \item the CCD gating strategy used to extend the magnitude range,
  \item the partitioning of the observations into 1268 independent calibration
  units.
\end{itemize}
A few additions are required in order to accommodate the new modelling described
in the present paper. 
In \paperI we adopted the terminology of \citet{anderson2000} to describe the
PSF, in which the distinction is made between the `instrumental' PSF,
which is never directly observed,
and the `effective' PSF, which accounts for pixelisation and is used to
model observations.
These terms are tailored towards traditional framing cameras, and in order to
properly describe \gaia's PSF, accounting for the TDI mode of operation, we need
to introduce a further distinction between the `instantaneous effective'
PSF and the `integrated effective' PSF.
According to these definitions, the instantaneous effective PSF is the 2D
distribution of photoelectron flux from a point source at a single location in
the focal plane, accounting for pixelisation but crucially not including
integration along the CCD.
The integrated effective PSF is the result of integrating the instantaneous
effective PSF along the CCD, and it is this that is used to model \gaia
observations.

\begin{figure}
\resizebox{\hsize}{!}{\includegraphics{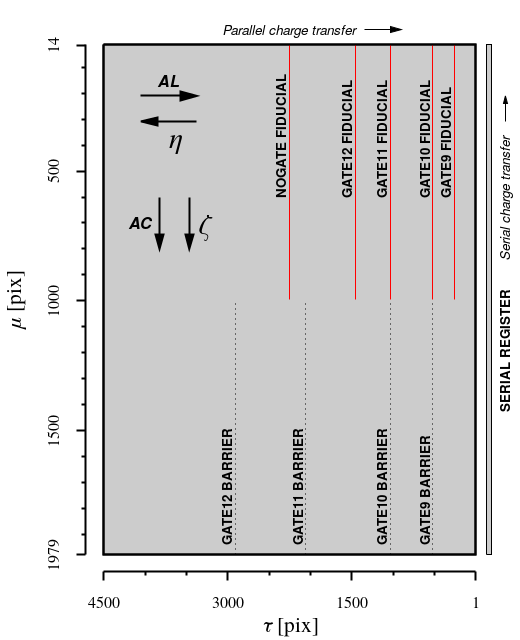}}
\caption{Schematic diagram of one of \gaia's CCDs, as seen from the
illuminated side and showing various features relevant to the PSF modelling. The
($\tau , \mu$) coordinates represent positions within the pixel grid. The images
of stars move from left to right during integration, with the serial register
lying on the right. The main AL and AC directions and their relation to the
field angles $\eta$ and $\zeta$ are indicated at the top left. Fiducial lines
and barriers for the five longest gates are indicated; NOGATE has no
corresponding barrier and uses the entire AL range.}
\label{fig:ccd_diagram}
\end{figure}

We also introduce some terms related to the design and operation of \gaia's
CCDs. Figure~\ref{fig:ccd_diagram} presents a diagram of one such CCD, with
various features labelled.
The images of stars travel from left to right, with the AL and AC directions
indicated at the top left. These are aligned with the `field angles'
$\eta$ and $\zeta$ that represent angular coordinates in the Field
of View Reference System for each telescope (the AL direction is reversed
relative to $\eta$). The field angles are defined in \citet{LL:BAS-003} and
play a fundamental role in the astrometric solution \citep{Lindegren2012}.
The CCD image section spans 1966 light sensitive pixels in the AC direction;
these are referred to as pixel columns and are indexed by the $\mu$
coordinate which ranges from 14 to 1979 inclusive\footnote{There are 14 prescan
pixels lying at $\mu = 0$ to $13$. These play no role in this paper.}.
In the AL direction there are 4500 pixel rows, referred to as TDI
lines.
These are indexed by the $\tau$ coordinate, which ranges from 1 to 4500. 
Both $\mu$ and $\tau$ are continuous variables, with pixel centres lying at
whole integer coordinates.
TDI lines 1, 2, 5, 6, 9 and 10 are masked and are not light sensitive. During
integration, charge is transferred in the parallel direction along pixel columns
from $\tau = 4500$ to $1$ at a fixed rate of 0.9828 milliseconds per TDI line,
for a total crossing time of \textasciitilde$4.42$ seconds.
Individual pixels measure $10\times30$ microns in the AL $\times$ AC
directions, with a nominal plate scale of $58.9\times176.8$ milliarcseconds
(mas).
Finally, the pixel columns and TDI lines within each CCD are not
perfectly aligned with the field angles.

At several locations in the AL direction there are electronic barriers referred
to as CCD `gates'. These are located between consecutive TDI lines and
are activated during the transit of a bright star to temporarily hold back the
transferred charge. This has the effect of reducing the effective CCD area and
corresponding integration time, thus reducing the signal level of the resulting
image and extending the magnitude range for bright sources. The eight gates
routinely in use (including no gate) are listed in Table~\ref{tab:ccdgates}.
Each gate has a corresponding `fiducial line' to which the observation
times of sources are referred. These lie at the mean $\tau$ coordinate of the
light sensitive TDI lines that form the gate, and are denoted $\tau_\text{F}$.
Note that 1D observations are always ungated (although see
footnote~\ref{fn:gated1d}).

\begin{figure}
\resizebox{\hsize}{!}{\includegraphics[trim={0 1cm 0 3cm},clip]{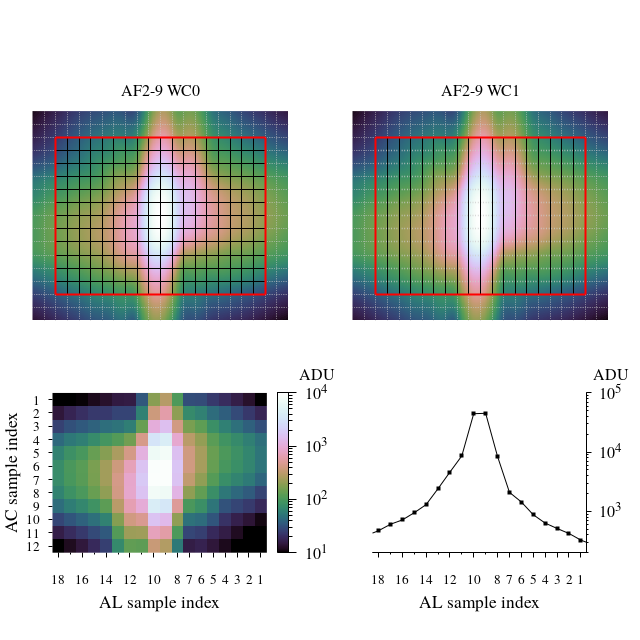}}
\caption{Example window geometry and sampling strategy for
a simulated $G\approx13$ source in window class 0 (WC0; upper left) and WC1
(upper right). These particular configurations apply to sources with onboard
estimated $G<13$ (WC0) and $13<G<16$ (WC1), in CCD strips AF2-9. Faint grey
lines mark the boundaries between individual samples. The red boxes mark the
extent of the window region, with the internal black lines indicating the
binning of individual samples. The lower panels depict the resulting downlinked
2D (left) and 1D (right) observations.}
\label{fig:window_diagram}
\end{figure}

In Fig.~\ref{fig:window_diagram} we depict the window geometry and
sampling strategy that define the 2D and 1D observations of sources, for a
particular subset of the data. These vary with CCD strip and onboard estimated
$G$ magnitude, and a complete description is given in Table 1 of \paperI.
However, in general each AL sample in a 1D observation is formed by the on-chip
binning of 12 AC pixels during readout, and thus encloses the same fraction of
the source flux as an equivalent 2D observation. Note that while each sample has
a unique $\mu$ coordinate, the use of TDI mode means there is no similar
association with the $\tau$ coordinate.
The PSF and LSF are calibrated independently using the 2D and 1D
observations, respectively, and therefore there is no guarantee that the LSF
equals the marginalised PSF\footnote{They are in fact expected to be
different, due at least to the use of different magnitude ranges for the
calibrating observations (resulting in different signal level dependent
effects), and to the varying AC source location in the window, a dependence
that is not yet incorporated in the LSF model (see Sect.~\ref{sec:srcAcLoc}).},
or that image parameters obtained with the PSF equal those obtained with the LSF
applied to a marginalised 2D window.

The angular velocities of stars in field angle coordinates are denoted
$\dot{\eta}$ and $\dot{\zeta}$. These are obtained from the satellite
attitude calibration, which is solved prior to the PLSF either by
AGIS or internal bootstrapping  (section 3.4.5.4 in the forthcoming DR4 official
documentation, Casta{\~n}eda et al. 2026 in prep.).
 $\dot{\eta}$ and $\dot{\zeta}$ vary with
position in the focal plane and between the two telescopes; while they vary in
time in response to the attitude rate, they are assumed to be constant for the
duration of a single CCD observation.
The equivalent AL and AC linear velocities on the CCD, denoted $\dot{\tau}$ and
$\dot{\mu}$, can be obtained from $\dot{\eta}$ and $\dot{\zeta}$ on division by
the AL and AC angular pixel scales $\alPixScale$ and
$\acPixScale$:
\begin{align}
\begin{split}
\label{eqn:alAcRates}
\dot{\tau} = & \, \dot{\eta} / \alPixScale \\
\dot{\mu} = & \, \dot{\zeta} / \acPixScale
\end{split}
\end{align}
While the nominal value of $\acPixScale=176.8$ mas\,pix$^{-1}$ is accurate
enough for our purposes, the sensitivity to $\alPixScale$ is much greater and a
calibrated in-flight value must be used. The algorithm used to estimate
$\alPixScale$ is presented in App.~\ref{app:al_pix_scale}.
The linear velocity of the charge packet in the AL and AC directions is constant
and is denoted $(\dot{\tau}_0, \dot{\mu}_0)$, where $\dot{\tau}_0 =
-0.0009828^{-1} \! \approx \! -1017.501$ pix\,sec$^{-1}$ is the (fixed) charge
transfer rate. Although the charge is transferred exclusively along the pixel columns,
it has a nonzero motion in the AC direction due to small rotations of the CCDs and
minor projection effects. This is quantified by $\dot{\mu}_0$, referred to
as the `native AC rate' elsewhere in \gaia documentation. $\dot{\mu}_0$ varies
per CCD and FOV but is otherwise constant. The algorithm used to estimate
$\dot{\mu}_0$ is presented in App.~\ref{app:native_ac_rate}.
Note that calibrated values of $\dot{\mu}_0$ and $\alPixScale$ were not
required prior to the modelling introduced in this paper.
With these definitions, the relative velocity of stellar images and the charge
packet is given by $(\dot{\tau} - \dot{\tau}_0, \dot{\mu} - \dot{\mu}_0)$, with
the total displacement in pixels during the exposure obtained on multiplication
by the exposure time for the corresponding CCD gate, denoted $t_{\mathrm{exp}}$.
\section{Drift-scan mode with precession \label{sec:driftscan}}
The great majority of astronomical telescopes and imaging systems operate as
framing cameras in point-and-stare mode, where the telescope is held stationary
or tracked to e.g. compensate for the rotation of the Earth, such that the stars
or other objects being imaged remain in a fixed position in the detector during
the exposure. This allows the received signal to accumulate in individual
pixels.
In contrast, \gaia spins continuously during operation such that the images of
stars drift smoothly across the focal plane as they are being observed,
travelling along the CCDs in almost exactly the direction of the pixel
columns at an almost constant rate.
At the same time, the accumulating charge is moved through the CCDs at a fixed
rate that matches the expected average drift rate of stellar images,
with the \textasciitilde$4.42$ second crossing time for individual CCDs placing
a fixed upper limit on the exposure time for all sources.
The combination of a drift-scanning telescope with CCDs operating in TDI mode
offers certain advantages over point-and-stare imaging. In particular, many
CCD level instrument calibrations collapse from two to one dimension, as the
variation in the direction along scan is marginalised. This includes the CCD
response, flatfield, background, and geometric calibration. Images are captured
as continuous strips of essentially arbitrary length in the direction along
scan, which is a useful strategy for survey instruments. For example, this
technique is employed by the HiRISE camera on Mars Reconnaissance Orbiter
\citep{hirise} and the Lunar Reconnaissance Orbiter Camera \citep{lroc}, both of
which are used to image long continuous swathes of terrain.
Drift-scan mode has also been used to great success in other astronomical
surveys such as the Sloan Digital Sky Survey \citep{1998AJ....116.3040G}.
\subsection{Stellar image motion \label{sec:alAcRate}}
However, the particular drift-scan strategy implemented by \gaia introduces some
complications in the PSF modelling. As it spins, \gaia's rotation axis precesses
at a rate of around $4^{\circ}$ per day relative to the stars, allowing
the whole sky to be observed over a period of around 63 days. The evolution of
the spacecraft pointing is known as the \gaia `scanning law'
\citep[see][Section 5.2]{2016A&A...595A...1G}.
The precessional motion induces a periodic across-scan motion of stellar images
as they transit the focal plane, which causes the integrated images to be
smeared out in the across-scan direction. The across scan
motion varies sinusoidally in time with a nominal period of six hours (one
revolution) and an amplitude of 173~mas\,sec$^{-1}$, which, given the
\textasciitilde$4.42$ second integration time and nominal across-scan pixel
scale of $\acPixScale=176.8$ mas\,pix$^{-1}$, implies a maximum smearing of around 4.3
pixels, with the majority of observations smeared across-scan by at least
\textasciitilde$3$ pixels.
This is a major disturbance in the PSF that must be incorporated into the
modelling.
In \paperI the smearing effect was modelled empirically, by introducing a
dependence of the basis component amplitudes on the across-scan rate (see
Section \ref{sec:dr3psf} in this paper). This was only partially successful, and
major systematic errors remained. As it was noted at the time, this was mainly
due to the along-scan rate of stellar images deviating systematically from the
charge transfer rate, leading to a shearing effect on the PSF that was not
reproduced by the model.

In fact, small but significant differences between the AL image rate and the
charge transfer rate are unavoidable, and arise due to several effects.
First, the two telescopes have slightly different focal lengths
\citep[\textasciitilde$34.9708$m for FOV1 and \textasciitilde$34.9688$m for
FOV2, see figure 4.24 in ][]{2022gdr3.reptE...4H},
so the projected images of stars move at slightly different linear speeds on the
CCDs even at the same angular rate; the difference is around $0.058$ pix\,sec$^{-1}$,
which is equivalent to \textasciitilde$3.4$ mas\,sec$^{-1}$.
The spin rate of the satellite is adjusted so that the average of the two FOVs
matches the charge transfer rate.
Variations in the along-scan pixel scale $\alPixScale$ across the focal plane
mean that no single value is appropriate for either telescope anyway.
Second, the scanning law itself induces a small sinusoidal variation in
$\dot{\eta}$ that has a one revolution period and an amplitude of up
to \textasciitilde1.1 mas\,sec$^{-1}$. This is analogous to the better-known
$\dot{\zeta}$ modulation, but out of phase by $\pi/2$ and with a smaller
amplitude \citep[see][]{LL:LL-056} that varies strongly depending on the CCD
row, being largest (and of opposite sign) in rows 1 and 7 and almost zero in row
4. The $\dot{\eta}$ variation is due to precession-induced field rotation and
not, for example, due to changes in the spin rate of the satellite. The precession
rate varies over the year due to the ellipticity of Earth's orbit and the need
to keep the sun at a constant aspect angle. This induces a small annual
modulation in the amplitude of the $\dot{\eta}$ variation.

\begin{figure}
\resizebox{\hsize}{!}{\includegraphics{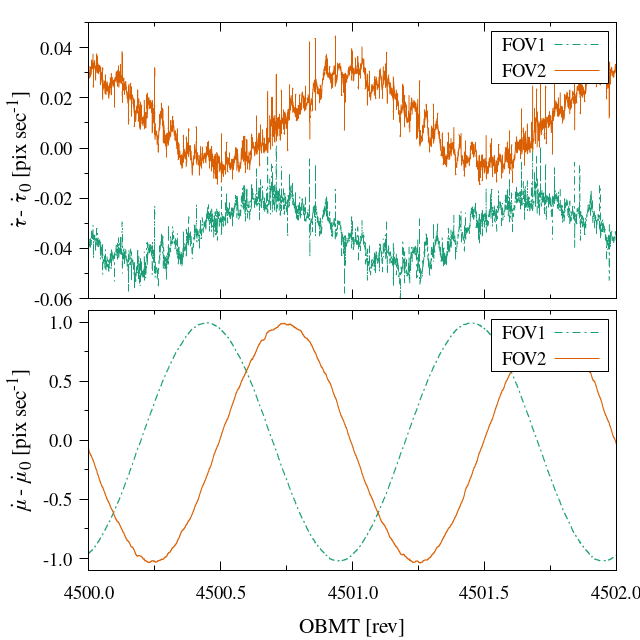}}
\caption{Example of the observed AL (upper) and AC (lower) stellar
image drift rates relative to the transferring charge over a two revolution (12
hour) period, for both FOVs and measured at the centre of a CCD in row 1.
}
\label{fig:alAcRateExample}
\end{figure}

\begin{figure*}
\centering
\includegraphics[width=1.0\textwidth]{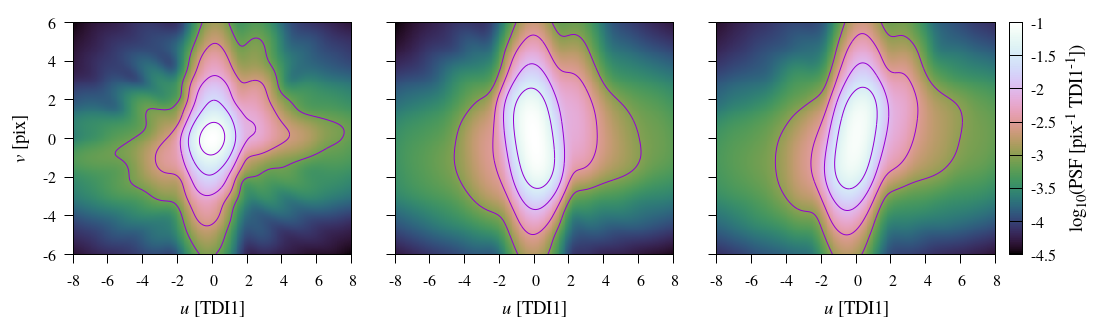}
\caption{Effects of AL and AC source motion on the PSF. These plots depict the
integrated effective PSF for three different regimes of $\dot{\tau}$ and
$\dot{\mu}$.
The $u$ and $v$ coordinates are relative to the PSF origin, as explained at the
start of Sect.~\ref{sec:dr3psf}.
These plots have been generated using the calibrated
PSF model presented later in this paper, and correspond to the FOV1 PSF for
NOGATE observations in the ROW2 AF4 device. In the left panel
$\dot{\tau}=\dot{\tau}_0$ and $\dot{\mu}=\dot{\mu}_0$ such that the stellar
image motion is perfectly matched to the charge transfer and no smearing occurs
in either dimension. In both the centre and right panels
$\dot{\tau}-\dot{\tau}_0=0.226$ pix\,sec$^{-1}$, such that the stellar image
lags 1 pixel behind the charge in the AL direction during the 4.42 second
exposure. In the centre and right panels $\dot{\mu}-\dot{\mu}_0=0.974$ and
$-0.974$ pix\,sec$^{-1}$ respectively, such that the stellar image moves
$\pm4.3$ pixels in the AC direction, orthogonally to the charge transfer. Note that
the $\dot{\tau}$ value is about 20 times larger than what is routinely observed
in the real data, in order to make the impact on the PSF more obvious for the
plots.
Throughout this paper we make use of the cubehelix colour scheme introduced in
\citet{2011BASI...39..289G}.}
\label{fig:alAcSmearing}
\end{figure*}

Finally, both $\dot{\eta}$ and $\dot{\zeta}$ are affected by frequent
disturbances in the attitude from a variety of phenomena, including
thermomechanical micro-clanks, micro-meteroid impacts, and fuel movements in the
propellant tanks. The onboard attitude control system detects and corrects
these, but this inevitably leads to short periods where the attitude rate is
compromised.
In Fig.~\ref{fig:alAcRateExample} we show an example of the observed variation
in $\dot{\tau}$ and $\dot{\mu}$ in the two fields of view over a period of two
revolutions, which is a combination of all of these effects.
Any offset from zero means that the stellar image moves relative to the
integrating charge, and the resulting integrated image is smeared out along a
line determined by the motion in each dimension.
In Fig.~\ref{fig:alAcSmearing} we demonstrate the apparent shearing effect that
this has on the integrated effective PSF, and its dependence on the magnitude
and sign of the motion in each of the AL and AC directions.
Note that in this work we assume the AC component of the smearing has no effect
on the LSF, and that it is sensitive only to $\dot{\tau}-\dot{\tau}_0$. This is
not entirely true: large AC smearing causes minor additional flux loss from the
(12 pixel wide) window which may have a very minor effect on the LSF shape,
and signal-level dependent effects will vary with the AC smearing since
this has a significant impact on the pixel occupancy in the core of the
charge packet. Both of these effects are very weak, and may be addressed in
future PLSF model developments.
\subsection{Along-scan variations and the `corner effect' \label{sec:corner}}
\begin{figure}
\resizebox{\hsize}{!}{\includegraphics{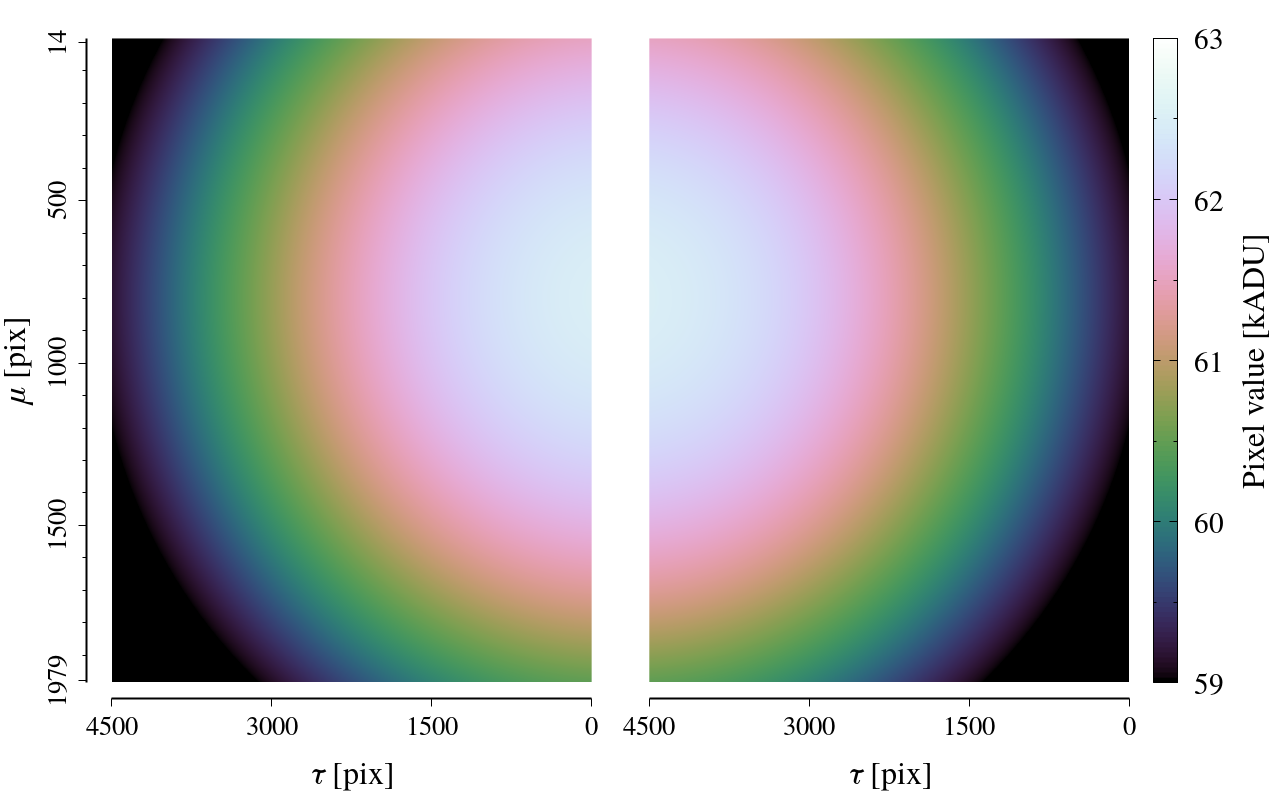}}
\caption{Illustrative example of (artificial) flatfield images for
a pair of CCDs manufactured from the same circular silicon wafer. This
figure approximately reproduces the effect seen in industrial pre-launch
flatfield data from \gaia's CCDs at 900nm, and which cannot be published.
The CCD on the left is of TYPE-01 and the CCD on the right is of TYPE-02.}
\label{fig:flatfields}
\end{figure}

\begin{figure}
\resizebox{\hsize}{!}{\includegraphics{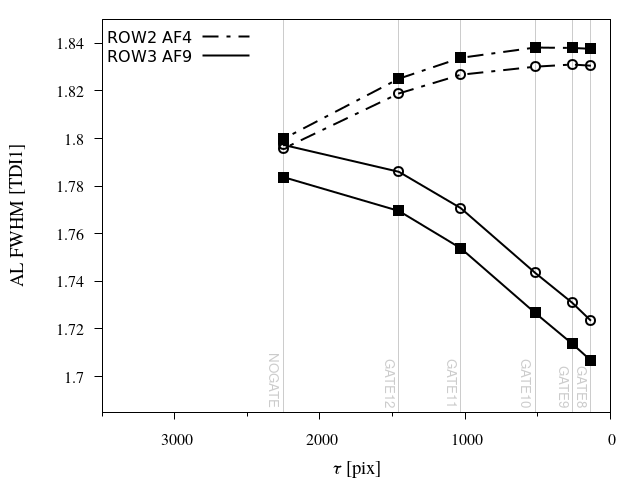}}
\caption{Effects of AL variations in the CCD response on the PSF, and the
dependence on CCD gate and device type. The AL full-width half-maximum (FWHM)
of the integrated effective PSF is plotted as a function of CCD gate, for two
devices of TYPE-01 (ROW2 AF4, dot-dashed line) and TYPE-02 (ROW3 AF9, solid
line). Each CCD gate spans a range in $\tau$, with the points plotted at the
fiducial line positions. The solid squares correspond to FOV1 and the open
circles to FOV2. These figures have been generated using the calibrated PSF
model rather than directly from the observations.}
\label{fig:alFwhmVsGate}
\end{figure}

In addition to the major smearing effect, the motion of the stellar image
relative to the transferring charge packet induces an additional modulation in
the resulting integrated image that is more subtle.
When the stellar image is significantly trailed, different samples in the image
are exposed over slightly different ranges of $\tau$, and will
therefore have a weak dependence on any spatial variations in the instantaneous
effective PSF in the AL direction within the CCD.
While purely optical variations in the PSF are generally insignificant over the
AL extent of a single CCD, it turns out that the CCDs used by \gaia have a 
systematic spatial variation in the pixel response nonuniformity that introduces
significant AL variation in the electronic component of the PSF,
originating in the detector itself.
This is ultimately caused by a characteristic circular pattern of thickness
variation arising from the way each device was manufactured
from either the left or right half of a circular silicon wafer. Thinner regions
have a lower quantum efficiency at red wavelengths, resulting in a lower overall
response. This is referred to as the `corner effect' elsewhere in \gaia
documentation,
due to the response being weakest towards the corners where the CCDs are
thinnest.
This is depicted in Fig.~\ref{fig:flatfields}.
All the CCDs naturally fall into two
types depending on whether they were manufactured from the left (TYPE-01) or
right (TYPE-02) half of the circular wafer. Across the SM and AF part of the
focal plane there are 35 TYPE-01 devices and 41 TYPE-02 devices, as listed in
App.~\ref{app:ccd_types}. There are also ten pairs of twin devices
that have been manufactured from each half of the same wafer; these can
sometimes have similar properties. Regardless of type, the devices are always
orientated in the focal plane with the serial register on the right.
Thinner regions of the CCD have a lower quantum efficiency at red wavelengths,
which results in a (polychromatic) PSF that is narrower due to diffraction
effects. There may also be some contribution from reduced charge diffusion due
to the shorter distance travelled by photoelectrons to reach the electrodes. The
main observational consequence of this is that the AL width of the PSF varies as
a function of $\tau$. This is clearly visible when inspecting the PSF for
different CCD gates, since each gate samples a different range of $\tau$ and is
subject to a different average CCD response. In Fig.~\ref{fig:alFwhmVsGate} we
present the PSF along-scan full-width half-maximum (AL FWHM) as a function of
CCD gate for two devices of different type. Within each device the behaviour is
very similar between the FOVs despite the optical PSFs being very different.
However, the devices diverge significantly in their gate-dependent behaviour,
since in TYPE-01 devices the pixel response plateaus close to the serial
register, so the short gates have similar PSFs, whereas in TYPE-02 devices the
pixel response decreases rapidly close to the serial register and successively
shorter gates have narrower PSFs.

\begin{figure}
\resizebox{\hsize}{!}{\includegraphics{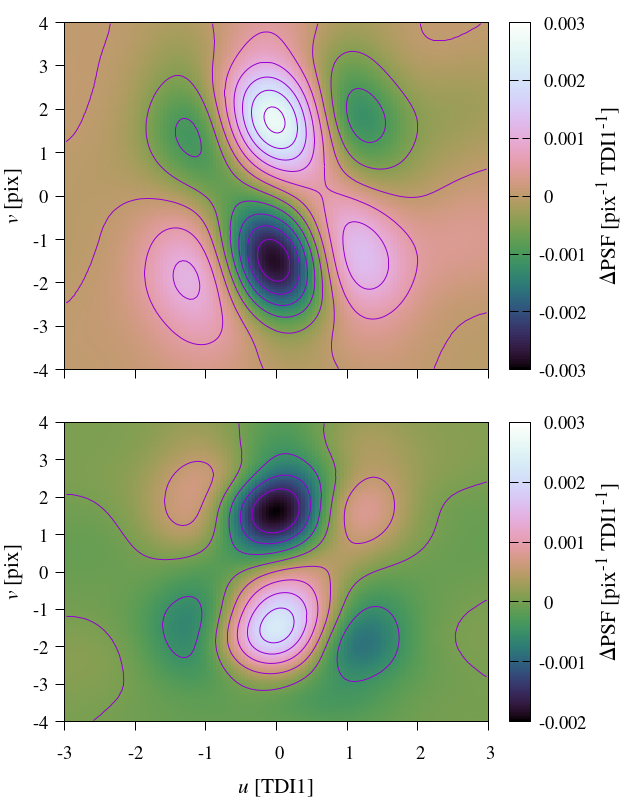}}
\caption{Effects of AL variations in the CCD response on the PSF, and the
dependence on source motion. Each plot shows the difference between two
integrated effective PSFs for $\dot{\tau}=\dot{\tau}_0$ and $\dot{\mu}-\dot{\mu}_0=\pm1.0$
pix\,sec$^{-1}$,
such that the PSFs are smeared exclusively in the AC direction.
The PSFs are for NOGATE, so the entire $\tau$ range of the CCD is covered.
Due to the AC motion, the lower half of each plot shows the approximate
difference between the instantaneous effective PSF at high $\tau$ minus the
instantaneous effective PSF at low $\tau$, and vice-versa in the upper half. The
lower panel is a TYPE-01 device (ROW2 AF4) and the upper panel is a TYPE-02
device (ROW3 AF9).}
\label{fig:tdiLineDependenceSystematics}
\end{figure}

The consequence of this is that whenever the stellar image is trailed, the
integrated effective PSF becomes sensitive to $\tau$ variations in the
instantaneous effective PSF, which are dominated by the corner effect described
above\footnote{Although the corner effect dominates, there are other isolated
anomalies in certain devices, and a pair of devices (AF5 and AF8 in ROW2) that
both have a low outlying AL FWHM due to being unusually thin. This is visible in
figure 12 in \paperI. What was not recognised at the time is that they are twin
devices manufactured from the same wafer (see App.~\ref{app:ccd_types}).}.
This manifests as a characteristic modulation in the PSF that is opposite in
sign between the two device types. This is demonstrated in
Fig.~\ref{fig:tdiLineDependenceSystematics}, in which the structure is
caused entirely by the dependence of the instantaneous effective PSF on $\tau$.
Note that since $\dot{\tau} - \dot{\tau}_0 \ll \dot{\mu} - \dot{\mu}_0$ this
phenomenon has a much weaker impact on the LSF and is not observed in the 1D
observations.
This dependence is different to the other major PLSF dependences on e.g.~source
colour or $\mu$ location, since observations are produced by integration over a
range of $\tau$ and do not sample a single value of it.
It is also somewhat unexpected for \gaia, since it contravenes the idea that
drift-scan mode marginalises instrumental variations in the along-scan
direction. However, it is present in the data and must be accounted for in the
PSF modelling.
\section{Brief review of the EDR3 PLSF models \label{sec:dr3plsf}}
In this section we present a brief review of the PSF and LSF models implemented
for \gaia EDR3, in order to introduce some notation and provide context for the
improvements presented in this paper. Further details can be found in section 3
of \paperI, although note that some of the nomenclature has been updated for
the present paper.
\subsection{The EDR3 PSF model \label{sec:dr3psf}}
The PSF model $P$ predicts the fractional charge contained in a sample located
at $(u,v)$ relative to the PSF origin.
The $u$ and $v$ axes are aligned with the AL and AC directions,
and oriented such that $v$ increases in the direction of increasing $\mu$, and
$u$ increases in the direction of increasing observation
time\footnote{A sample with larger $u$ reaches the serial register later.}.
Note that $u$ is formally in units of the TDI period, denoted TDI1, and $v$ is
in units of pixels. $P$ is dependent also on the effective
wavenumber\footnote{\label{fn:nuEff} The \nueff value for each source
is computed by PhotPipe from the mean BP and RP spectra, and is by necessity
taken from the previous data release, in this case DR3. For DR4 processing, the
\nueff value actually used in the PLSF calibration is a weighted combination of
the measured value and a prior, as explained in 
Hern{\'a}ndez et al.~(2026, in prep.). See also
\citet{LL:JDB-028}.} $\nu_{\text{eff}}$, the AC position in the 
device\footnote{\label{fn:winAcPos} When processing \gaia observations, for both 
PLSF calibration and window modelling purposes, we use
the $\mu$ value of the window centre rather than that of the source contained
within it, since this may not be known in advance. This may differ from the
$\mu$ of the source by up to a few pixels, which has no significant effect on
the PLSF.} $\mu$, and the AC rate $\dot{\mu}$. No correction for the native AC
rate $\dot{\mu}_0$ was required.
Interpreted as a probability density function conditional on the
$\nu_{\text{eff}},\mu, $ and $\dot{\mu}$ variables, it can be written
$P(u,v|\nu_{\text{eff}},\mu,\dot{\mu})$.
The PSF model is composed as the linear combination of a 2D mean PSF, denoted
$G_0$, and $N$ weighted basis components, denoted $G_n$, with associated weight
factors $\gsw_n$ and $N=30$ in EDR3.
The full expression for the PSF model is written
\begin{equation}\label{eq:genshapelets}
P(u,v|\nu_{\text{eff}},\mu,\dot{\mu}) = G_0(u, v) + \sum_{n=1}^{N}
\gsw_n(\nu_{\text{eff}},\mu,\dot{\mu}) \, G_n(u, v) \,.
\end{equation}
$G_0$ is normalised such that its integral over all $(u,v)$ is 1.0, and it has a
fixed weight of 1.0. The other components are normalised such that their
integrals over all $(u,v)$ are 0.0. This guarantees that the full PSF model is
normalised to 1.0 regardless of the weighting of the components. The basis
components are by construction mutually orthogonal as far as possible (though
see Sect.~\ref{sec:h00}), which improves numerical stability and ensures a
unique solution.
The weight factors $\gsw_n(\nu_{\text{eff}},\mu,\dot{\mu})$ are represented as
multidimensional splines in the dependent variables (using the implementation
described in \citet{2007ASSL..350.....V} appendix B, extended to multiple
dimensions), with appropriately configured spline orders and knot sequences in
each dimension. The full set of spline coefficients over all $N$ basis
components defines the complete set of parameters of the PSF model.

The $G_0$ and $G_n$ functions are constructed as the linear combination of outer
products of 1D basis functions in the AL and AC direction, denoted $\al_i$ and
$\ac_j$, where $i$ and $j$ index functions of different order in each
dimension\footnote{In \paperI the symbol $H$ was used to denote both $\al$ and
$\ac$; in the present paper these functions need to be distinguished.}, so that
\begin{align}\label{eqn:gn_exc_smear}
\begin{split}
G_0(u, v) & = \al_0(u)\ac_0(v) + \sum_{i+j>0}^{\substack{i=I-1\\j=J-1}} 
 \, \alpha^0_{ij} \al_i(u) \, \ac_j(v) \\
G_n(u,v) & = \sum_{i+j>0}^{\substack{i=I-1\\j=J-1}} \, \alpha^n_{ij} \al_i(u) \,
\ac_j(v) \qquad \textrm{for } n>0,
\end{split}
\end{align}
with $I=21$ and $J=21$ in DR3.
Each product $\al_i(u) \ac_j(v)$ is referred to as a `pseudo shapelet'
\citep[to distinguish it from the shapelets model presented
in][]{Refregier_2003} and the full model was named `compound shapelets'
in \paperI. The (constant) matrix $\alpha^n_{ij}$ defines the construction of
$G_0$ and $G_n$ from linear combinations of the 1D functions of order $i$ and
$j$.
The functions $\al$ and $\ac$ are physically motivated, and are
derived in advance from simulations of \gaia's optical system as described in
\citet{LL:LL-084}. They account for pixelisation in each dimension and are
represented using the S-spline function presented in \citet[][section
3.3.5]{2022gdr3.reptE...3C}. The S-spline is formulated to satisfy the
shift-invariant-sum requirement, which expresses the conservation of flux under
sub-pixel shifts of the source, and is important for the photometry.
The matrix $\alpha$ is then obtained by training on real \gaia observations
using the principal components analysis algorithm described in
\citet{LL:LL-090}.
This produces $G_0$ and $G_n$ functions that are tailored to the in-flight PSF,
and which for a limited number ($N=30$) of components provides the smallest
expected RMS error among all linear models.

Finally, while the true PSF is strictly positive everywhere, this property is
not enforced in our model either by construction or by the calibration of the
parameters. As such, it represents a noisy estimate of the true PSF and may be
negative at locations where the true PSF is very small, or in poorly constrained
regions of the parameter space. It is expected that users of the model within
DPAC handle these (rare) situations appropriately.
\subsection{The EDR3 LSF model \label{sec:dr3lsf}}
The LSF model is denoted $L$, and is composed as the linear combination of a 1D
mean LSF, denoted $H_0$, and $N$ weighted basis components, denoted $H_n$, with
associated weight factors $\gsw_n$ and $N=25$ in EDR3. It is analogous to the
PSF model but spans only the AL dimension. As such, it is a function only of the
$u$ coordinate, and the dependent variables do not include the AC rate.
The full expression for the LSF model is written
\begin{equation} \label{eqn:edr3_lsf}
L(u|\nu_{\text{eff}},\mu) = H_0(u) + \sum_{n=1}^{N} \gsw_n(\nu_{\text{eff}},\mu)
\, H_n(u) \,.
\end{equation}
In EDR3 the 1D functions $H$ are identical to the $\al$ functions derived from
simulations, i.e.\ without training them on real \gaia observations, so $H_0
\equiv f_0$, $H_1 \equiv f_1$ etc. This was found to be sufficient for EDR3.
However, in principle they could be composed of linear combinations of $\al$ in
order to tailor them to the in-flight LSF modes, so that e.g.\
\begin{align}\label{eqn:hn}
\begin{split}
H_0(u) & = \al_0(u) + \sum_{i=1}^{I-1} \, \alpha^0_{i} \al_i(u) \\
H_n(u) & = \sum_{i=1}^{I-1} \, \alpha^n_{i} \al_i(u) \qquad \textrm{for } n>0.
\end{split}
\end{align}
This was not done in EDR3 but is in DR4, so we prefer to use the
$H$ symbol to generalise the model.
\section{The DR4 PLSF model \label{sec:dr4psf}}
The PLSF models used in DR4 have undergone several major advances relative to
the EDR3 models presented in the previous section. Here we present a complete
derivation of the updated models, the calibration algorithms used in the
operational processing and the PLSF configurations chosen for modelling
different subsets of the data.
Note that we have also updated and improved the LSF and PSF basis components
$H$ and $G$, the 1D functions $\al$ and $\ac$ used to compose them, and the
fundamental S-spline function used to interpolate $\al$ and $\ac$. These are
somewhat secondary to the PLSF models themselves, and their presentation is
deferred to Casta{\~n}eda et al.~(2026 in prep, section 3.3.5).
\subsection{Derivation of the DR4 PLSF models}
In this section we describe three major advances in the PLSF model, which are
the switch from a limited (AC-only) empirical model of the source motion to a
complete (AL and AC) analytic model (Sect.~\ref{sec:analyticAlAcDeriv}), the
incorporation of a crucial dependence on TDI line number
(Sect.~\ref{sec:tdiLineDepDeriv}), and the introduction of constraints between
the PLSF origin and the geometric instrument calibration (Sect.~\ref{sec:h00}).
The complete derivation is presented in terms of the PSF model; the updated LSF
model, to which only a subset of these effects apply, is presented briefly in
Sect.~\ref{sec:dr4Lsf}.
\subsubsection{Analytic modelling of the AL and AC source motion
\label{sec:analyticAlAcDeriv}}
The EDR3 PSF model included a limited modelling of the source motion through the
dependence of the basis component amplitudes $\gsw_n$ on the $\dot{\mu}$
parameter, which was in turn parameterised using a third order polynomial
(see table 3 in \paperI) with coefficients calibrated empirically by fitting to
observations.
Unlike the other empirically calibrated dependences (on $\nu_{\text{eff}}$
and $\mu$) the effects of AL and AC source motion can be modelled from first
principles as a simple smearing of the integrated PSF along the direction of
motion. The first step is to remove $\dot{\mu}$ from the dependent variables in
Eq.~\ref{eq:genshapelets} to obtain
\begin{equation} \label{eqn:psfNoMuDot}
P(u,v|\nu_{\text{eff}},\mu) = G_0(u, v) + \sum_{n=1}^{N}
\gsw_n(\nu_{\text{eff}},\mu) \, G_n(u, v) \,.
\end{equation}
By substituting the expressions for $G_0$ and $G_n$ from
Eq.~\ref{eqn:gn_exc_smear} and specifying $\gsw_0(\nu_{\text{eff}},\mu)=1$,
$\alpha^0_{00}=1$ and $\alpha^{1+}_{00}=0$ we obtain an expression directly
in terms of the 1D functions $\al_i$ and $\ac_j$:
\begin{equation}
P(u,v|\nu_{\text{eff}},\mu) = \sum_{n=0}^{N} \gsw_n(\nu_{\text{eff}},\mu)
\sum_{i=0}^{I-1} \sum_{j=0}^{J-1} \alpha^n_{ij} \al_i(u) \ac_j(v) \,,
\end{equation}
where $I=25$ and $J=25$ in DR4.
We now introduce a new variable $\tau$ that represents the TDI line number
during the exposure, and introduce correction terms to $u$ and $v$ that
represent the displacement of the stellar image from the charge image as a
function of $\tau$ and the field angle rates $\dot{\eta}$ and $\dot{\zeta}$, to
obtain an expression for the instantaneous effective PSF at TDI line
number $\tau$:
\begin{multline} \label{eqn:instEffPsfExcTdi}
P(u,v|\nu_{\text{eff}},\mu,\dot{\eta},\dot{\zeta},\tau) = \sum_{n=0}^{N}
\gsw_n(\nu_{\text{eff}},\mu) \, \times \\
\sum_{i=0}^{I-1} \sum_{j=0}^{J-1} \alpha^n_{ij} \al_i(u + \Delta
u(\tau,\dot{\eta})) \, \ac_j(v + \Delta v(\tau,\dot{\zeta})) \,.
\end{multline}
The displacement terms $\Delta u(\tau,\dot{\eta})$ and $\Delta
v(\tau,\dot{\zeta})$ have the following forms, making use of the expressions in
Eq.~\ref{eqn:alAcRates} and the exposure time $t_{\mathrm{exp}}$:
\begin{align}
\Delta u(\tau,\dot{\eta}) & = -\left(\frac{\tau_\text{F} - \tau}{\Delta
\tau}\right) \, (\dot{\tau} - \dot{\tau}_0) \, t_{\mathrm{exp}} \nonumber \\
                          & = -\left(\frac{\tau_\text{F} - \tau}{\Delta
                          \tau}\right) \, \left(\frac{\dot{\eta}}{\alPixScale} -
                          \dot{\tau}_0\right) t_{\mathrm{exp}}
                          \label{eqn:deltaU}
                          \\
\Delta v(\tau,\dot{\zeta}) & = \left(\frac{\tau_\text{F} - \tau}{\Delta \tau}\right)
\, (\dot{\mu} - \dot{\mu}_0) \, t_{\mathrm{exp}} \nonumber \\
                           & = \left(\frac{\tau_\text{F} - \tau}{\Delta \tau}\right)
                           \, \left(\frac{\dot{\zeta}}{\acPixScale} -
                           \dot{\mu}_0\right) t_{\mathrm{exp}}
\end{align}
where the sign change in Eq.~\ref{eqn:deltaU} reflects the fact that the AL
direction is opposite to the direction of increasing $\eta$.
In these expressions $\Delta \tau = \tau_{\text{min}} - \tau_{\text{max}}$,
where $\tau_{\text{min}}$ and $\tau_{\text{max}}$ are the minimum and maximum
value of the TDI line number. The value of $\tau_\text{max}$ depends on the CCD
gate used to observe the source, with longer gates having a larger value.
In contrast, $\tau_\text{min}$ has the same value of 1 for all gates,
corresponding to the final TDI line before the serial register is reached. In
adopting this value we have ignored the fact that TDI lines 1, 2, 5, 6, 9 and 10
are masked and are not light sensitive; as the AL/AC motion is very small over
such a short extent the impact of this approximation is insignificant
\footnote{An alternative value of $\tau_\text{min}=6$ would give a $\Delta \tau$
consistent with the exposure time, though this would still be an approximation
for the integral over $\tau$. The particular choice makes no significant
difference.}.
The quantity $\tau_\text{F}$ is the TDI line number of the CCD gate
fiducial line, which corresponds to the mean of the light sensitive TDI lines and is slightly
more than half of $\tau_\text{max}$ due to the 6 masked TDI lines. $\tau_\text{F}$ is
the reference coordinate used in the astrometric solution, and for the purposes
of modelling the PSF we define the displacement of the stellar image away from
the integrating charge image to be zero at $\tau_\text{F}$. The values of
$\tau_{\text{max}}$, $\tau_\text{F}$ and $t_{\mathrm{exp}}$ for all gates routinely in
use is listed in Table~\ref{tab:ccdgates}.
\begin{table}
\caption{\label{tab:ccdgates}CCD gate constants.}
\centering
\begin{tabular}{lrrrD{,}{}{0}}
\toprule
CCD gate & $\tau_{\text{max}}$ & $\tau_\text{F}$ & $t_{\text{exp}}$ [ms] & \text{Order} \\
\midrule 
NOGATE & 4500 & 2253.497 & 4416.70 & 3, \\
GATE12 & 2906 & 1456.496 & 2850.12 & 3,\tablefootmark{a}  \\
GATE11 & 2054 & 1030.494 & 2012.77 & 2, \\
GATE10 & 1030 & 518.488 & 1006.39 & 2, \\
GATE9 & 518 & 262.477 & 503.19 & 1, \\
GATE8 & 262 & 134.453 & 251.60 & 1, \\
GATE7 & 134 & 70.406 & 125.80 & 1, \\
GATE4 & 22 & 13.750 & 15.72 & 1, \\
\bottomrule
\end{tabular}
\tablefoot{The final column lists the order of the spline used to model the
$\tau$ dependence, described in Sect.~\ref{sec:config}.\\
\tablefoottext{a}{SM WC1 model uses first order despite using
GATE12}}
\end{table}
The instantaneous effective PSF is never actually observed, and in order to
model the PSF for a particular observation we compute the integrated
effective PSF by marginalising the nuisance parameter $\tau$:
\begin{multline} \label{eqn:intEffPsfExcTdi}
P(u,v|\nu_{\text{eff}},\mu,\dot{\eta},\dot{\zeta}) = 
 \sum_{n=0}^{N} \gsw_n(\nu_{\text{eff}},\mu) \, \times \\
\sum_{i=0}^{I-1} \sum_{j=0}^{J-1} \frac{\alpha^n_{ij}}{\Delta \tau}
\int\limits_{\tau=\tau_{\text{max}}}^{\tau_{\text{min}}}
 \al_i(u + \Delta u(\tau,\dot{\eta})) \, \ac_j(v + \Delta v(\tau,\dot{\zeta})) \,\text{d}\tau
\end{multline}
The integration limits depend on the CCD gate of the observation according to
Table~\ref{tab:ccdgates}. Note that the integral range is reversed, to reflect
the sense of the TDI line coordinate on the device: exposure starts at high TDI
line number and ends at low TDI line number.
We use the optimization $\beta_{ij}(\nu_{\text{eff}},\mu) = \sum_{n=0}^{N}
\gsw_n(\nu_{\text{eff}},\mu) \, \alpha^n_{ij}$ to eliminate the summation over
$n$ since $\gsw_n$ has the same value for all $(u,v)$ samples in the PSF model for a
particular observation. This results in the expression
\begin{multline} \label{eqn:analyticIntEffPsfExcTdi}
P(u,v|\nu_{\text{eff}},\mu,\dot{\eta},\dot{\zeta}) = \\
\sum_{i=0}^{I-1} 
\sum_{j=0}^{J-1} 
\frac{\beta_{ij}(\nu_{\text{eff}},\mu) }{\Delta \tau}
\int\limits_{\tau=\tau_{\text{max}}}^{\tau_{\text{min}}}
\al_i(u + \Delta u(\tau,\dot{\eta})) \, \ac_j(v + \Delta v(\tau,\dot{\zeta}))
\,\text{d}\tau
\end{multline}
Note that the computation can be further optimised by caching and reusing the
sampled values of $\al_i(u + \Delta u(\tau,\dot{\eta}))$ and $\ac_j(v + \Delta
v(\tau,\dot{\zeta}))$ when the $(u,v)$ locations fall on a regular grid, which
occurs in routine processing when a 2D window is being modelled.
The integration is performed numerically using
Gauss-Legendre quadrature, according to which the integral becomes a
weighted sum over $K$ steps in the $\tau$ coordinate,
\begin{multline} \label{eqn:numIntEffPsfExcTdi}
P(u,v|\nu_{\text{eff}},\mu,\dot{\eta},\dot{\zeta}) = \\
\sum_{i=0}^{I-1} 
\sum_{j=0}^{J-1} 
\frac{\beta_{ij}(\nu_{\text{eff}},\mu)}{2} 
\sum_{k=0}^{K-1} w_k 
 \al_i(u + \Delta
u(\tau_k,\dot{\eta})) \, \ac_j(v + \Delta v(\tau_k,\dot{\zeta})) \, ,
\end{multline}
where the leading $\Delta \tau$ factor is absorbed into the weights $w_k$. This
is the equation that is implemented in the \gaia data processing and used to
model the PSF for observations in GATE4 to GATE9, for which the dependence 
of the PSF shape on $\tau$ can be ignored (order = 1 in
Table~\ref{tab:ccdgates}).
We have selected $K=9$ as an optimal tradeoff between numerical accuracy and
execution time; for details of the numerical integration, values of the $w_k$
weight factors and their associated $\tau_k$ coordinates, see appendix
\ref{app:num_int}.
The first derivatives of the PSF model with respect to the $u$ and $v$
parameters are required when fitting the model to an observation; these have the
simple forms
\begin{multline}
\frac{\partial P(u,v|\nu_{\text{eff}},\mu,\dot{\eta},\dot{\zeta})}{\partial u} = \\
\sum_{i=0}^{I-1} 
\sum_{j=0}^{J-1} 
\frac{\beta_{ij}(\nu_{\text{eff}},\mu)}{2} 
\sum_{k=0}^{K-1} w_k 
\al_i^{\prime}(u + \Delta u(\tau_k,\dot{\eta})) \, \ac_j(v + \Delta
v(\tau_k,\dot{\zeta}))
\end{multline}
and
\begin{multline}
\frac{\partial P(u,v|\nu_{\text{eff}},\mu,\dot{\eta},\dot{\zeta})}{\partial v} = \\
\sum_{i=0}^{I-1} 
\sum_{j=0}^{J-1}
\frac{\beta_{ij}(\nu_{\text{eff}},\mu)}{2} 
\sum_{k=0}^{K-1} w_k 
\al_i(u + \Delta u(\tau_k,\dot{\eta})) \, \ac_j^{\prime}(v + \Delta
v(\tau_k,\dot{\zeta})) \, ,
\end{multline}
where the prime denotes the first derivative.
Finally, when the exposure time or either of the $\dot{\eta}$ and $\dot{\zeta}$
terms are very small\footnote{Specifically, when
$|(\dot{\tau} - \dot{\tau}_0) \, t_{\mathrm{exp}}|$ or
$|(\dot{\mu}  - \dot{\mu}_0 ) \, t_{\mathrm{exp}}|$ are below a threshold,
chosen to be $10^{-5}$ pixels.}
the corresponding $\Delta u$ and/or $\Delta v$ terms in
Eq.~\ref{eqn:analyticIntEffPsfExcTdi} are eliminated. These conditions are
met for very short gates or observations whose AL/AC motion is closely matched
to the charge transfer.
In these circumstances the integration is either avoided entirely or can be
performed by analytic integration of whichever of the $\al_i$ or $\ac_j$
functions still retain a dependence on $\tau$.
Recall that these functions are represented using the S-spline described in
Casta{\~n}eda et al.~(2026 in prep, section 3.3.5), which is ultimately based on
b-splines and can be integrated analytically.
\subsubsection{PSF dependence on TDI line number \label{sec:tdiLineDepDeriv}}
As explained in Sect.~\ref{sec:corner}, spatial variations in the PSF
shape within each device are not restricted to the AC direction (parameterised
by $\mu$), and the variation in the AL direction, parameterised by the TDI line
number $\tau$, is also significant. The dependence on $\tau$ is similar in
principle to the dependences on $\mu$ and $\nu_{\text{eff}}$, with the important
difference that individual observations span a range in TDI line number during
exposure rather than a single value.
This can be incorporated into our model by expanding the parameterisation of the
weight factors $\gsw_n$ to include $\tau$, so that they are modified to:
\begin{equation*}
\gsw_n(\nu_{\text{eff}},\mu) \rightarrow \gsw_n(\nu_{\text{eff}},\mu,\tau)\,.
\end{equation*}
Under this approach, the instantaneous effective PSF
(Eq.~\ref{eqn:instEffPsfExcTdi}) is modified to
\begin{multline} \label{eqn:instEffPsfIncTdi}
P(u,v|\nu_{\text{eff}},\mu,\dot{\eta},\dot{\zeta},\tau) = \sum_{n=0}^{N}
\gsw_n(\nu_{\text{eff}},\mu,\tau) \, \times \\
\sum_{i=0}^{I-1} \sum_{j=0}^{J-1} \alpha^n_{ij} \al_i(u + \Delta
u(\tau,\dot{\eta})) \, \ac_j(v + \Delta v(\tau,\dot{\zeta})) \,.
\end{multline}
The expression for the integrated effective PSF
(Eq.~\ref{eqn:intEffPsfExcTdi}) is adjusted to bring the
weight factors inside the integral, since they now depend on $\tau$. Applying
the Gauss-Legendre quadrature scheme, and redefining $\beta$ to include $\tau$
so that $\beta_{ij}(\nu_{\text{eff}},\mu,\tau) = \sum_{n=0}^{N}
\gsw_n(\nu_{\text{eff}},\mu,\tau) \, \alpha^n_{ij}$, we finally obtain the
following expression
\begin{multline} \label{eqn:intEffPsfIncTdi}
P(u,v|\nu_{\text{eff}},\mu,\dot{\eta},\dot{\zeta}) = \\
\sum_{i=0}^{I-1} 
\sum_{j=0}^{J-1} 
\sum_{k=0}^{K-1} w_k 
\frac{\beta_{ij}(\nu_{\text{eff}},\mu,\tau_k)}{2} 
 \al_i(u + \Delta
u(\tau_k,\dot{\eta})) \, \ac_j(v + \Delta v(\tau_k,\dot{\zeta})) \, .
\end{multline}
This is the equation that is implemented in the \gaia data processing and used
to model the PSF for observations in GATE10 to NOGATE, for which the dependence 
of the PSF shape on $\tau$ cannot be ignored (order > 1 in
Table~\ref{tab:ccdgates}).
The first derivatives with respect to the $u$ and $v$ parameters can be formed
in the same way as for Eq.~\ref{eqn:numIntEffPsfExcTdi}, i.e. by replacing the
$\al$ and $\ac$ functions with their first derivatives, respectively.
The need to compute $\beta$ at each of the $K$ numerical integration
steps leads to a modest drop in computational performance relative to
Eq.~\ref{eqn:numIntEffPsfExcTdi}, and the dependence of $\beta$ on $\tau$
means that the numerical integration cannot be avoided in situations where the
$\Delta u$ or $\Delta v$ factors are eliminated.

Observations for which $\Delta u$ and/or $\Delta v$ are close zero, i.e. when 
$|\dot{\tau} - \dot{\tau}_0|$ or $|\dot{\mu} - \dot{\mu}_0|$ are small, offer
little constraint on the $\tau$ dependence in $\beta$. This has some
implications for the calibration. For example, during the first month or so of
science data collection \gaia followed the Ecliptic Pole Scanning Law, in which
the $\dot{\mu}$ distribution of the observations is around four times narrower
than the Nominal Scanning Law. This led to a somewhat underconstrained $\tau$
dependence in the longer gates over revolutions $1078$ to $1192$, which required
some adjustments in the calibration pipeline during the late stages of the data
processing for DR4.
Finally, we note that it would in principle be possible to obtain the model for
a short gate observation by integrating the model for a longer gate over a
restricted range in $\tau$. This was considered but not explored in depth.
\subsubsection{Constraints on the LSF/PSF origin \label{sec:h00}}
In the context of the \gaia global astrometric solution there is a degeneracy
between the PLSF origin and the geometric instrument calibration, the latter
being solved separately to the PLSF as part of the AGIS processing
\citep[see][section 3.4]{Lindegren2012}. In the complete instrument model,
purely optical shifts in the PLSF origin, caused for example by evolving
wavefront tip-tilt or ice contamination, are indistinguishable from physical
displacements of the devices. Breaking this degeneracy is vital to separate the
roles of the different calibrations. This is done by enforcing some constraints
on the origin of the PLSF model, as explained in this section.

Up to this point the PLSF origin has not been explicitly defined, and is by
default coincident with the origin of the 1D functions $\al$ and $\ac$.
Early representations of the \gaia LSF included a translation parameter applied
to the sample location, allowing the LSF model to shift by an arbitrary
amount in the AL direction (see \paperI section 6.7.2).
In the present paper we generalise this idea to the PSF, and introduce $u_0$ and
$v_0$ to denote the shifts of the PSF in the AL and AC direction. To incorporate
these parameters in the PSF model, we first modify Eq.~\ref{eqn:psfNoMuDot}
to
\begin{equation} \label{eqn:fullPsfModelWithShifts}
P(u,v|\nu_{\text{eff}},\mu) = G_0(u-u_0,v-v_0) 
         + \sum_{n=1}^{N} \gsw_n(\nu_{\text{eff}},\mu)
         G_n(u-u_0,v-v_0) \,.
\end{equation}
Before DR4 the $u_0$ and $v_0$ parameters were implicitly fixed at zero, with
small shifts of the PLSF origin handled by appropriate weighting of the basis
components.  This is undesirable for two reasons. First, it requires
the basis components to reproduce pure shifts of the PLSF, which increases the
dimensionality. Second, it precludes the explicit calibration of the PLSF
shifts, which is necessary in order to break the degeneracy with the geometric
calibration. Therefore, in the DR4 PLSF model we treat $u_0$ and $v_0$ as free
parameters and make an explicit calibration of them. Like the other PSF
parameters they have dependences on $\nu_{\text{eff}}$ and $\mu$, so $u_0
\equiv u_0(\nu_{\text{eff}},\mu)$ etc.
However, the introduction of $u_0$ and $v_0$ is a major inconvenience for the
PLSF calibration because
Eq.~\ref{eqn:fullPsfModelWithShifts} is nonlinear in terms of these
parameters, e.g. $\partial^2 P / \partial u_0^2 \neq 0$. This requires the
fitting to be iterated, and complicates the implementation of the running
solution that is used to account for variation with time.
Instead, we have adopted the following linearised form of the model
\begin{multline} \label{eqn:truncatedPsfModelWithShifts}
P(u,v|\nu_{\text{eff}},\mu) = G_0(u,v) - u_0(\nu_{\text{eff}},\mu)
\frac{\partial G_0(u,v)}{\partial u} \\ - v_0(\nu_{\text{eff}},\mu)
\frac{\partial G_0(u,v)}{\partial v} + \sum_{n=1}^{N} \gsw_n(\nu_{\text{eff}},\mu) \, G_n(u,v)
\,,
\end{multline}
which is derived from Eq.~\ref{eqn:fullPsfModelWithShifts} by Taylor
expansion, ignoring terms of order $O(u_0 \gsw_n)$ and $O(v_0 \gsw_n)$ and
higher \citep{LL-134}.
This form of the model is much more convenient as it can be made equivalent
to Eq.~\ref{eqn:psfNoMuDot} by representing the derivatives of $G_0$ as
linear combinations of the pseudo shapelets (Eq.~\ref{eqn:gn_exc_smear})
and introducing them as additional basis components with amplitudes
$-u_0(\nu_{\text{eff}},\mu)$ and $-v_0(\nu_{\text{eff}},\mu)$.
In this formulation we have the identities
\begin{align}
\label{eqn:identities}
\begin{split}
G_1(u,v) & = \frac{\partial G_0(u,v)}{\partial u} \\
G_2(u,v) & = \frac{\partial G_0(u,v)}{\partial v} \\
\gsw_1(\nu_{\text{eff}},\mu) & = - u_0(\nu_{\text{eff}},\mu) \\
\gsw_2(\nu_{\text{eff}},\mu) & = - v_0(\nu_{\text{eff}},\mu)
\end{split}
\end{align}
The basis components $G_1$ and $G_2$ and their amplitudes $\gsw_1$ and $\gsw_2$
therefore model small shifts of the PSF origin in the AL and AC
directions, whereas the basis components $G_{3+}$ and their amplitudes
$\gsw_{3+}$ model the shape of the PSF.
In order to guarantee the independence of the shift and shape parameters the
basis components $G$ must be mutually orthogonal. This is enforced during the
production of the basis components: once the mean PSF $G_0$ and the first
derivatives $G_1$ and $G_2$ are established, the higher order basis components
$G_{3+}$ are post-processed to subtract their projections onto $G_1$ and $G_2$
in a process of Gram-Schmidt orthogonalisation, i.e.
\begin{equation}
G_n^{\perp} = G_n - \left(\frac{G_n \cdot G_1}{|G_1|^2}G_1\right) -
\left(\frac{G_n \cdot G_2}{|G_2|^2}G_2\right)
\end{equation}
for $n \geq 3$. These steps are ultimately carried out by manipulating the
elements of the matrix $\alpha_{ij}^n$.
The resulting PLSF basis components are presented in 
Casta{\~n}eda et al.~(2026 in prep, section 3.3.5.2).

The explicit calibration of the PSF shifts can now be used to
break the degeneracy with the geometric instrument calibration. This involves
shifting the PSF model by correcting the $g_1$ and $g_2$ terms by the their
values at the reference wavenumber $\nu_{\text{eff}}^{\text{ref}}$, where
$\nu_{\text{eff}}^{\text{ref}}=1.43\mu\text{m}^{-1}$. In practice this amounts
to making the transformation
\begin{align}
\label{eqn:achromaticConstraints}
\begin{split}
\gsw'_1(\nu_{\text{eff}},\mu) & = \gsw_1(\nu_{\text{eff}},\mu) -
\gsw_1(\nu_{\text{eff}}^{\text{ref}},\mu) \\
\gsw'_2(\nu_{\text{eff}},\mu) & = \gsw_2(\nu_{\text{eff}},\mu) -
\gsw_2(\nu_{\text{eff}}^{\text{ref}},\mu)
\end{split}
\end{align}
then using $\gsw'_1$ and $\gsw'_2$ in the PSF model. Note that this step is only
applied during Astrometric Detection and Image Parameter Determination
(DIPD; see Casta{\~n}eda et al.~2026 in prep, section 3.3.7)
 when the PSF model is used to
estimate the locations of sources in the \gaia data
stream\footnote{\label{fn:dipd}In previous data releases this procedure was
called simply Image Parameter Determination (IPD). The processing for DR4
includes a new step of on-ground source detection in order to improve the image
parameters of sources in situations where one or more secondary sources appear
in the same window, which risks biasing the image parameters of the target
source if unaccounted for in the model. This is now referred to as Astrometric
Detection and Image Parameter Determination, with the acronym DIPD.}; formally,
the estimated locations are defined as relative to the location of an equivalent
source with $\nu_{\text{eff}} = \nu_{\text{eff}}^{\text{ref}}$. This procedure
is referred to elsewhere in \gaia documentation as the non-chromatic
constraints---see \citet{LL-134} for further details\footnote{Note that
\citet{LL-134} omits the dependence on $\tau$ in Sect.4.2 since it was written
before the $\tau$ dependence was recognised; this does not affect the
conclusions.}.
Finally, when the PSF model includes a dependence on TDI line number $\tau$
such that e.g. $\gsw_1 \equiv \gsw_1(\nu_{\text{eff}},\mu, \tau)$, then the
constraints in Eq.~\ref{eqn:achromaticConstraints} are adjusted to
\begin{align}
\label{eqn:achromaticConstraintsIncTdi}
\begin{split}
\gsw'_1(\nu_{\text{eff}},\mu,\tau) & = \gsw_1(\nu_{\text{eff}},\mu,\tau) -
\gsw_1(\nu_{\text{eff}}^{\text{ref}},\mu,\tau_\text{F}) \\
\gsw'_2(\nu_{\text{eff}},\mu,\tau) & = \gsw_2(\nu_{\text{eff}},\mu,\tau) -
\gsw_2(\nu_{\text{eff}}^{\text{ref}},\mu,\tau_\text{F}) \,,
\end{split}
\end{align}
i.e. we use the value of the shift at the CCD fiducial line $\tau_\text{F}$. 

Note that there are several known limitations in this model that complicate its
use. First, the linearised form of the model in
Eq.~\ref{eqn:truncatedPsfModelWithShifts} is a truncated expansion and is
only valid for relatively small values of $u_0$ and $v_0$. Second, the mean PSF
$G_0$ may not be accurate for a given CCD and/or time. Finally, structure in the
PSF prevents $G_1$ and $G_2$ from being fully orthogonal.
These imply that the calibration of the PSF shift and shape are not fully
decoupled, and that the $\gsw_1$ and $\gsw_2$ parameters absorb a small amount
of shape change, and vice-versa. The application of the non-chromatic
constraints then results in some undesired change in the PSF model shape,
degrading the IPD performance. However, successive iterations of the PSF
calibration and global astrometric solution drive the correction terms
$\gsw_1(\nu_{\text{eff}}^{\text{ref}},\mu)$ and
$\gsw_2(\nu_{\text{eff}}^{\text{ref}},\mu)$ towards zero, which eliminates this
problem.
\subsubsection{The DR4 LSF model \label{sec:dr4Lsf}}
The vast majority of \gaia observations are 1D and are modelled using the LSF.
The marginalisation of the AC dimension greatly simplifies the modelling by
reducing the dimensionality and eliminating (to a very good approximation) the
dependences on AC rate and TDI line number.
The LSF model developed for DR4 can be derived from the EDR3 model presented in
Eq.~\ref{eqn:edr3_lsf} by applying a similar derivation as to the PSF, albeit
simpler, and making use of the relations in equations~\ref{eqn:hn} to first
write down the instantaneous effective LSF,
\begin{equation}
\begin{split} \label{eqn:instEffLsf}
L(u|\nu_{\text{eff}},\mu,\dot{\eta},\tau)
& = \sum_{n=0}^{N} \gsw_n(\nu_{\text{eff}},\mu) \sum_{i=0}^{I-1} \alpha^n_{i}
\al_i(u + \Delta u(\tau,\dot{\eta})) \\
& = \sum_{i=0}^{I-1} \beta_{i}(\nu_{\text{eff}},\mu) \al_i(u + \Delta
u(\tau,\dot{\eta})) \,,
\end{split}
\end{equation}
where $\beta_{i}(\nu_{\text{eff}},\mu) = \sum_{n=0}^{N}
\gsw_n(\nu_{\text{eff}},\mu) \, \alpha^n_{i}$, and with the identities
$\gsw_0(\nu_{\text{eff}},\mu)=1$, $\alpha^0_{0}=1$ and $\alpha^{1+}_{0}=0$.
As for the PSF, the matrix $\alpha$ is constant and defines the construction of
the $N$ 1D basis components from linear combinations of the 1D functions $f$.
To generate the model for a particular observation we integrate over $\tau$ to
obtain the integrated effective LSF,
\begin{equation}\label{eqn:intEffLsf}
L(u|\nu_{\text{eff}},\mu,\dot{\eta}) = \,
\sum_{i=0}^{I-1} \frac{\beta_{i}(\nu_{\text{eff}},\mu)}{\Delta \tau}
\int\limits_{\tau=\tau_{\text{max}}}^{\tau_{\text{min}}}
 \al_i(u + \Delta u(\tau,\dot{\eta})) \,\text{d}\tau
\end{equation}
The 1D observations always use NOGATE\footnote{\label{fn:gated1d}Technically, 1D
observations can use a shorter gate (or indeed, multiple gates) if they coincide with the transit
of a brighter star which has a higher priority in the gate selection. These
accidentally-gated observations are not used in the LSF calibration.}, so
$\tau_{\text{min}}=1$ and $\tau_{\text{max}}=4500$ according to Table~\ref{tab:ccdgates}.
The integration now accounts solely for the smearing in the AL direction, and is
equivalent to a convolution of $\al_i$ with a top hat of width $|(\dot{\tau}
- \dot{\tau}_0) \, t_{\mathrm{exp}}|$ and amplitude $|(\dot{\tau} -
\dot{\tau}_0) \, t_{\mathrm{exp}}|^{-1}$ centred on the origin. In the case of
the LSF this integration can be computed analytically, which avoids the need to
adopt the Gauss-Legendre quadrature scheme necessary for the PSF. The LSF model
is also invariant to changes in the sign of $(\dot{\tau} -\dot{\tau}_0)$, which is
not the case for the PSF model.

Note that the LSF origin is calibrated in the same way as for the PSF,
as described in Sect.~\ref{sec:h00}. Briefly, the $N$ 1D basis components are
constructed (via the elements of $\alpha$) such that component $n=1$ is the
derivative of the mean ($n=0$), with the $n=2+$ components being orthogonalised
in order to separate the parameters of the LSF into shift ($g_1$) and shape
($g_{2+}$) terms. When using the LSF to measure AL locations of observations we
correct the $g_1$ term as shown in Eq.~\ref{eqn:achromaticConstraints}.
\subsection{Calibration algorithm \label{sec:calibration}}
The PLSF models described above are calibrated for use in the science data
processing using regular science observations, specifically windows containing
single isolated stars, rather than any dedicated calibration data. The
eligibility, selection and preprocessing of the windows follows the same basic
procedure as described in \paperI sections 4.1 and 4.2, and will not be repeated
here. A few minor improvements and adaptations to the new calibration models
have been made; these will be described in detail in Casta{\~n}eda et al.~(2026
in prep.).
Note that while time variation in the PLSF is very significant over the course
of the mission, due to varying levels of ice contamination, focus evolution,
thermal instabilities and other factors, the PLSF models presented here contain
no time dependence.
Instead, the evolving state of the instrument is tracked by making independent
`partial solutions' for all 1268 PLSF calibration units for every one
revolution (\textasciitilde6 hour) interval throughout the mission. These
partial solutions, which may not be well constrained over the whole parameter
space, are then smoothed using a square root information filter.
This improves constraint while allowing gradual time evolution in the solution,
albeit with small discontinuities between consecutive revolutions.
This is the same `running solution' methodology used in the DR3
processing, and which is described in \paperI section 3.4.5. For DR4 we have
increased the time interval for the partial solutions from 0.5 to 1.0
revolution, mainly to reduce the data volume, and reduced the time constant in
the filter from $80$ revs$^{-1}$ to $40$ revs$^{-1}$ for the PSF and $20$
revs$^{-1}$ for the LSF, in order to better track rapid changes.
The trending over time is not explored further here, and instead we present the
updated calibration equations for the partial solution, which have been revised
considerably since DR3 due to the new PLSF model formulation.
We focus on the construction of the design matrix based on the the relations
between individual calibration samples and the coefficients of the PLSF models.

The weight factors $\gsw_n$ are modelled using multidimensional splines in
$(\nu_{\text{eff}},\mu, \tau)$ or $(\nu_{\text{eff}},\mu)$, depending on the
type of model. We will include the $\tau$ dependence in the presentation that
follows. The spline value is computed as the inner product of the $P$ spline
parameters $\vec{a}$, where
\begin{equation*}
\vec{a}^T=[a_1, a_2, \ldots, a_P] \, ,
\end{equation*}
and the spline coefficients $\vec{y}(\nu_{\text{eff}},\mu, \tau)$, where
\begin{equation*}
\vec{y}(\nu_{\text{eff}},\mu, \tau)^T=[y(\nu_{\text{eff}},\mu, \tau)_1,
y(\nu_{\text{eff}},\mu, \tau)_2, \ldots, y(\nu_{\text{eff}},\mu, \tau)_P] \, ,
\end{equation*}
so that for example
\begin{equation*}
\gsw_n = \vec{a}_n^T\vec{y}_n(\nu_{\text{eff}},\mu, \tau)  \, .
\end{equation*}
By convention, the spline coefficients are determined by the spline
configuration (the knot sequence and polynomial order) and the coordinate
$(\nu_{\text{eff}},\mu, \tau)$, while the spline parameters are the quantities
solved for when fitting the spline to observations. In this work we have adopted
the spline implementation described in appendix B of
\citet{2007ASSL..350.....V}, generalised to multiple dimensions.
The weight factor associated with each basis component has an independent set of
parameters, and in general may use a different spline configuration and
therefore have a different number of parameters, so that $P \equiv P_n$. The
full set of parameters for the PSF model can be represented in a single column vector $\vec{x}$ as follows:
\begin{equation*}
\vec{x}^T = [\vec{a}_1^T, \vec{a}_2^T, \ldots, \vec{a}_N^T] \, ,
\end{equation*}
where $\vec{a}_n$ contains the $P_n$ spline parameters corresponding to the
weight factor $\gsw_n$, with corresponding coefficients $\vec{y}_n$.
Note that $n = 1, 2, \ldots, N$ and that the amplitude of the
mean ($n=0$) is fixed at $1.0$ and excluded from the calibration. The
$p^{\text{th}}$ spline parameter for the $n^{\text{th}}$ weight factor is
denoted $a_n^p$, with corresponding coefficient $y_n^p$.

The calibration algorithm involves solving for the parameters $\vec{x}$
that provide the best fit to the observations in a least squares sense.
This can be expressed in the usual manner as a system of linear equations
\begin{equation}
\label{eq:partial}
A \vec{x} = \vec{b} \, ,
\end{equation}
where the vector $\vec{b}$ contains all of the observations used to constrain
the PSF model.
The observations correspond to individual samples drawn from carefully selected
and preprocessed windows, and are normalised by the source flux.
The windows are indexed using $m$, where $m = 1, 2, \ldots, M$. Each
window has an associated source colour $\nu_{\text{eff},m}$, AC position
$\mu_m$ and field angle rates $\dot{\eta}_m$ and $\dot{\zeta}_m$, and provides
multiple individual samples $s$ that are indexed using $r$, where $r = 1, 2,
\ldots, R$. Each sample has an associated location relative to the PSF origin,
denoted $(u^r,v^r)$, which is obtained from the predicted location of the
source\footnote{The predicted location is obtained using the source astrometry,
satellite attitude and geometric instrument calibrations.} in the window and the
position of the sample. The vector $\vec{b}$ can therefore be written as
$\vec{b}^T=[\vec{s}_1^T, \vec{s}_2^T, \ldots, \vec{s}_M^T]$
where $\vec{s}_m$ represents the $R$ samples from window $m$, and
$\vec{s}_m^T=[s_m^1, s_m^2, \ldots, s_m^R]$ where $s_m^r$ represents the
$r^{\text{th}}$ sample from the $m^{\text{th}}$ window.
Finally, the mean $G_0$ is subtracted from each sample; as the amplitude of the
mean is fixed at 1.0 it is excluded from the fit, and only the amplitudes of the
basis components are solved for by fitting to the sample residuals. In terms of
these quantities, the complete model for each sample is constructed as
\begin{equation*}
\begin{split}
\label{eqn:sampleModel}
s_m^r = & \,
\sum_{n=1}^{N} \sum_{p=0}^{P_n-1}
a^p_n
\sum_{i=0}^{I-1} \sum_{j=0}^{J-1} \frac{\alpha^n_{ij}}{2}
\sum_{k=0}^{K-1} w_k \,
y^p_n (\nu_{\text{eff},m},\mu_m, \tau_k) \\
& \times \, 
\al_i(u^r+\Delta u(\tau_k,\dot{\eta}_m))\,
\ac_j(v^r+\Delta v(\tau_k,\dot{\zeta}_m))
\end{split}
\end{equation*}
The rows of the design matrix $A$ are indexed by $m,r$ and the columns are
indexed by $n,p$. The element with indices $m,r,n,p$ is constructed
\begin{equation*}
\begin{split}
A_{m,r,n,p} = & \,
\sum_{i=0}^{I-1} \sum_{j=0}^{J-1} \frac{\alpha^n_{ij}}{2}
\sum_{k=0}^{K-1}
w_k \,
y^p_n (\nu_{\text{eff},m},\mu_m, \tau_k) \\
& \times \, 
\al_i(u^r+\Delta u(\tau_k,\dot{\eta}_m))\,
\ac_j(v^r+\Delta v(\tau_k,\dot{\zeta}_m))
\end{split}
\end{equation*}
This is used to construct the design matrix in the calibration of the PSF model
in Eq.~\ref{eqn:intEffPsfIncTdi}, in which the PSF shape has an explicit
dependence on $\tau$.
The variant of the PSF model presented in Eq.~\ref{eqn:numIntEffPsfExcTdi}
has no explicit dependence on $\tau$, and the corresponding expression for the
design matrix elements is
\begin{equation*}
\begin{split}
A_{m,r,n,p} = & \,
 \,y^p_n (\nu_{\text{eff},m},\mu_m)
\sum_{i=0}^{I-1} \sum_{j=0}^{J-1} \frac{\alpha^n_{ij}}{2}
\sum_{k=0}^{K-1}
w_k \, \\
& \times \, 
\al_i(u^r+\Delta u(\tau_k,\dot{\eta}_m))\,
\ac_j(v^r+\Delta v(\tau_k,\dot{\zeta}_m)) .
\end{split}
\end{equation*}
Finally, the LSF model presented in Eq.~\ref{eqn:intEffLsf} has the
following expression for the design matrix elements:
\begin{equation*}
A_{m,r,n,p} = \,
y^p_n (\nu_{\text{eff},m},\mu_m)
\sum_{i=0}^{I-1} 
\frac{\alpha^n_{i}}{\Delta \tau}
\int\limits_{\tau=\tau_{\text{max}}}^{\tau_{\text{min}}}
 \al_i(u^r + \Delta u(\tau,\dot{\eta}_m)) \,\text{d}\tau
\end{equation*}

Each element in $\vec{b}$ and the associated row in the design matrix are
weighted by the uncertainty on the observation, which is estimated assuming
Poisson statistics and accounting for the uncertainties in the various auxiliary
calibrations.
The solution for the parameters $\vec{x}$ is obtained by applying Householder
orthogonal transformations to Eq.~\ref{eq:partial} that reduces matrix $A$
to a particular upper triangular form. For further details, see \citet[appendix
C]{2007ASSL..350.....V} and \citet[chapter 4]{Bierman_1977}.
We also obtain the formal uncertainty on the parameters solution in terms of
the square root of the covariance matrix, denoted $U_{\vec{x}}$ \citep[based on
the notation from][]{2007ASSL..350.....V},
which can be used to obtain by transformation the covariance matrix for the PLSF
model samples associated with a particular observation.
\subsection{Configuration \label{sec:config}}
The configuration of the PLSF models refers to the choice of the number of basis
components and the configuration of the multidimensional splines used to
interpolate the amplitude of each component independently, as well as the range
for each dimension.
In Table~\ref{tab:lsfpsf_1d_config} and~\ref{tab:lsfpsf_2d_config} we present
the configurations of the main LSF and PSF dependences that were selected for
operational use; we find that splines of low order\footnote{\label{fn:spline}By
convention a spline of order $N$ has polynomial segments of degree $N-1$, such
that a spline of order 2 has linear segments, etc.} in each dimension are
generally sufficient for the majority of observations, with only the \nueff
dimension requiring a single internal knot (although see
Sect.~\ref{ssec:config_issues} for further discussion). This was largely
determined empirically through trial and error, although AGIS requires that the
shift parameters have a linear dependence on \nueff with no knots
\citep[see][section 7.1]{LL-134}, which explains why the $g_1$ and $g_1,g_2$
functions have separate entries. The range of $\mu$ and $\tau$ are determined by
the CCD light sensitive area and gate length.
Recall from footnote~\ref{fn:winAcPos} that for practical reasons we use the
$\mu$ of the window centre rather than the source within it; this is a good
enough estimate.
The \nueff range of $1.08 \leq \nu_{\text{eff}} \leq 1.9$ $\mu$m$^{-1}$ spans
the vast majority of stars\footnote{As described in 
Casta{\~n}eda et al.~(2026 in prep, section 3.2.4.1)
 the \nueff value used throughout the data processing
is a weighted combination of the \nueff computed by PhotPipe, from the mean XP
spectrum, and a default value of $1.43\mu$m$^{-1}$ that serves as a prior.
Extreme unphysical values of \nueff are replaced entirely with the default. This
is applied upstream of the PLSF.}, and corresponds roughly to $-0.53 \leq
G_{\rm BP} - G_{\rm RP} \leq 7.6$ according to E.4 in \citet{EDR3-DPACP-128}.
Sources with \nueff lying outside of this range are not used in the PLSF
calibration, and their observations are processed using the PLSF model for the
closest limiting value ($1.08$ or $1.9$ $\mu$m$^{-1}$), a procedure referred to
as `clamping' the \nueff. None of the other empirically calibrated dependences
require this intervention.
\begin{table}
\caption{\label{tab:lsfpsf_1d_config}Spline configuration for 1D LSF dependences}
\centering
\begin{tabular}{llrrrr}
    \toprule
    Parameter        & Units           & Min   & Max   & Order & Knots \\
    \midrule \multicolumn{6}{c}{$g_1$ function} \vspace{0.2cm} \\
    \nueff           & $\mu$m$^{-1}$   & 1.08  & 1.9   & 2     & [-]  \\
    $\mu$            & pixels          & 14    & 1979  & 3     & [-] \\
    \midrule
    \multicolumn{6}{c}{$g_{2}$--$g_N$ functions} \vspace{0.2cm} \\
    $\nu_\text{eff}$ & $\mu$m$^{-1}$   & 1.08  & 1.9   & 3     & [1.49] \\
    $\mu$            & pixels          & 14    & 1979  & 3     & [-] \\
    \bottomrule
\end{tabular}
\end{table}
\begin{table}
\caption{\label{tab:lsfpsf_2d_config}Spline configuration for 2D PSF dependences}
\centering
\begin{tabular}{llrrrr}
    \toprule
    Parameter        & Units           & Min   & Max   & Order & knots \\
    \midrule \multicolumn{6}{c}{$g_1,g_2$ functions} \vspace{0.2cm} \\
    \nueff           & $\mu$m$^{-1}$   & 1.08  & 1.9   & 2     & [-] \\
    $\mu$            & pixels          & 14    & 1979  & 3     & [-] \\ 
    $\tau$			 & pixels   	   & 1 & \multicolumn{2}{c}{[~see
    Tab.~\ref{tab:ccdgates}~]} & [-] \\
    \midrule
    \multicolumn{6}{c}{$g_{3}$--$g_N$ functions} \vspace{0.2cm} \\
    \nueff           & $\mu$m$^{-1}$   & 1.08  & 1.9   & 3     & [1.49] \\
    $\mu$            & pixels          & 14    & 1979  & 3     & [-] \\
    $\tau$			 & pixels		   & 1 & \multicolumn{2}{c}{[~see
    Tab.~\ref{tab:ccdgates}~]} & [-] \\
    \bottomrule
\end{tabular}
\end{table}
The combination of the number of basis components and the spline configuration
determines the total number of free parameters in the model.
For the LSF model we use 25 1D basis components ($N=25$) which corresponds to
294 free parameters per calibration unit. This is reduced to 17 basis components
and 198 free parameters for the calibration units corresponding to the AF1 CCD
strip; these use shorter windows that require fewer bases.
For the PSF model we use 30 2D basis components for all calibration units. This
corresponds to 348 free parameters for the calibration units with no $\tau$
dependence, which are those that use CCD gates 4,7,8 and 9 in AF, and all the
WC1 calibration units in SM. Calibration units with a linear $\tau$ dependence
(gates 10 and 11) have 696 free parameters, and those with a quadratic $\tau$
dependence (gates 0 and 12, including SM WC0) have 1044 free parameters.

Each of the 1268 calibration units in the SM and AF focal plane has a dedicated
LSF or PSF model that is solved independently, with two minor exceptions. The
WC2 LSF solutions in AF (i.e. corresponding to 1D observations with $G \gtrsim
16$) are discarded and replaced with the corresponding WC1 solutions (calibrated using
$13\lesssim G \lesssim16$ observations) from the same device and FOV, and the
WC1 PSF solutions in SM (i.e. corresponding to 2D observations with $G \gtrsim
13$) are discarded and replaced with the corresponding WC0 solutions (calibrated
using $G \lesssim13$ observations) from the same device (each SM device observes
only one FOV) but with the dependence on $\tau$ eliminated. The motivation for
these choices is that in both cases the PLSF dependences are expected to be the
same, since the observations use the same CCD area and FOV and no magnitude
dependent effects are currently included in the model, and the brighter
observations allow a higher signal to noise solution. The $\tau$ dependence is
eliminated in the SM WC1 model for performance reasons and because the larger
sample binning of the WC1 data ($4\times4$ pixels) makes the effects much
weaker. Note that in EDR3 the SM calibrations were discarded entirely (see
section 5.3 in \paperI); for DR4 we now achieve an independent calibration of
the SM PSF.

Finally, all of \gaia's nominal SM and AF science observations
correspond to one of the 1268 calibration units, and have a logical choice of
PLSF model to use in the processing. However, a small fraction of observations
are captured in non-nominal circumstances, for which there is no obvious choice
of PLSF. For example, 1D windows are occasionally observed with a CCD gate
activated, which can occur when the transit of a faint source coincides with
that of a bright source in the same CCD. Indeed, different AL samples in a
single window may have used different gates. In dense regions a 1D window may
be wholly or partly truncated in the AC direction (i.e.~have AL samples that
summed fewer AC pixels than expected) if it overlaps with a neighbouring window,
resulting in 1D AL samples that enclose different fractions of the AC flux
distribution. Other non-nominal situations can arise given the complexity of
\gaia.
These problematic observations are not used in the PLSF calibration, and the
resulting PLSF models are not strictly applicable to them. In some cases it may
be possible to assemble a suitable approximate model for the observation from
one or more of the nominal PLSFs. However, their processing by downstream
systems within DPAC differs depending on the particular requirements of the
task, and is not documented here.
\subsection{Covariance propagation algorithm}
In some circumstances it is useful to obtain estimates of the uncertainty
on the model PLSF samples by propagation of the formal uncertainty on the
parameters solution.
In poorly constrained regions of the parameter space, for example at extreme
values of the effective wavenumber or large distances from the PLSF origin, the
uncertainty on the PLSF model can be significant.
This should in principle be accounted for when performing statistical tests,
such as when assessing the goodness-of-fit of the PLSF model to a particular
observation.
Note that while the errors on the observed samples arising from Poisson
fluctuations are uncorrelated \citep[ignoring effects of photon transfer curve
nonlinearity, e.g.][]{2014JInst...9C3048A}, the uncertainties on the model PLSF
samples from propagation of the formal calibration uncertainty can have
significant non-zero covariance terms.

The following derivation is presented in terms of the PSF model, and includes
the $\tau$ dependence. The derivation for the LSF model can be obtained in a
similar fashion.
The formal uncertainty on the PSF model parameters $\vec{x}$ is represented by
the covariance matrix $\Sigma_{\vec{x}}$, where $\Sigma_{\vec{x}} =
U_{\vec{x}}U_{\vec{x}}^T$ and $U_{\vec{x}}$ is the square root of the covariance
matrix as explained in Sect.~\ref{sec:calibration}.
Using this to estimate the uncertainty on the PSF model samples involves two
steps. The first is to transform the covariance from the PSF model parameters
space $\vec{x}$ to the basis component amplitudes space $\vec{\gsw}$, given the
$(\nu_{\text{eff}},\mu, \tau)$ values of the observation being modelled.
This is done according to the standard tensor transformation law
\begin{align}\label{eqn:covXG}
\begin{split}
\Sigma_{\vec{\gsw}} & = Y \, \Sigma_{\vec{x}}  Y^T  \\
             & = (Y U_{\vec{x}}) (Y U_{\vec{x}})^T  \,,
\end{split}
\end{align}
where $\Sigma_{\vec{\gsw}}$ is the covariance matrix for the basis component
amplitudes. The transformation matrix $Y$ contains the spline coefficients for
each basis component at the observation parameters $(\nu_{\text{eff}},\mu,
\tau)$, and has the structure
\begin{equation}
Y = 
\begin{bmatrix}
\vec{y}_1(\nu_{\text{eff}},\mu, \tau)^T & \vec{0} & \ldots &  \vec{0} \\
\vec{0} & \vec{y}_2(\nu_{\text{eff}},\mu, \tau)^T & \ldots &  \vec{0} \\
        & \vdots                                    &        &          \\
\vec{0} & \vec{0} & \ldots & \vec{y}_N(\nu_{\text{eff}},\mu, \tau)^T \\
\end{bmatrix}.
\end{equation}
The sparsity of $Y$ can be exploited to compute Eq.~\ref{eqn:covXG} in an
efficient manner.

The next stage is to transform the covariance matrix for the basis component
amplitudes $\Sigma_{\vec{\gsw}}$ to the space of the PSF model samples, given
the set of $(u,v)$ sample locations. We denote the resulting covariance matrix
$\Sigma_{\vec{P}}$, and it is computed by
\begin{equation}
\Sigma_{\vec{P}} = G \, \Sigma_{\vec{\gsw}} G^T \,.
\end{equation}
The transformation matrix $G$ contains the values of each basis component at
each of the sample locations, and it has the structure
\begin{equation}
G = 
\begin{bmatrix}
G_1(u_1,v_1) & G_2(u_1,v_1) & \ldots & G_N(u_1,v_1) \\
G_1(u_2,v_1) & G_2(u_2,v_1) & \ldots & G_N(u_2,v_1) \\
\vdots  \\
G_1(u_1,v_2) & G_2(u_1,v_2) & \ldots & G_N(u_1,v_2) \\
G_1(u_2,v_2) & G_2(u_2,v_2) & \ldots & G_N(u_2,v_2) \\
\vdots  \\
G_1(u_U,v_V) & G_2(u_U,v_V) & \ldots & G_N(u_U,v_V) \\
\end{bmatrix}
\end{equation}
where
\begin{equation}
G_n(u,v) = \sum_{i=0}^{I-1} \sum_{j=0}^{J-1} \alpha^n_{ij} \al_i(u + \Delta
u(\tau,\dot{\eta})) \, \ac_j(v + \Delta v(\tau,\dot{\zeta}))\,.
\end{equation}
In implementation it is usually more efficient to perform these two stages
separately
($\Sigma_{\vec{x}} \rightarrow \Sigma_{\vec{\gsw}} \rightarrow
\Sigma_{\vec{P}}$) , because iterative fitting of the PSF to an observation
involves updating the matrix $G$ while keeping the matrix $Y$ constant. However, the
transformations can be combined as follows:
\begin{align}
\Sigma_{\vec{P}} & = G \, \Sigma_{\vec{\gsw}} G^T \nonumber \\
                 & = G (Y U_{\vec{x}}) (Y U_{\vec{x}})^T G^T \nonumber \\
                 & = G \, Y U_{\vec{x}} (G \, Y U_{\vec{x}})^T\,.
\end{align}
Finally, the TDI line dimension must be integrated over.
This is implemented using the same quadrature scheme as before, where
we perform a summation over the $K$ TDI line steps $\tau_k$ with
associated weights $w_k$. Note that both the matrices $G$ and $Y$ depend on the
TDI line number $\tau_k$ and are brought inside the sum. The full equation for
the covariance matrix of the PSF model samples is therefore
\begin{align}
\Sigma_{\vec{P}} & = 
\sum_{k=0}^{K-1} w_k 
G_k Y_k U_{\vec{x}} (G_k Y_k U_{\vec{x}})^T
\,.
\end{align}

However, note that in the calibrations carried out for DR4 we found that
the formal uncertainty, quantified via $\Sigma_{\vec{P}}$, was much smaller than
the remaining systematic errors in the PLSF models, and was therefore not a
useful estimate of the true uncertainty.
$\Sigma_{\vec{P}}$ is also very time consuming to process for every observation,
and it could not fit into the available computing resources.
For these reasons, the data processing for DR4 made no use of the PLSF model
covariance information during IPD, and this section is included for the sake of
completeness and in case the covariance information is used in future data
processing.
\section{Results \label{sec:results}}
In this section we present some results of applying our PLSF models to real
\gaia observations from the DR4 input data. This is a very rich and varied
dataset, and there are many different PLSF configurations and calibration
units to choose from.
We can only show a limited selection, and our focus is on demonstrating the new
features of the PLSF models described in this paper, the reduction of
systematic errors in the reconstruction of the observations,
and improvements in the estimated source locations relative to the
previous PLSF models that were applied in DR3.
Improvements in the derived \gaia DR4 data products, particularly the source
astrometry and $G$-band photometry, will be deferred to other papers and/or the
official documentation. Due to many updates in the production of these data
relative to DR3 it is impossible to isolate the changes associated purely with
the improved PLSF modelling.
We also defer to the official documentation any discussion of the time evolution
and overall performance of the PLSF modelling over the complete DR4 data.

We have selected \gaia observations over the onboard mission timeline (OBMT)
range 4400 to 4410 revolutions\footnote{Corresponding roughly to
1$^{\text{st}}$--4$^{\text{th}}$ November 2016; a tool for
transformation between OBMT and other time systems is available at
\texttt{https://gaia.esac.esa.int/decoder/obmtDecoder.jsp}},
which is a stable period of good image quality and low ice contamination.
Each observation corresponds to a window containing a single, isolated star that 
has been prepared for use in the PLSF calibration as described in section~4.1 of
\paperI (with minor improvements in the background subtraction and uncertainty
estimation). For each observation, the predicted location within the window and
the field angle rates were obtained from an intermediate astrometric solution
(AGIS-4.1) generated as part of the iterative calibrations carried out during
the production of DR4 (see Hern{\'a}ndez et al.~2026, in prep.).
In some examples we also make use of the official PLSF calibrations (ELSF-4.2).
Other examples involve refitting the model with different configurations in
order to demonstrate certain features; these are standalone calibrations
produced only for the purposes of this paper.
\subsection{AL and AC source motion in 1D and 2D observations}
\begin{figure}
\resizebox{\hsize}{!}{\includegraphics{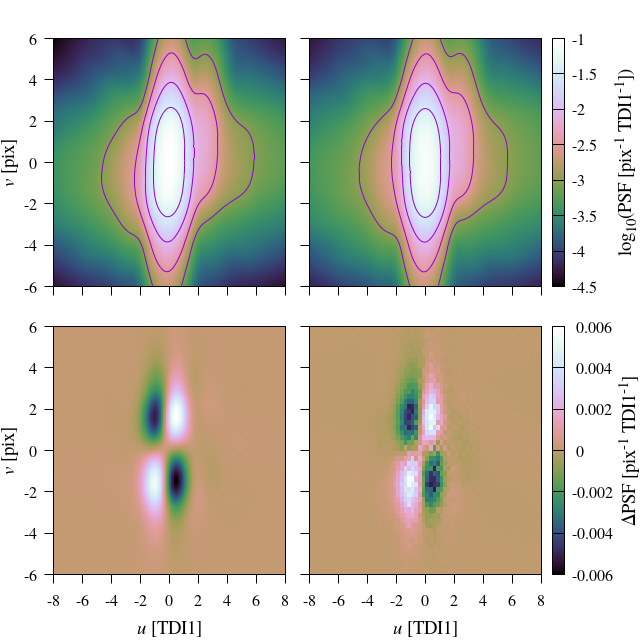}}
\caption{Reproduction of the effects of AL and AC source motion in the PSF
model. The upper two panels show two model PSFs that are identical apart from
the configured AL and AC source motion: in both cases $\dot{\tau} - \dot{\tau}_0
= -0.045$ pix\,sec$^{-1}$ while $\dot{\mu} - \dot{\mu}_0=-0.975$ (left) and
$0.975$ (right) pix\,sec$^{-1}$.
The lower left panel shows the difference between the two, which is dominated by
the reversed sign of the shearing effect due to the change in sign of the AC
rate.
The lower right panel depicts the equivalent effect seen directly in the
data, and is produced by stacking many observations at $\dot{\mu} - \dot{\mu}_0
> 0.975$ and $\dot{\mu} - \dot{\mu}_0 < -0.975$ pix\,sec$^{-1}$ and plotting the
difference.}
\label{fig:alAcRatePsfModel}
\end{figure}

In Fig.~\ref{fig:alAcRatePsfModel} we present the model PSF for one particular
calibration (FOV1 ROW1 AF6 NOGATE at revolution 4400) and two configurations of
$\dot{\tau} - \dot{\tau}_0$ and $\dot{\mu} - \dot{\mu}_0$ that are commonly
encountered in the data, with $\dot{\tau} - \dot{\tau}_0 = 0.045$ pix\,sec$^{-1}$ and
$\dot{\mu} - \dot{\mu}_0 = \pm 0.975$ pix\,sec$^{-1}$. The \nueff and $\mu$ parameters
are set to nominal values of $1.5\mu$m$^{-1}$ and $1000$ pix respectively. These
PSFs are roughly representative of observations captured three hours apart, when
the AC rate has changed sign but the AL rate is the same. The two PSFs are
related by a shearing effect that is apparent when their difference is plotted.
For completeness we include a panel depicting the equivalent effect directly in
the observations.
This figure is related to figure~22 in \paperI, which presented the effect as an
example of a feature missing from the EDR3 PSF model. This effect is now fully
incorporated in the PSF model for DR4.

\begin{figure}
\resizebox{\hsize}{!}{\includegraphics{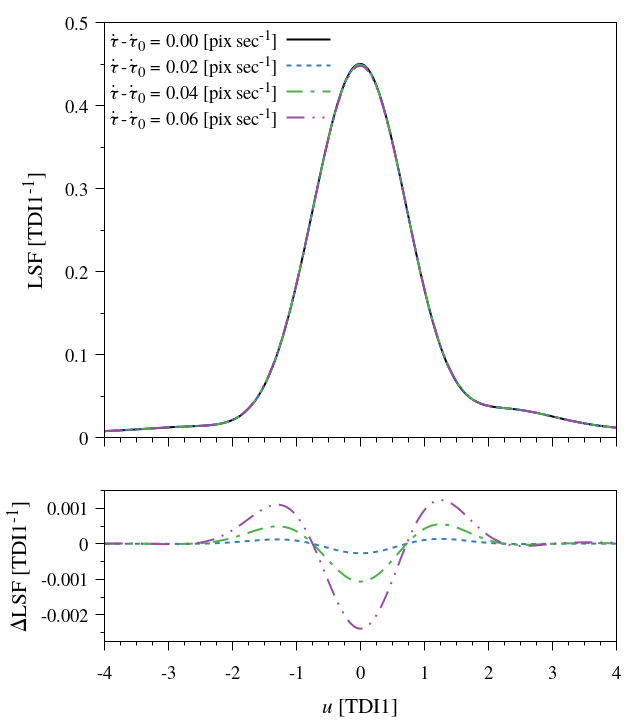}}
\caption{Reproduction of the effect of AL source motion in the LSF model. 
The upper panel shows four model LSFs for four different values of
$\dot{\tau} - \dot{\tau}_0$ ranging from 0.0 to 0.06 pix\,sec$^{-1}$.
The lower panel shows the differences relative to the LSF for $\dot{\tau} -
\dot{\tau}_0 = 0.0$ pix\,sec$^{-1}$.}
\label{fig:alRateLsfModel}
\end{figure}

\begin{figure}[h!]
\resizebox{\hsize}{!}{\includegraphics{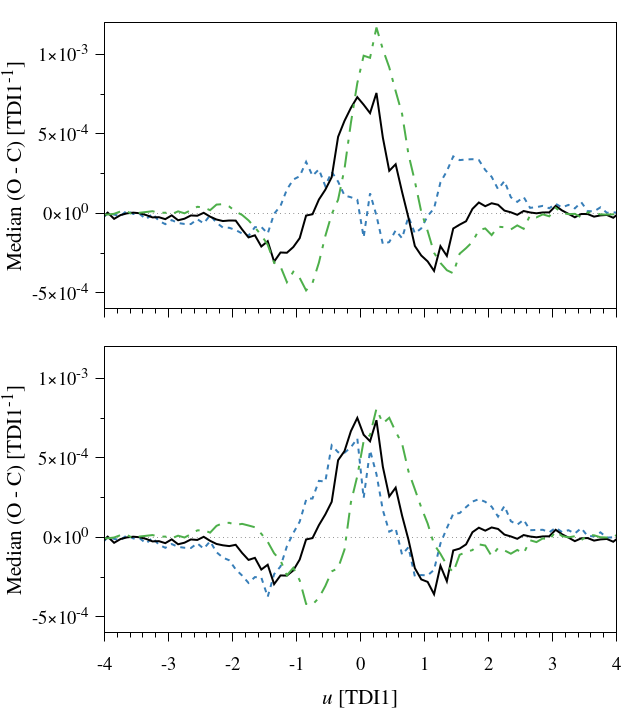}}
\caption{Both the upper and lower panels show the median residuals between the
LSF model and selected 1D observations split into three equal percentiles of
$\dot{\tau} - \dot{\tau}_0$. The solid black line depicts the central third
($33$-$66$\%), with the lower and upper thirds depicted by the dashed blue
and dot-dashed green lines, respectively. In the upper panel, the LSF model is
computed using the median value of $\dot{\tau} - \dot{\tau}_0$ for the
observations, whereas in the lower plot the LSF model is computed using the
correct $\dot{\tau} - \dot{\tau}_0$ for each observation.}
\label{fig:alRateLsfResid}
\end{figure}

In Fig.~\ref{fig:alRateLsfModel} we present the model LSF for one particular
calibration (FOV1 ROW3 AF5 NOGATE at revolution 4400) and four values of
$\dot{\tau} - \dot{\tau}_0$ that are representative of the range of values
encountered in the data. Recall that the LSF model is invariant to a change in
sign of $\dot{\tau} - \dot{\tau}_0$, so we plot only the positive values. The
\nueff and $\mu$ parameters are set to nominal values of $1.5\mu$m$^{-1}$ and
$1000$ pix respectively. A nonzero value of $\dot{\tau} - \dot{\tau}_0$ causes a
smearing of the LSF that is symmetric about the origin. The overall impact on
the LSF is smaller than for the PSF due to the lack of additional dependences on
$\dot{\mu}$ and $\tau$.
While the effect of $\dot{\tau} - \dot{\tau}_0$ variation in the model LSF is
the same for all CCDs etc, different regions of the focal plane observe quite
different distributions in $\dot{\tau} - \dot{\tau}_0$ due to the varying
amplitude of the scan law contribution, which is maximum in rows 1 and 7 and
minimum in row 4.
In Fig.~\ref{fig:alRateLsfResid} we present the residuals to the LSF model for
1D observations in FOV1 ROW7 AF5 NOGATE over revolutions 4400-4410, using both a
fixed median value of $\dot{\tau} - \dot{\tau}_0$ to compute the LSF model for
all observations, and the correct per-observation value. In the latter case the
residuals are reduced in size and show much weaker dependence on $\dot{\tau} -
\dot{\tau}_0$, indicating that the effect is successfully reproduced by the
model. The remaining structure in the residuals is likely due to a combination
of several unmodelled effects that are known about and described in
Sect.~\ref{sec:unmodelled_deps}.
\subsection{TDI line dependence in 2D observations}
\begin{figure*}[h!]
\centering
  \includegraphics[width=0.95\textwidth]{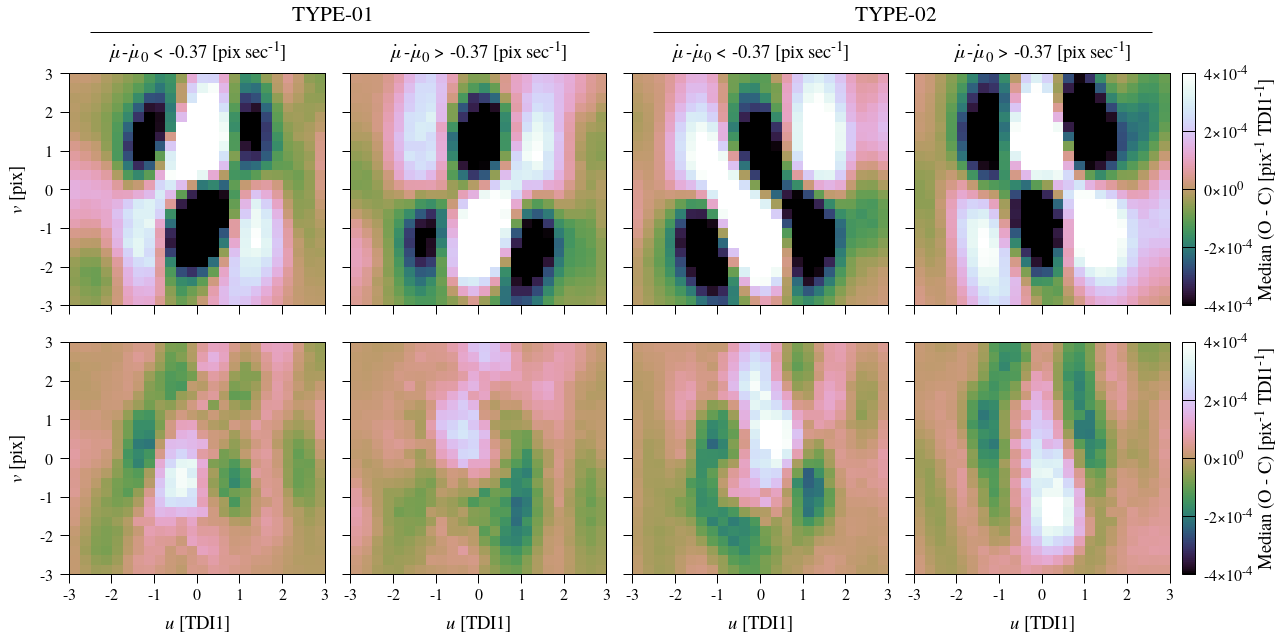}
  \caption{Median residuals to the PSF model for NOGATE observations in a
  TYPE-01 device (ROW2 AF4; left two columns) and a TYPE-02 device (ROW3 AF9;
  right two columns). For each device, the left and right columns correspond to
  observations with $\dot{\mu} - \dot{\mu}_0 < -0.37$ and $\dot{\mu} -
  \dot{\mu}_0 > 0.37$ pix\,sec$^{-1}$ respectively. In the top row the PSF model $\tau$ 
  dependence uses a first order spline with no knots. In the bottom row the
  PSF model $\tau$ dependence uses a third order spline with no knots.}
    \label{fig:tdiLineDepResid}
\end{figure*}

Figure~\ref{fig:tdiLineDependenceSystematics} demonstrates the $\tau$ dependence
in the calibrated PSF model, and its codependence on AC rate and device type due
to the corner effect.
It is also instructive to inspect the residuals to the 2D NOGATE observations
with and without the activation of the $\tau$ dependence in the PSF model, in
order to demonstrate its impact and justify the choice of a third order
dependence on $\tau$ in the model configuration.
In Fig.~\ref{fig:tdiLineDepResid} we present a series of plots depicting the
spatial structure in the residuals between the PSF model and the 2D NOGATE
observations, for two different device types and with the observations split by
$\dot{\mu} - \dot{\mu}_0$. These are derived from recalibrations of the PSF
model using two different configurations of the $\tau$ dependence. In the top
row, the $\tau$ dependence uses a first order spline with no knots, which
essentially disables it. The residuals to this model show major spatial
structure that changes in sign with both the AC rate (due to the inverted
relationship between the sample locations and the range of $\tau$ over which
they were exposed) and the device type (due to the inverted corner effect).
In the bottom row, the $\tau$ dependence uses a third order spline with no
knots, which is the configuration used operationally. In this case the residuals
show much less spatial structure that has a much weaker dependence on AC rate
and device type.
The remaining systematic structure in the residuals is likely dominated by the
brighter-fatter effect, with contributions from several other unmodelled effects
and known limitations in the model. These are discussed further in
Sect.~\ref{sec:unmodelled_deps}.
\subsection{Calibration of AL and AC shifts}
\noindent
In Sect.~\ref{sec:h00} we introduced the AL and AC shifts of the PLSF origin,
denoted $u_0$ and $v_0$, and described how they are calibrated by the adoption
of a linearised form of the PLSF models and the identities in
Eq.~\ref{eqn:identities}.
The linearised model is an approximation that is only valid over a limited
range of $u_0$ and $v_0$, and relies on an accurate mean $G_0$.
Here we present some tests designed to investigate the accuracy of the
linearised model and the valid range of $u_0$ and $v_0$.

\begin{figure}[h!]
\resizebox{\hsize}{!}{\includegraphics{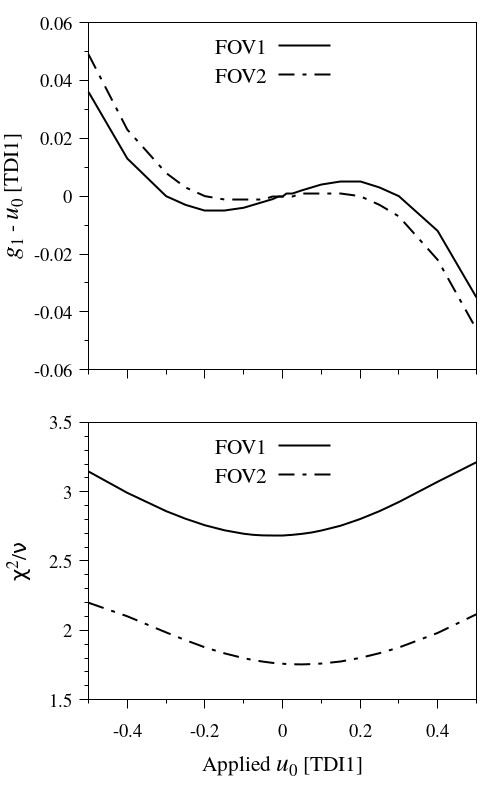}}
\caption{Tests of the calibration of the AL shift using the linearised
LSF model. The upper panel depicts the difference between the estimated shift
($g_1$) and the applied value ($u_0$) as a function of $u_0$. The lower panel
depicts the goodness-of-fit of the solution as a function of $u_0$, in terms of
the $\chi^2$ per degree of freedom.}
\label{fig:alShiftCalibration}
\end{figure}

For selected calibration units we have performed many trial calibrations of
the PLSF models, and for each calibration we have 
adjusted the predicted AL and AC locations of the observations
by a fixed amount, in order to mimic shifts of the PLSF origin.
We then inspect the solution for the $g_1$ and $g_2$ parameters, which should
be tightly correlated with the applied adjustments within the valid range of the
model.
Figure~\ref{fig:alShiftCalibration} presents the results for the LSF model, for
both FOVs in the device in row 4 strip AF5. The upper panel plots the difference
between the applied $u_0$ and the recovered $g_1$, as a function of $u_0$.  Note
that $g_1$ has dependences on \nueff and $\mu$ and has been sampled at
$\nu_{\text{eff}}=1.43\mu$m$^{-1}$ and $\mu=1000$ pix respectively; this choice
is arbitrary since the applied $u_0$ shift has no dependence on these. The
divergence from zero for $|u_0| \gtrsim 0.2$ TDI1 indicates breakdown of the
model and the extent of the valid range of $u_0$.
For both FOVs $g_1-u_0$ is roughly linear within this region, with $g_1/u_0
\approx 1.04$ for FOV1 and $g_1/u_0 \approx 1.01$ for FOV2. Ideally we would
have $g_1/u_0 = 1$; the observed offset from unity is due to differences between
the true mean LSF and the $G_0$ function used in the model, which in the DR4
processing is a fixed function for all devices in the same FOV. This is known to
be suboptimal since the LSF for each telescope varies significantly across the
focal plane and in time.
The lower panel in Fig.~\ref{fig:alShiftCalibration} indicates a gradual
degradation in the LSF solution as a function of the applied shift.

\begin{figure}[h!]
\resizebox{\hsize}{!}{\includegraphics{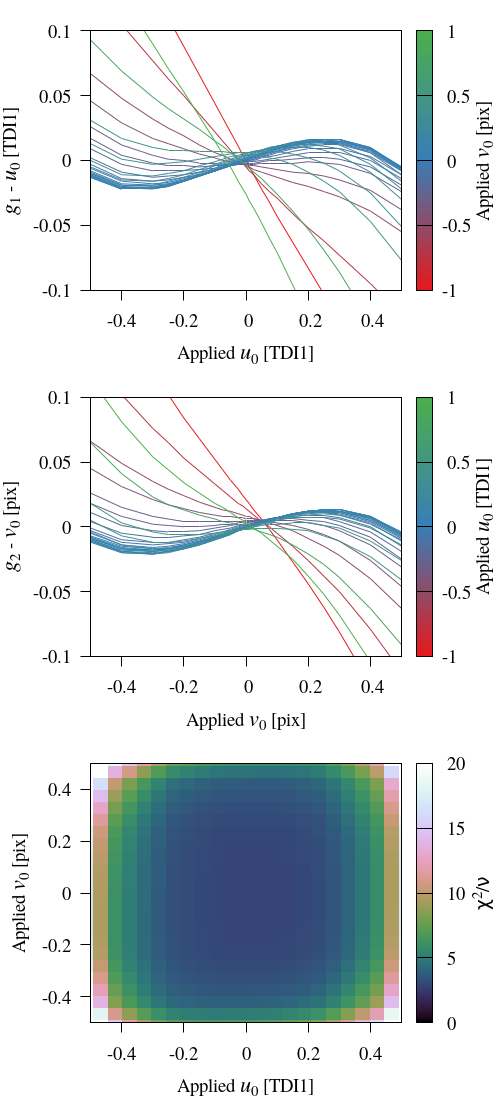}}
\caption{Tests of the calibration of the AL and AC shifts using the linearised
PSF model. The upper two panels depict the differences between the estimated AL
($g_1$) and AC ($g_2$) shifts and the applied values ($u_0$ and $v_0$) as a
function of $u_0$ and $v_0$. The lower panel depicts the goodness-of-fit of the
solution as a function of $u_0$ and $v_0$, in terms of the $\chi^2$ per degree
of freedom.}
\label{fig:alAcShiftCalibration}
\end{figure}

Figure~\ref{fig:alAcShiftCalibration} presents the equivalent results for the
FOV1 NOGATE PSF calibrations in the same device, now including the AC terms and
analysis of the combined shifts.
For the PSF we have the additional complication that the $g_1$ and $g_2$
functions have a dependence on $\tau$ (see Table~\ref{tab:lsfpsf_2d_config}),
with the shift being defined as the value at the gate fiducial line
$\tau_\text{F}$.
We find that the valid ranges of $u_0$ and $v_0$ are similar to the LSF case
($|u_0| \lesssim 0.2$ TDI1, $|v_0| \lesssim 0.2$ pix), and that within this
range both $g_1/u_0$ and $g_2/v_0$ exhibit a similar difference from unity. This
is partly due to a suboptimal $G_0$ function, which for the PSF model is
split by FOV and additionally by CCD strip.
However it is also partly due to a slight lack of constraint in the dependence
of $g_1$ and $g_2$ on $\tau$, which uses a third order spline for the longer
gates. In recent processing (post-DR4) we have reduced this by adjusting the
configuration of the $g_1$ and $g_2$ functions to always use a first order
spline in $\tau$.

Finally, the application of the constraints described in Sect.~\ref{sec:h00}
ensures that successive iterations between the PLSF calibration and the attitude
and geometric instrument calibrations have the effect of driving the shifts $u_0$
and $v_0$ present in the real observations towards zero. For the DR4 processing
several such iterations were carried out, and the final shift values were well
within the valid range of the model (see Casta{\~n}eda et al.~2026 in prep,
section 3.5.5.5).
\subsection{Comparison against DR3 PLSF models\label{sec:dr3VsDr4}}
Within the overall \gaia data processing, the critical role of the PLSF models
is in the determination of the source location (AL, and AC for 2D)
and instrumental flux for every observation, quantities collectively referred to as
the image parameters. These are estimated using a maximum likelihood
algorithm presented in \citet{LL:LL-078}, with adaptations for DR4 described in
footnote~\ref{fn:dipd} in this paper. The image parameters are the basic
observations used in the astrometric and photometric solutions for each source.

We have carried out tests to roughly quantify how our improved PLSF models have
affected the image parameters relative to those obtained using the DR3 PLSF
models described in \paperI.
This has been performed by calibrating both our new PLSF models and the
equivalent DR3 models to the same set of observations described at the start of
this Section, and using them both to solve for the image parameters for all the 
observations. We then compare the estimated source location(s) against the known
location for each source, which is obtained from the astrometric solution and
provides a suitable ground truth. A similar test of the instrumental flux is
unfortunately not possible, due to the inability to invert the photometric
calibration (see section 6.4 of \paperI). We therefore focus on the AL and AC
source locations in our analysis.
Note that for practical reasons we cannot make a direct comparison with the DR3
image parameters for the same set of observations, and must instead
approximately recreate the processing that was carried out for DR3 but using the
improved DR4 auxiliary calibrations, astrometric solution and PLSF basis
components. To imitate the DR3 PLSF we have disabled the analytic model of
AL and AC source motion and the dependence on TDI line number, and re-introduced
an empirical dependence on AC source motion with the same configuration used in
DR3.
This allows us to cleanly isolate the effects of the new PLSF model features,
while providing only a lower limit on the improvements in the estimated source
locations relative to DR3.

\begin{figure}[h!]
\resizebox{\hsize}{!}{\includegraphics{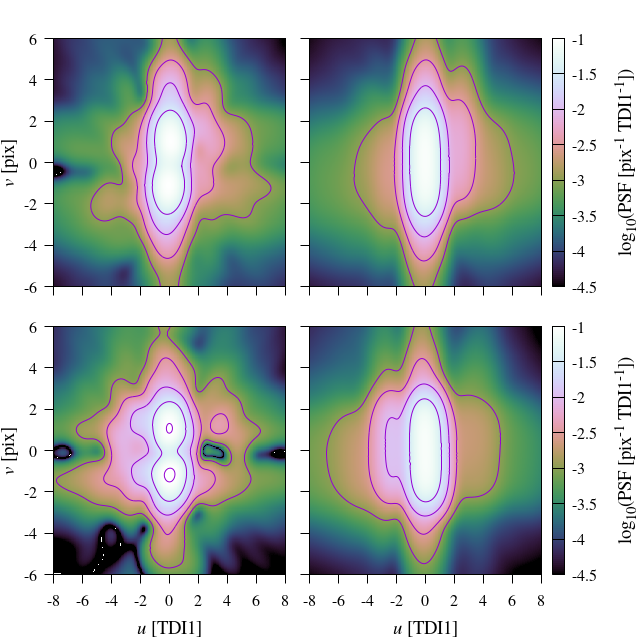}}
\caption{Comparison of the new DR4 (right) and recalibrated DR3 (left)
PSF models for a particular calibration unit (NOGATE in ROW5 AF5), recalibrated
to selected observations as described in Sect.~\ref{sec:dr3VsDr4}. The upper
panels correspond to FOV1 and the lower panels to FOV2. The models have been
computed for $\dot{\mu} - \dot{\mu}_0=-1$ pix\,sec$^{-1}$.}
\label{fig:dr3VsDr4Psf}
\end{figure}

\begin{figure}[h!]
\resizebox{\hsize}{!}{\includegraphics{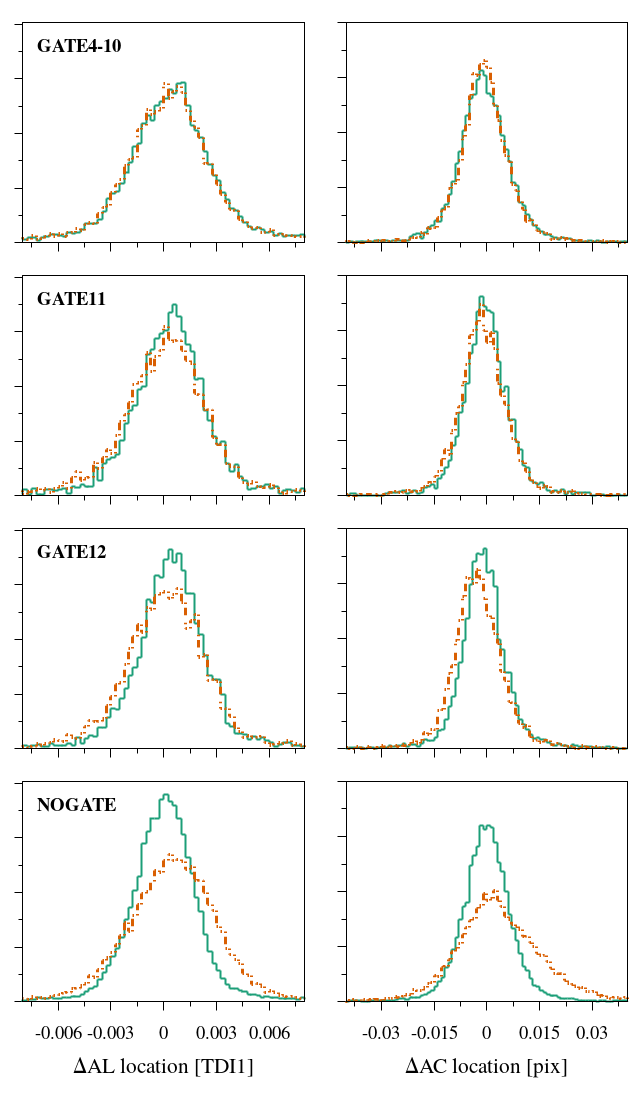}}
\caption{Distribution of the estimated AL (left) and AC (right) source
locations relative to their ground truth values, obtained from selected 2D
observations in the CCD in row 5 AF5 and corresponding to FOV1 (FOV2 is depicted
in Fig.~\ref{fig:dr3VsDr4Ipd2dFov2}). The orange dot-dashed line corresponds to
the recalibrated DR3 PSF model and the green line corresponds to the DR4 PSF
model.
In the top row, observations from CCD gates 4 to 10 have been merged; these are
the least affected by the updated PSF model. The second, third and fourth rows
correspond to observations from gates 11, 12 and 0, respectively, as indicated
in the upper left corner. The plots are normalised to unit area, so the vertical
scale is arbitrary and omitted for clarity.}
\label{fig:dr3VsDr4Ipd2dFov1}
\end{figure}

\begin{figure}[h!]
\resizebox{\hsize}{!}{\includegraphics{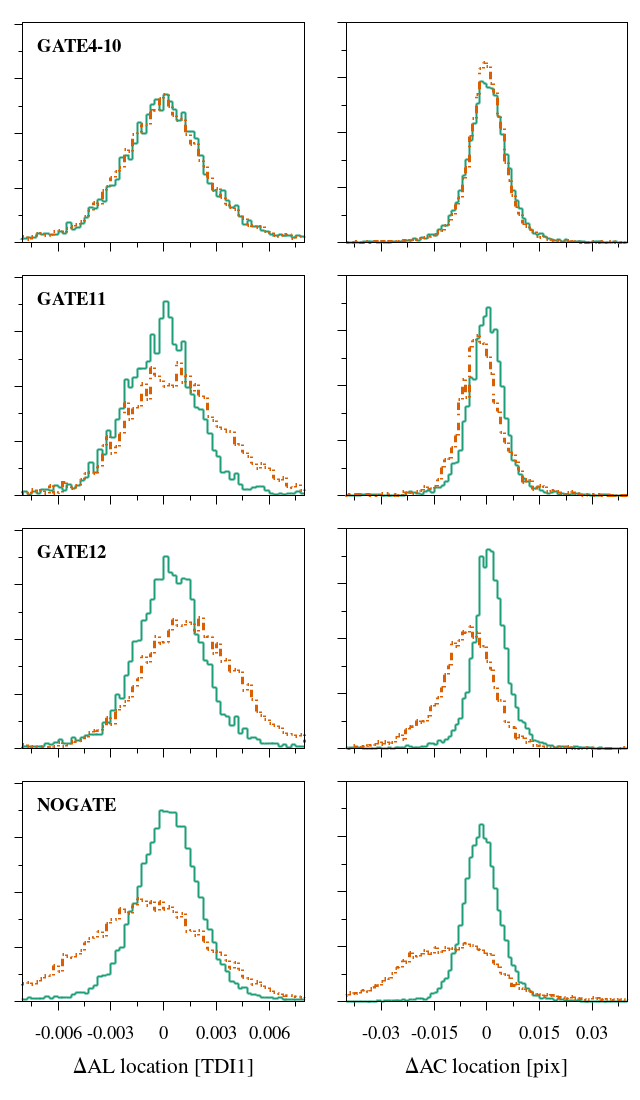}}
\caption{Equivalent to Fig.~\ref{fig:dr3VsDr4Ipd2dFov1} for FOV2.}
\label{fig:dr3VsDr4Ipd2dFov2}
\end{figure}

\begin{figure}[h!]
\resizebox{\hsize}{!}{\includegraphics{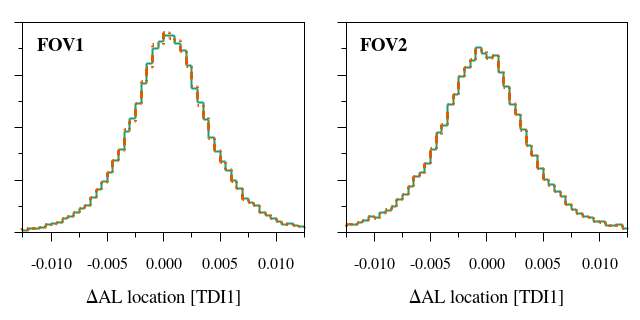}}
\caption{
Distribution of the estimated AL source locations relative to their ground truth
values, obtained from selected 1D window class 1 observations in the CCD in row
5 AF5 and corresponding to FOV1 (left) and FOV2 (right). The orange dot-dashed
line corresponds to the recalibrated DR3 LSF model and the green line
corresponds to the DR4 LSF model. The plots are normalised to unit area, so the
vertical scale is arbitrary and omitted for clarity.}
\label{fig:dr3VsDr4Ipd1d}
\end{figure}

Figure~\ref{fig:dr3VsDr4Psf} presents some selected PSF models from this
activity. These clearly demonstrate the limitation of the DR3 model at
reproducing the effects of high AC source motion in the CCD gates with longer
exposure times. This has been completely overcome in our new work by the
transition to an analytic model of this effect.
The instabilities evident in the DR3 PSF models shown in
Fig.~\ref{fig:dr3VsDr4Psf} affect the estimation of the source location. This
is depicted in Figs.~\ref{fig:dr3VsDr4Ipd2dFov1} and
\ref{fig:dr3VsDr4Ipd2dFov2}, where we present results for one particular
representative CCD.
As expected, the change in the estimated source location is greatest for 2D
observations and proportional to the gate length. Significant improvements are
obtained for ungated 2D observations, corresponding to sources with $11.5
\lesssim G \lesssim 13$, where the distribution of estimated AL and AC source
location relative to the ground truth shows a significantly lower scatter
and bias. The improvements diminish over gates 12 ($11.2 \lesssim G \lesssim
12.5$) and 11 ($11 \lesssim G \lesssim 12$), and for gates 10 and shorter ($G
\lesssim 11$) the differences are insignificant. Note that the observations in
different gates overlap considerably in magnitude, and the ranges given here are
approximate.
Figure~\ref{fig:dr3VsDr4Ipd1d} presents the equivalent results for 1D
observations in window class 1 ($13 \lesssim G \lesssim 16$) in the same CCD;
these show no significant change in the AL source location. This is consistent
with the minor impact of the AL source motion on the LSF, which is not
included in the DR3 model. The main advantage of the updated LSF model is the
calibration of AL shifts in the LSF origin, rather than any improvements in the
estimated source location.
\section{Discussion \label{sec:disc}}
In this section, we briefly discuss certain features and limitations of the
modelling that have not been explicitly covered so far. These relate to the
processing carried out for DR4 and also to potential improvements for DR5.
\subsection{Expectations for DR4 \label{ssec:expectDr4}}
Relative to EDR3, we expect improvements in the PLSF modelling for all
observations due to the introduction of new 1D and 2D basis components, the
adjusted configuration of the \nueff dependence and improvements to many of the
auxiliary instrument calibrations.
However, based on the results presented in Sect.~\ref{sec:dr3VsDr4}
and elsewhere the largest improvements are expected to be achieved for 2D
observations in the longest gates, since these benefit the most from both the
drastically improved modelling of the AL/AC smearing effect and the completely
new dependence of the PSF shape on $\tau$.
This corresponds to sources with (apparent) $G$ magnitude roughly in the
$11$~to~$13$ range.
Comparing figure~16 in \paperI with Fig.~\ref{fig:tdiLineDepResid} in this
paper suggests that the systematic residuals to the model for this subset of the
data have reduced in amplitude by at least a factor of ten. The modelling of
brighter observations in shorter gates will have improved to a lesser extent.
For 1D observations, corresponding to sources with $G\gtrsim13$, we expect more
modest improvement, since the AL rate has a relatively minor effect on the LSF
compared to other unmodelled dependences described later in this section.
Note that this is purely in terms of the residuals to the PLSF models
and the measured source locations. Quantitative predictions for the
improvements in the derived data products are difficult to make, although we
note that in Hern{\'a}ndez et al.~(2026, in prep.) the $11$~to~$13$ magnitude
range clearly shows a greater reduction in the astrometric biases and residuals compared to
earlier releases, due at least in part to our improvements in the PSF model.
\subsection{Configuration issues \label{ssec:config_issues}}
The PLSF models described in this paper have a large number of free parameters
that control their joint dependences on the \nueff, $\mu$ and $\tau$ dimensions,
each of which has different characteristics. These are fitted across a large
number of independent calibration units representing different subsets of the
observations that are split broadly by magnitude, detector, window geometry and
telescope, and which have varying degrees of sensitivity to different physical
effects.
This requires a careful choice of the configuration to ensure that the models
are able to reproduce the observations across all regions of the parameter space
while remaining well constrained.
The global configuration presented in Table~\ref{tab:lsfpsf_1d_config}
and~\ref{tab:lsfpsf_2d_config} is known to be suboptimal in several ways.

First, the dependence on $\mu$ is fixed across all devices and both FOVs, when
in reality the observations exhibit quite complex behaviour.
Within certain devices there are localised anomalies in the CCD response that
introduce variations in the electronic component of the PLSF that depend on
$\mu$. One example is the CCD in row 6 strip AF3, which appears to be thinner
over $\mu=1100$--$1300$ pix and $\tau=2055$--$4500$ pix, i.e.~affecting the
GATE12 and NOGATE observations which utilise this region of the device. The
LSF for both FOVs (and the PSF for NOGATE and GATE12) is significantly narrower
in the AL direction over the affected $\mu$ range.
All the CCDs in row 1 show a rapid and complex variation in the PSF with $\mu$
that affects FOV2 much more strongly than FOV1, and is therefore optical in
nature. This might be caused by vignetting from the twin periscopes of the Basic
Angle Monitor (BAM), which obscure part of the entrance pupil of each telescope,
or it could be related to the greater distance of the FOV2 field coordinate
origin from row 1 \citep{LL:BAS-003}.
The main impact is on the AC distribution of flux, so the LSF is relatively
unaffected.
Finally, the CCD in row 5 strip AF2 has a deep charge trap in the serial
register around $\mu=1258$ pix that causes a deficit of flux in all observations
at higher $\mu$ coordinates.
None of these phenomena are well reproduced by the PLSF models due to the lack
of flexibility in the $\mu$ configuration.

The configuration of the \nueff dependence has improved since EDR3, due to the
addition of a spline knot and the widening of the calibration range from
1.24-1.72 $\mu$m$^{-1}$ to 1.08-1.9 $\mu$m$^{-1}$, which now
encompasses the vast majority of stars. However, there are lingering issues at
both the red and blue ends of the range, in the former due to insufficient
flexibility in the spline, and in the latter due to limitations in the 2D basis
components, which struggle to fit the narrow PSFs of blue sources due to a lack
of high spatial frequency components. Both of these issues are limited to the
modelling of 2D windows.
The lack of high spatial frequency basis components also affects the modelling
of 2D observations at low values of $|\dot{\mu} - \dot{\mu}_0|$, which have
narrow profiles in the AC direction.

As well as extending the basis components to higher spatial frequencies, the 1D
bases could be improved by using a different set for each CCD strip, as is
currently the case for the 2D bases. For both types, the use of time dependent
bases that are updated after each refocus or decontamination event would help
reduce time variation in the model performance (see Casta{\~n}eda et al.~2026 in
prep, section 3.5.5.5).

All of the issues described in this section have already been overcome in the
early data processing for DR5, and will not be discussed further.
\subsection{Remaining unmodelled dependences \label{sec:unmodelled_deps}}
There are several physical effects present in the data that are completely
unaccounted for in the modelling for DR4, which we discuss briefly in this
section.
\subsubsection{Source AC location for 1D observations
\label{sec:srcAcLoc}}
Although the 1D observations do not sample the AC distribution of source flux
due to the on-chip binning, the finite extent of the window in the AC direction
(12 pixels) means that some (small) fraction of the flux is not captured by the
window.
This flux loss has an impact on the photometry and since EDR3 is accounted for
in the photometric calibration \citep[see][section~4.4]{EDR3-DPACP-117}.
The window placement, which happens onboard in real time, is designed to put
each source at the centre of the window in the AC direction. However, there is
some scatter due to the source subpixel location and other factors. These
variations in the source AC location in the window have an impact on the
shape of the LSF, since a different region of the PSF is being sampled
by the window. The shape change can be quite complex since the PSF has some
asymmetry in the AC direction.

\begin{figure}
\resizebox{\hsize}{!}{\includegraphics{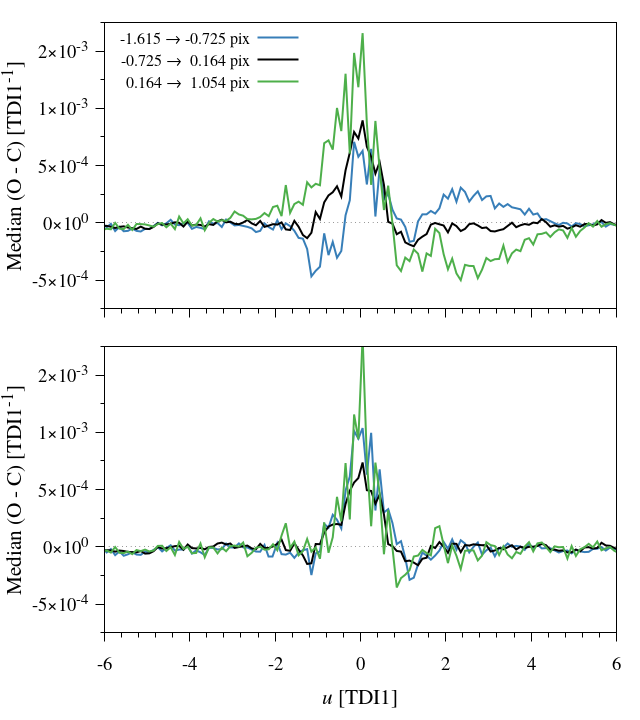}}
\caption{Both the upper and lower panels show the median residuals between the
LSF model and selected 1D observations split into three equal percentiles of
source AC location relative to the window centre as indicated in the key, in
units of pixels. The solid black line depicts the central third ($33$-$66$\%),
with the lower and upper thirds depicted by the blue and green lines,
respectively. In the upper panel, the LSF model has no dependence on the source
AC location, whereas in the lower panel this is modelled using a
second order spline with no knots (i.e. a simple linear dependence on
source AC location; see footnote~\ref{fn:spline}).}
\label{fig:srcAcLocLsfResid}
\end{figure}

Figure~\ref{fig:srcAcLocLsfResid} demonstrates this by plotting the median
residuals to the LSF model for a set of 1D observations, with the observations
split into three equal percentiles of the source AC location relative to the
window centre (the distribution of which differs slightly per CCD and FOV). In
the upper panel we have used the nominal DR4 LSF model, and the different
percentiles reveal quite strong variations in the data that are unaccounted for
in the model.
In the lower panel, we have recalibrated the LSF model and introduced a new
linear dependence on the source AC location. This simple configuration greatly
reduces the variation in the residuals.
Note that the residuals are net positive in the core due largely to
the nonlinear effects described in Sect.~\ref{sec:bfe}. Specifically, the
solution for the LSF model is weighted towards the brighter, higher
signal-to-noise observations which have broader profiles, while the majority of
observations are fainter and sharper.
There may also be a small bias in the normalisation of the calibrating
observations due to the challenges in doing this step accurately, as explained
in section 6.4 of \paperI.

This extension to the LSF model was not developed in time for the
DR4 processing, but has already been integrated into the software in preparation for
the production of DR5.
For DR4, the bias in the measured AL image locations arising from this effect
has been compensated for in the astrometric solution by the introduction of new
calibration terms (see Hern{\'a}ndez et al.~2026, in prep.).
\subsubsection{Brighter-Fatter effect \label{sec:bfe}}
\begin{figure}
\resizebox{\hsize}{!}{\includegraphics{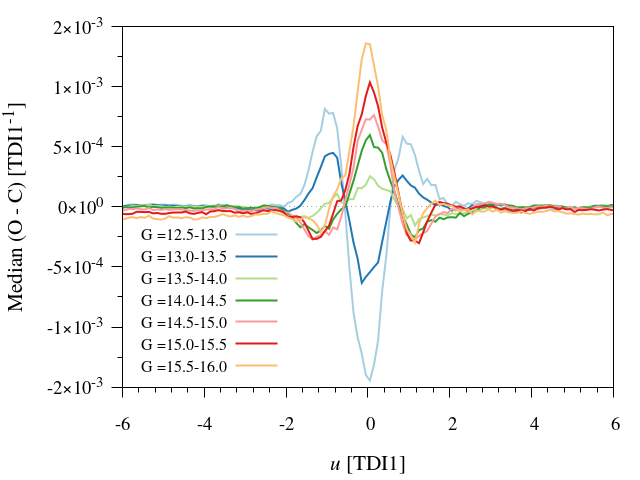}}
\caption{Median residuals between the LSF model and selected 1D observations
split into seven equal ranges of source G magnitude.}
\label{fig:bfLsf}
\end{figure}

The brighter-fatter (BF) effect \citep{2014JInst...9C3048A} is one of several
effects present in \gaia's PLSF that are nonlinear in the source flux, and which
manifest themselves as a dependence of the PLSF shape on the source magnitude.
Note that we prefer the term `signal level dependent' since the effects are
actually determined by the photoelectron counts and pixel occupancy, which for
\gaia is correlated with the source magnitude only within the range of each CCD
gate. Even within a single gate two observations of the same magnitude can have
very different pixel occupancy statistics due to the AC smearing and subpixel
location effects.
The BF effect is the tendency for observations of brighter sources (those with
higher signal levels) to be systematically wider. It is caused by the deflection
of incoming photoelectrons into neighbouring pixels, due to electrostatic
repulsion from the charge already present in a pixel.
As demonstrated in Fig.~\ref{fig:bfLsf}, the BF effect is very clearly
detected in \gaia observations across a wide range in signal level, where it
manifests as a systematic trend for observations of brighter sources to have
wider profiles in the AL direction. Note that since the LSF model has no
existing dependence on signal level it reproduces roughly the average BF present
in the calibrating observations, so observations fainter than the average are
narrower than the model. Very similar behaviour is seen for 1D and 2D
observations, all devices, both FOVs and all CCD gates.
It is also observed in the AC direction for 2D observations but at a much lower
level, perhaps due to the potential barriers that prevent AC charge transfer,
and is dwarfed by serial CTI (Sect.~\ref{sec:cti}).

No attempt was made to include this effect in the PLSF modelling for DR4.
Since the shape distortion is roughly symmetric in AL the impact on the
astrometry will be relatively limited. Existing pixel-level models of the BF
effect designed for other surveys
\citep[e.g.][]{2015JInst..10C5032G,2023A&A...670A.118A} have limited
applicability to \gaia data due to the use of TDI mode and the complex
association between physical pixels and samples in the observations.
However, significant work has been done within our group to develop a
sample-level model of signal dependent charge redistribution effects in \gaia
observations that accounts for BF and potentially other effects, and it remains
an ambition to deploy this in the modelling for DR5.
\subsubsection{Charge Transfer Inefficiency \label{sec:cti}}
Charge transfer inefficiency (CTI) refers to the trapping and release of
photoelectrons during the transfer of charge between pixels.
It affects \gaia observations in two distinct ways.
Parallel CTI (pCTI) occurs in the CCD image section as a source is being
observed, and causes a distortion of the image in the AL direction.
Serial CTI (sCTI) occurs in the serial register as the image is being read out,
and causes a distortion of the image in the AC direction.
The various defects that cause charge trapping operate on different timescales,
and the two types of CTI are sensitive to different types of defect due to the
very different rates at which the charge is clocked in the parallel direction
(982.8 $\mu$s transfers) versus the serial direction \citep[a complex sequence of
0.1 / 10 / 80 $\mu$s transfers; ][]{pemnu}.
While sCTI is dominated by defects introduced during manufacture and evolves
slowly throughout the mission \citep{DPACP-214}, pCTI becomes significantly
stronger over time since it is much more sensitive to the types of defect that
are generated in-flight by radiation damage in the space environment
(\citealt{2022JATIS...8a6003A}, Crowley et al.~2026 in prep.).
Both types of CTI distort the images of stars by causing a deficit of charge on
the leading edge of the profile, i.e at earlier observation times (pCTI) and
closer to the readout node (sCTI), and an excess on the trailing side.

\begin{figure}
\resizebox{\hsize}{!}{\includegraphics{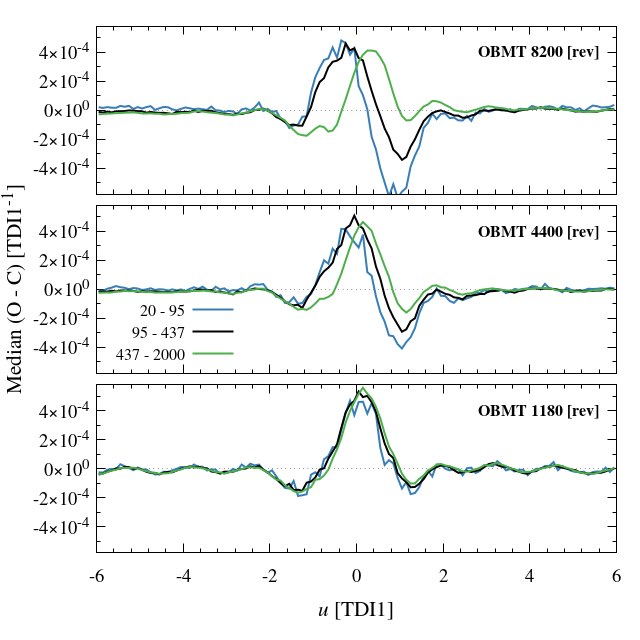}}
\caption{Each panel depicts the residuals between the LSF model and selected 1D
observations split into three ranges of charge injection distance, as indicated
in the key in units of TDI1. The three panels span three different mission times
that cover almost the full DR4 time range, with the time indicated in the top
right of each panel.}
\label{fig:pCtiWc1}
\end{figure}

However, pCTI presents a greater risk to the science data due to the resulting
systematic bias in the estimated AL image location \citep{2012MNRAS.419.2995P,
2012MNRAS.422.2786H}.
Hardware level mitigations include the use of periodic charge injections to keep
traps filled, and a supplementary buried channel to confine small charge packets
to a smaller physical volume of silicon \citep{2013MNRAS.430.3155S}.
As shown in Fig.~\ref{fig:pCtiWc1}, unmitigated pCTI is clearly detected in
the data where it manifests as a dependence of the residuals to the LSF model on
both the time since the last charge injection and the mission time.
The analysis is complicated by the presence of other unmodelled dependences,
particularly BF, and the fact that the LSF model itself is calibrated to a
nonzero level of pCTI and does not represent an undamaged observation.
Considerable work was carried out before launch on the analysis and modelling of
pCTI effects given \gaia's unique observing mode and windowed data
\citep[e.g.][]{2011MNRAS.414.2215P,2013MNRAS.430.3078S}, with the aim of
accounting for the residual effects of unmitigated charge distortion in the data
processing via a forward modelling approach, rather than attempting to
reconstruct the undamaged observations \citep[e.g.~][]{2026MNRAS.546f2186M}.
Although the image location bias is compensated for in the astrometric
processing in a statistical sense via appropriate calibration terms that depend
on the charge injection distance,
it is our ultimate goal to include a sample-level model of nonlinear charge
redistribution effects applied on top of the nominal PLSF, in order to account
for pCTI on a per-observation basis. However, this has not been integrated into
the data processing for DR4, due to a combination of the lower-than-predicted
radiation damage levels and the priority given to the modelling of other more
significant effects, such as those described in this paper.

\begin{figure}
\resizebox{\hsize}{!}{\includegraphics{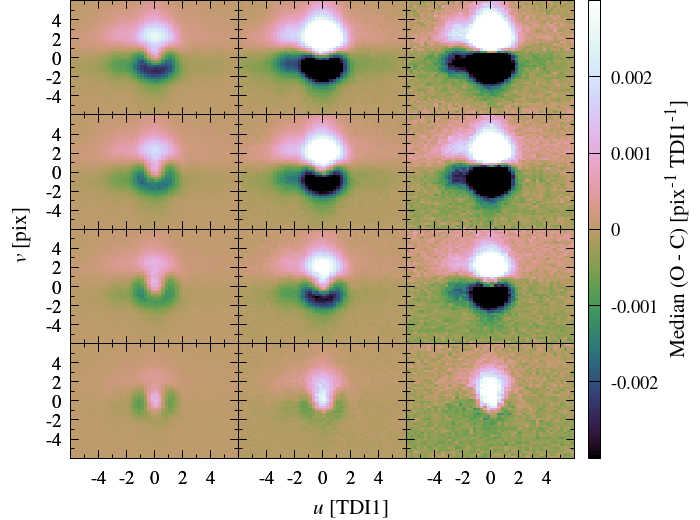}}
\caption{Each panel depicts the residuals between the NOGATE PSF model (C) and
selected 2D Calibration Faint Star observations (O), split by $G$ magnitude and
$\mu$ coordinate. The columns from left to right correspond to $G=15-16$ mag,
$17-18$ mag and $19-20$ mag respectively, and the rows from bottom to top
correspond to $\mu=14-505$ pix, $505-997$ pix, $997-1488$ pix and $1488-1979$
pix respectively.}
\label{fig:sCtiCfs}
\end{figure}

Serial CTI affects only the AC distribution of flux within an observation. While
it may have a minor impact on the 1D observations by modifying the AC flux loss
from the window, it has a large impact on the 2D observations where it manifests
as a systematic distortion that depends strongly on both the magnitude and $\mu$
coordinate of the observation.
This is demonstrated in Fig.~\ref{fig:sCtiCfs}, which depicts the residuals
between the NOGATE PSF model and selected Calibration Faint Star (CFS)
observations\footnote{A small, random fraction of faint ($G>13$) observations
that would normally be assigned 1D windows are instead assigned 2D windows, for the
purposes of calibrating various instrumental effects at the faint end, including
sCTI. These are known as Calibration Faint Stars.}, which also use NOGATE and
therefore have an identical linear component of the PSF.
The NOGATE PSF model is calibrated to observations over the $11\lesssim G
\leq13$ magnitude range, which are significantly brighter than the CFS observations and
less affected by sCTI. The CFS observations therfore exhibit quite extreme sCTI
symptoms relative to the model, although lower level effects are present in all
magnitude ranges.
The charge redistribution causes a bias in the estimated AC image location,
shifting it to larger $\mu$. This has little impact on the science data since
the AC locations have very little dependence on the parallax factor and are not
currently included in the astrometric solution for each source.
They are, however, useful for both the attitude and geometric instrument
calibrations carried out by AGIS.
As with both BF and pCTI, while no sample-level modelling of sCTI has been
performed for DR4, it may be included in the data processing for future releases
as part of our modelling of nonlinear charge redistribution effects. Recent work
to characterise the trap populations in the serial register, presented in
\citet{DPACP-214}, will assist in this, although their model cannot be directly
applied to science observations for the reasons described therein.
\subsection{The GaiaNIR mission \label{sec:gaiaNir}}
Our work demonstrates the importance of including subtle instrumental effects
when producing realistic image models. Since the PSF distortions described in
this paper are caused mostly by the spinning and precessing motion of the
satellite, rather than by e.g.~the particular detector technology, other
missions that use similar observing principles to \gaia will need to consider
them.
This is critical both during the operational phase for the routine processing of
data, but it also plays a vital role during the planning stages for use in
deriving accurate performance predictions and error budgeting. Simulations that
neglect such effects will necessarily produce optimistic or misleading results.
This is particularly true for astrometric surveys due to the complex dependence
of the PSF on the AC rate and the strong correlation with the AL parallax
factor.

The proposed GaiaNIR mission is currently under consideration for inclusion in
the ESA's Voyage 2050 science programme~\citep{2021ExA....51..783H,hobbs2021}.
This spaceborne astrometric survey mission would operate using the same
well-established scanning strategy as \gaia and Hipparcos, observing instead at
near-infrared (NIR) wavelengths.
Due to the lack of NIR-sensitive CCDs capable of operating in TDI mode,
alternative detector technologies are being investigated. A promising option are
the linear mode avalanche photodiode arrays \citep[LM-APDs; see][]{rixon2023}.
In such a scenario, the TDI mode would be implemented in onboard software by the
stacking of many short exposures. This could enable the partial mitigation of
the effects of the stellar image motion at the observation level, by shifting
the individual exposures prior to stacking, an idea explored in more detail in
\citet{2023arXiv231200488H}.
\section{Conclusion \label{sec:conc}}
We have presented our improved models of \gaia's PLSF that have been implemented
and deployed in the production of DR4. This will lead to improvements in the
astrometry and $G$-band photometry relative to DR3 that are greater than
expected from the increased number of observations alone.
\gaia's unique point spread function has taken some time to investigate and
model. With the improvements presented in this paper, we now consider the linear
part of the PLSF to be well understood.
The same model, with certain improvements as discussed, will be deployed in the
production of the final catalogue, along with planned additional modelling of
several nonlinear effects that are at this point still ignored in the PLSF.
This work will provide a useful reference both to users of \gaia data, and also
to future astrometric survey missions that use the same observing principles as
\gaia.

\bibliographystyle{aa}
\bibliography{refs}

@MISC{2022gdr3.reptE...3C,
       author = {{Casta{\~n}eda}, J. and {Hobbs}, D. and {Fabricius}, C. and {Davidson}, M. and {Rowell}, N. and {Lindegren}, L. and {Hambly}, N. and 
                 {Bastian}, U. and {Portell}, J. and {Torra}, F. and {Clotet}, M. and {Smart}, R. and {Mora}, A. and {Biermann}, M. and {L{\"o}ffler}, W. and 
                 {Brown}, A. and {Busonero}, D. and {Riva}, A.},
        title = "{Gaia DR3 documentation Chapter 3: Pre-processing}",
         year = 2022,
        month = jun,
          eid = {3},
        pages = {3},
       adsurl = {https://ui.adsabs.harvard.edu/abs/2022gdr3.reptE...3C}
}

@MISC{2022gdr3.reptE...4H,
       author = {{Hobbs}, D. and {Lindegren}, L. and {Bastian}, U. and {Klioner}, S. and {Butkevich}, A. and {Stephenson}, C. and 
                 {Hern{\'a}ndez}, J. and {Lammers}, U. and {Bombrun}, A. and {Mignard}, F. and {Altmann}, M. and {Davidson}, M. and {de Bruijne}, J. and 
                 {Fern{\'a}ndez-Hern{\'a}ndez}, J. and {Siddiqui}, H. and {Utrilla}, E.},
        title = "{Gaia DR3 documentation Chapter 4: Astrometric data}",
         year = 2022,
        month = jun,
          eid = {4},
        pages = {4},
       adsurl = {https://ui.adsabs.harvard.edu/abs/2022gdr3.reptE...4H}
}

@UNPUBLISHED{LL-134,
author = {L.~Lindegren},
title={{C}alibration constraints in {I}{D}{U}},
institution={Lund Observatory},
year={2025},
month={October},
url={https://dms.cosmos.esa.int/COSMOS/doc_fetch.php?id=1736895},
note={{GAIA}-C3-TN-LU-LL-134},
type={Technical Note} }

@UNPUBLISHED{LL:LL-056,
author = {L.~Lindegren},
title={{T}he speed of a star image in the {G}aia field of view from general attitude motion or scanning law},
institution={},
year={2025},
month={January},
url={https://dms.cosmos.esa.int/COSMOS/doc_fetch.php?id=1729223},
note={{GAIA}-CU3-TN-LU-LL-056},
type={Technical Note} }

@ARTICLE{DPACP-214,
   author = {{Pagani}, C. and {Hambly}, N. and {Davidson}, M. and {Rowell}, N. and {Crowley}, C. and {Collins}, R. and {van Leeuwen}, F. and {Seabroke}, G. M. and {Holland}, A. and {Barstow}, M. A. and {Evans}, D. W. },
    title = "{\textit{Gaia} serial CTI modelling and radiation damage study}",
  journal = {\aap, Forthcoming article},
     year = 2026,
       url= {https://doi.org/10.1051/0004-6361/202557540},
      doi = {https://doi.org/10.1051/0004-6361/202557540}
}

@UNPUBLISHED{LL:JDB-028,
author = {J.~{de~Bruijne} and L.~Lindegren and O.~Svensson and others},
title={{C}hromaticity in {G}aia-3},
year={2006},
month={August},
url={https://dms.cosmos.esa.int/COSMOS/doc_fetch.php?id=2694426},
note={{GAIA}-CA-TN-ESA-JDB-028},
type={Technical Note} }

@UNPUBLISHED{LL:BAS-003,
	author = {U.~Bastian},
	title={{R}eference systems, conventions and notations for {G}aia},
	institution={Astronomisches Rechen-Institut, Heidelberg (part of ZAH, Zentrum fuer Astronomie, Heidelberg)},
	year={2020},
	month={May},
	url={https://dms.cosmos.esa.int/COSMOS/doc_fetch.php?id=358698},
	note={{GAIA}-CA-SP-ARI-BAS-003},
	type={Specification}
}

@ARTICLE{2016A&A...595A...1G,
   author = {{Gaia Collaboration} and {Prusti}, T. and {de Bruijne}, J.~H.~J. and 
	{Brown}, A.~G.~A. and {Vallenari}, A. and {Babusiaux}, C. and 
	{Bailer-Jones}, C.~A.~L. and {Bastian}, U. and {Biermann}, M. and 
	{Evans}, D.~W. and others},
    title = "{The Gaia mission}",
  journal = {\aap},
archivePrefix = "arXiv",
   eprint = {1609.04153},
 primaryClass = "astro-ph.IM",
     year = 2016,
    month = nov,
   volume = 595,
      eid = {A1},
    pages = {A1},
      doi = {10.1051/0004-6361/201629272},
   adsurl = {http://adsabs.harvard.edu/abs/2016A%26A...595A...1G}
}

@article{DR1-DPACP-8,
	author = {{Gaia Collaboration} and {Brown, A. G. A.} and {Vallenari, A.} and {Prusti, T.} and {de Bruijne, J. H.J.} and {Mignard, F.} and {Drimmel, R.} and {Babusiaux, C.} and {Bailer-Jones, C. A.L.} and {Bastian, U.} and {Biermann, M.} and {Evans, D. W.} and {Eyer, L.} and {Jansen, F.} and {Jordi, C.} and {Katz, D.} and {Klioner, S. A.} and {Lammers, U.} and {Lindegren, L.} and {Luri, X.} and {O’Mullane, W.} and {Panem, C.} and {Pourbaix, D.} and {Randich, S.} and {Sartoretti, P.} and {Siddiqui, H. I.} and {Soubiran, C.} and {Valette, V.} and {van Leeuwen, F.} and {Walton, N. A.} and {Aerts, C.} and {Arenou, F.} and {Cropper, M.} and {Høg, E.} and {Lattanzi, M. G.} and {Grebel, E. K.} and {Holland, A. D.} and {Huc, C.} and {Passot, X.} and {Perryman, M.} and {Bramante, L.} and {Cacciari, C.} and {Castañeda, J.} and {Chaoul, L.} and {Cheek, N.} and {De Angeli, F.} and {Fabricius, C.} and {Guerra, R.} and {Hernández, J.} and {Jean-Antoine-Piccolo, A.} and {Masana, E.} and {Messineo, R.} and {Mowlavi, N.} and {Nienartowicz, K.} and {Ordóñez-Blanco, D.} and {Panuzzo, P.} and {Portell, J.} and {Richards, P. J.} and {Riello, M.} and {Seabroke, G. M.} and {Tanga, P.} and {Thévenin, F.} and {Torra, J.} and {Els, S. G.} and {Gracia-Abril, G.} and {Comoretto, G.} and {Garcia-Reinaldos, M.} and {Lock, T.} and {Mercier, E.} and {Altmann, M.} and {Andrae, R.} and {Astraatmadja, T. L.} and {Bellas-Velidis, I.} and {Benson, K.} and {Berthier, J.} and {Blomme, R.} and {Busso, G.} and {Carry, B.} and {Cellino, A.} and {Clementini, G.} and {Cowell, S.} and {Creevey, O.} and {Cuypers, J.} and {Davidson, M.} and {De Ridder, J.} and {de Torres, A.} and {Delchambre, L.} and {Dell’Oro, A.} and {Ducourant, C.} and {Frémat, Y.} and {García-Torres, M.} and {Gosset, E.} and {Halbwachs, J.-L.} and {Hambly, N. C.} and {Harrison, D. L.} and {Hauser, M.} and {Hestroffer, D.} and {Hodgkin, S. T.} and {Huckle, H. E.} and {Hutton, A.} and {Jasniewicz, G.} and {Jordan, S.} and {Kontizas, M.} and {Korn, A. J.} and {Lanzafame, A. C.} and {Manteiga, M.} and {Moitinho, A.} and {Muinonen, K.} and {Osinde, J.} and {Pancino, E.} and {Pauwels, T.} and {Petit, J.-M.} and {Recio-Blanco, A.} and {Robin, A. C.} and {Sarro, L. M.} and {Siopis, C.} and {Smith, M.} and {Smith, K. W.} and {Sozzetti, A.} and {Thuillot, W.} and {van Reeven, W.} and {Viala, Y.} and {Abbas, U.} and {Abreu Aramburu, A.} and {Accart, S.} and {Aguado, J. J.} and {Allan, P. M.} and {Allasia, W.} and {Altavilla, G.} and {Álvarez, M. A.} and {Alves, J.} and {Anderson, R. I.} and {Andrei, A. H.} and {Anglada Varela, E.} and {Antiche, E.} and {Antoja, T.} and {Antón, S.} and {Arcay, B.} and {Bach, N.} and {Baker, S. G.} and {Balaguer-Núñez, L.} and {Barache, C.} and {Barata, C.} and {Barbier, A.} and {Barblan, F.} and {Barrado y Navascués, D.} and {Barros, M.} and {Barstow, M. A.} and {Becciani, U.} and {Bellazzini, M.} and {Bello García, A.} and {Belokurov, V.} and {Bendjoya, P.} and {Berihuete, A.} and {Bianchi, L.} and {Bienaymé, O.} and {Billebaud, F.} and {Blagorodnova, N.} and {Blanco-Cuaresma, S.} and {Boch, T.} and {Bombrun, A.} and {Borrachero, R.} and {Bouquillon, S.} and {Bourda, G.} and {Bouy, H.} and {Bragaglia, A.} and {Breddels, M. A.} and {Brouillet, N.} and {Brüsemeister, T.} and {Bucciarelli, B.} and {Burgess, P.} and {Burgon, R.} and {Burlacu, A.} and {Busonero, D.} and {Buzzi, R.} and {Caffau, E.} and {Cambras, J.} and {Campbell, H.} and {Cancelliere, R.} and {Cantat-Gaudin, T.} and {Carlucci, T.} and {Carrasco, J. M.} and {Castellani, M.} and {Charlot, P.} and {Charnas, J.} and {Chiavassa, A.} and {Clotet, M.} and {Cocozza, G.} and {Collins, R. S.} and {Costigan, G.} and {Crifo, F.} and {Cross, N. J.G.} and {Crosta, M.} and {Crowley, C.} and {Dafonte, C.} and {Damerdji, Y.} and {Dapergolas, A.} and {David, P.} and {David, M.} and {De Cat, P.} and {de Felice, F.} and {de Laverny, P.} and {De Luise, F.} and {De March, R.} and {de Martino, D.} and {de Souza, R.} and {Debosscher, J.} and {del Pozo, E.} and {Delbo, M.} and {Delgado, A.} and {Delgado, H. E.} and {Di Matteo, P.} and {Diakite, S.} and {Distefano, E.} and {Dolding, C.} and {Dos Anjos, S.} and {Drazinos, P.} and {Duran, J.} and {Dzigan, Y.} and {Edvardsson, B.} and {Enke, H.} and {Evans, N. W.} and {Eynard Bontemps, G.} and {Fabre, C.} and {Fabrizio, M.} and {Faigler, S.} and {Falcão, A. J.} and {Farràs Casas, M.} and {Federici, L.} and {Fedorets, G.} and {Fernández-Hernández, J.} and {Fernique, P.} and {Fienga, A.} and {Figueras, F.} and {Filippi, F.} and {Findeisen, K.} and {Fonti, A.} and {Fouesneau, M.} and {Fraile, E.} and {Fraser, M.} and {Fuchs, J.} and {Gai, M.} and {Galleti, S.} and {Galluccio, L.} and {Garabato, D.} and {García-Sedano, F.} and {Garofalo, A.} and {Garralda, N.} and {Gavras, P.} and {Gerssen, J.} and {Geyer, R.} and {Gilmore, G.} and {Girona, S.} and {Giuffrida, G.} and {Gomes, M.} and {González-Marcos, A.} and {González-Núñez, J.} and {González-Vidal, J. J.} and {Granvik, M.} and {Guerrier, A.} and {Guillout, P.} and {Guiraud, J.} and {Gúrpide, A.} and {Gutiérrez-Sánchez, R.} and {Guy, L. P.} and {Haigron, R.} and {Hatzidimitriou, D.} and {Haywood, M.} and {Heiter, U.} and {Helmi, A.} and {Hobbs, D.} and {Hofmann, W.} and {Holl, B.} and {Holland, G.} and {Hunt, J. A.S.} and {Hypki, A.} and {Icardi, V.} and {Irwin, M.} and {Jevardat de Fombelle, G.} and {Jofré, P.} and {Jonker, P. G.} and {Jorissen, A.} and {Julbe, F.} and {Karampelas, A.} and {Kochoska, A.} and {Kohley, R.} and {Kolenberg, K.} and {Kontizas, E.} and {Koposov, S. E.} and {Kordopatis, G.} and {Koubsky, P.} and {Krone-Martins, A.} and {Kudryashova, M.} and {Kull, I.} and {Bachchan, R. K.} and {Lacoste-Seris, F.} and {Lanza, A. F.} and {Lavigne, J.-B.} and {Le Poncin-Lafitte, C.} and {Lebreton, Y.} and {Lebzelter, T.} and {Leccia, S.} and {Leclerc, N.} and {Lecoeur-Taibi, I.} and {Lemaitre, V.} and {Lenhardt, H.} and {Leroux, F.} and {Liao, S.} and {Licata, E.} and {Lindstrøm, H. E.P.} and {Lister, T. A.} and {Livanou, E.} and {Lobel, A.} and {Löffler, W.} and {López, M.} and {Lorenz, D.} and {MacDonald, I.} and {Magalhães Fernandes, T.} and {Managau, S.} and {Mann, R. G.} and {Mantelet, G.} and {Marchal, O.} and {Marchant, J. M.} and {Marconi, M.} and {Marinoni, S.} and {Marrese, P. M.} and {Marschalkó, G.} and {Marshall, D. J.} and {Martín-Fleitas, J. M.} and {Martino, M.} and {Mary, N.} and {Matijevič, G.} and {Mazeh, T.} and {McMillan, P. J.} and {Messina, S.} and {Michalik, D.} and {Millar, N. R.} and {Miranda, B. M. H.} and {Molina, D.} and {Molinaro, R.} and {Molinaro, M.} and {Molnár, L.} and {Moniez, M.} and {Montegriffo, P.} and {Mor, R.} and {Mora, A.} and {Morbidelli, R.} and {Morel, T.} and {Morgenthaler, S.} and {Morris, D.} and {Mulone, A. F.} and {Muraveva, T.} and {Musella, I.} and {Narbonne, J.} and {Nelemans, G.} and {Nicastro, L.} and {Noval, L.} and {Ordénovic, C.} and {Ordieres-Meré, J.} and {Osborne, P.} and {Pagani, C.} and {Pagano, I.} and {Pailler, F.} and {Palacin, H.} and {Palaversa, L.} and {Parsons, P.} and {Pecoraro, M.} and {Pedrosa, R.} and {Pentikäinen, H.} and {Pichon, B.} and {Piersimoni, A. M.} and {Pineau, F.-X.} and {Plachy, E.} and {Plum, G.} and {Poujoulet, E.} and {Prša, A.} and {Pulone, L.} and {Ragaini, S.} and {Rago, S.} and {Rambaux, N.} and {Ramos-Lerate, M.} and {Ranalli, P.} and {Rauw, G.} and {Read, A.} and {Regibo, S.} and {Reylé, C.} and {Ribeiro, R. A.} and {Rimoldini, L.} and {Ripepi, V.} and {Riva, A.} and {Rixon, G.} and {Roelens, M.} and {Romero-Gómez, M.} and {Rowell, N.} and {Royer, F.} and {Ruiz-Dern, L.} and {Sadowski, G.} and {Sagristà Sellés, T.} and {Sahlmann, J.} and {Salgado, J.} and {Salguero, E.} and {Sarasso, M.} and {Savietto, H.} and {Schultheis, M.} and {Sciacca, E.} and {Segol, M.} and {Segovia, J. C.} and {Segransan, D.} and {Shih, I.-C.} and {Smareglia, R.} and {Smart, R. L.} and {Solano, E.} and {Solitro, F.} and {Sordo, R.} and {Soria Nieto, S.} and {Souchay, J.} and {Spagna, A.} and {Spoto, F.} and {Stampa, U.} and {Steele, I. A.} and {Steidelmüller, H.} and {Stephenson, C. A.} and {Stoev, H.} and {Suess, F. F.} and {Süveges, M.} and {Surdej, J.} and {Szabados, L.} and {Szegedi-Elek, E.} and {Tapiador, D.} and {Taris, F.} and {Tauran, G.} and {Taylor, M. B.} and {Teixeira, R.} and {Terrett, D.} and {Tingley, B.} and {Trager, S. C.} and {Turon, C.} and {Ulla, A.} and {Utrilla, E.} and {Valentini, G.} and {van Elteren, A.} and {Van Hemelryck, E.} and {van Leeuwen, M.} and {Varadi, M.} and {Vecchiato, A.} and {Veljanoski, J.} and {Via, T.} and {Vicente, D.} and {Vogt, S.} and {Voss, H.} and {Votruba, V.} and {Voutsinas, S.} and {Walmsley, G.} and {Weiler, M.} and {Weingrill, K.} and {Wevers, T.} and {Wyrzykowski, Ł.} and {Yoldas, A.} and {Žerjal, M.} and {Zucker, S.} and {Zurbach, C.} and {Zwitter, T.} and {Alecu, A.} and {Allen, M.} and {Allende Prieto, C.} and {Amorim, A.} and {Anglada-Escudé, G.} and {Arsenijevic, V.} and {Azaz, S.} and {Balm, P.} and {Beck, M.} and {Bernstein, H.-H.} and {Bigot, L.} and {Bijaoui, A.} and {Blasco, C.} and {Bonfigli, M.} and {Bono, G.} and {Boudreault, S.} and {Bressan, A.} and {Brown, S.} and {Brunet, P.-M.} and {Bunclark, P.} and {Buonanno, R.} and {Butkevich, A. G.} and {Carret, C.} and {Carrion, C.} and {Chemin, L.} and {Chéreau, F.} and {Corcione, L.} and {Darmigny, E.} and {de Boer, K. S.} and {de Teodoro, P.} and {de Zeeuw, P. T.} and {Delle Luche, C.} and {Domingues, C. D.} and {Dubath, P.} and {Fodor, F.} and {Frézouls, B.} and {Fries, A.} and {Fustes, D.} and {Fyfe, D.} and {Gallardo, E.} and {Gallegos, J.} and {Gardiol, D.} and {Gebran, M.} and {Gomboc, A.} and {Gómez, A.} and {Grux, E.} and {Gueguen, A.} and {Heyrovsky, A.} and {Hoar, J.} and {Iannicola, G.} and {Isasi Parache, Y.} and {Janotto, A.-M.} and {Joliet, E.} and {Jonckheere, A.} and {Keil, R.} and {Kim, D.-W.} and {Klagyivik, P.} and {Klar, J.} and {Knude, J.} and {Kochukhov, O.} and {Kolka, I.} and {Kos, J.} and {Kutka, A.} and {Lainey, V.} and {LeBouquin, D.} and {Liu, C.} and {Loreggia, D.} and {Makarov, V. V.} and {Marseille, M. G.} and {Martayan, C.} and {Martinez-Rubi, O.} and {Massart, B.} and {Meynadier, F.} and {Mignot, S.} and {Munari, U.} and {Nguyen, A.-T.} and {Nordlander, T.} and {Ocvirk, P.} and {O’Flaherty, K. S.} and {Olias Sanz, A.} and {Ortiz, P.} and {Osorio, J.} and {Oszkiewicz, D.} and {Ouzounis, A.} and {Palmer, M.} and {Park, P.} and {Pasquato, E.} and {Peltzer, C.} and {Peralta, J.} and {Péturaud, F.} and {Pieniluoma, T.} and {Pigozzi, E.} and {Poels, J.} and {Prat, G.} and {Prod’homme, T.} and {Raison, F.} and {Rebordao, J. M.} and {Risquez, D.} and {Rocca-Volmerange, B.} and {Rosen, S.} and {Ruiz-Fuertes, M. I.} and {Russo, F.} and {Sembay, S.} and {Serraller Vizcaino, I.} and {Short, A.} and {Siebert, A.} and {Silva, H.} and {Sinachopoulos, D.} and {Slezak, E.} and {Soffel, M.} and {Sosnowska, D.} and {Straižys, V.} and {ter Linden, M.} and {Terrell, D.} and {Theil, S.} and {Tiede, C.} and {Troisi, L.} and {Tsalmantza, P.} and {Tur, D.} and {Vaccari, M.} and {Vachier, F.} and {Valles, P.} and {Van Hamme, W.} and {Veltz, L.} and {Virtanen, J.} and {Wallut, J.-M.} and {Wichmann, R.} and {Wilkinson, M. I.} and {Ziaeepour, H.} and {Zschocke, S.}},
	title = {Gaia Data Release 1 - Summary of the astrometric, photometric, and survey properties},
	DOI= "10.1051/0004-6361/201629512",
	url= "https://doi.org/10.1051/0004-6361/201629512",
	journal = {\aap},
	year = 2016,
	volume = 595,
	pages = "A2",
}

@article{DR2-DPACP-36,
	author = {{Gaia Collaboration} and {Brown, A. G. A.} and {Vallenari, A.} and {Prusti, T.} and {de Bruijne, J. H. J.} and {Babusiaux, C.} and {Bailer-Jones, C. A. L.} and {Biermann, M.} and {Evans, D. W.} and {Eyer, L.} and {Jansen, F.} and {Jordi, C.} and {Klioner, S. A.} and {Lammers, U.} and {Lindegren, L.} and {Luri, X.} and {Mignard, F.} and {Panem, C.} and {Pourbaix, D.} and {Randich, S.} and {Sartoretti, P.} and {Siddiqui, H. I.} and {Soubiran, C.} and {van Leeuwen, F.} and {Walton, N. A.} and {Arenou, F.} and {Bastian, U.} and {Cropper, M.} and {Drimmel, R.} and {Katz, D.} and {Lattanzi, M. G.} and {Bakker, J.} and {Cacciari, C.} and {Castañeda, J.} and {Chaoul, L.} and {Cheek, N.} and {De Angeli, F.} and {Fabricius, C.} and {Guerra, R.} and {Holl, B.} and {Masana, E.} and {Messineo, R.} and {Mowlavi, N.} and {Nienartowicz, K.} and {Panuzzo, P.} and {Portell, J.} and {Riello, M.} and {Seabroke, G. M.} and {Tanga, P.} and {Thévenin, F.} and {Gracia-Abril, G.} and {Comoretto, G.} and {Garcia-Reinaldos, M.} and {Teyssier, D.} and {Altmann, M.} and {Andrae, R.} and {Audard, M.} and {Bellas-Velidis, I.} and {Benson, K.} and {Berthier, J.} and {Blomme, R.} and {Burgess, P.} and {Busso, G.} and {Carry, B.} and {Cellino, A.} and {Clementini, G.} and {Clotet, M.} and {Creevey, O.} and {Davidson, M.} and {De Ridder, J.} and {Delchambre, L.} and {Dell’Oro, A.} and {Ducourant, C.} and {Fernández-Hernández, J.} and {Fouesneau, M.} and {Frémat, Y.} and {Galluccio, L.} and {García-Torres, M.} and {González-Núñez, J.} and {González-Vidal, J. J.} and {Gosset, E.} and {Guy, L. P.} and {Halbwachs, J.-L.} and {Hambly, N. C.} and {Harrison, D. L.} and {Hernández, J.} and {Hestroffer, D.} and {Hodgkin, S. T.} and {Hutton, A.} and {Jasniewicz, G.} and {Jean-Antoine-Piccolo, A.} and {Jordan, S.} and {Korn, A. J.} and {Krone-Martins, A.} and {Lanzafame, A. C.} and {Lebzelter, T.} and {Löffler, W.} and {Manteiga, M.} and {Marrese, P. M.} and {Martín-Fleitas, J. M.} and {Moitinho, A.} and {Mora, A.} and {Muinonen, K.} and {Osinde, J.} and {Pancino, E.} and {Pauwels, T.} and {Petit, J.-M.} and {Recio-Blanco, A.} and {Richards, P. J.} and {Rimoldini, L.} and {Robin, A. C.} and {Sarro, L. M.} and {Siopis, C.} and {Smith, M.} and {Sozzetti, A.} and {Süveges, M.} and {Torra, J.} and {van Reeven, W.} and {Abbas, U.} and {Abreu Aramburu, A.} and {Accart, S.} and {Aerts, C.} and {Altavilla, G.} and {Álvarez, M. A.} and {Alvarez, R.} and {Alves, J.} and {Anderson, R. I.} and {Andrei, A. H.} and {Anglada Varela, E.} and {Antiche, E.} and {Antoja, T.} and {Arcay, B.} and {Astraatmadja, T. L.} and {Bach, N.} and {Baker, S. G.} and {Balaguer-Núñez, L.} and {Balm, P.} and {Barache, C.} and {Barata, C.} and {Barbato, D.} and {Barblan, F.} and {Barklem, P. S.} and {Barrado, D.} and {Barros, M.} and {Barstow, M. A.} and {Bartholomé Muñoz, S.} and {Bassilana, J.-L.} and {Becciani, U.} and {Bellazzini, M.} and {Berihuete, A.} and {Bertone, S.} and {Bianchi, L.} and {Bienaymé, O.} and {Blanco-Cuaresma, S.} and {Boch, T.} and {Boeche, C.} and {Bombrun, A.} and {Borrachero, R.} and {Bossini, D.} and {Bouquillon, S.} and {Bourda, G.} and {Bragaglia, A.} and {Bramante, L.} and {Breddels, M. A.} and {Bressan, A.} and {Brouillet, N.} and {Brüsemeister, T.} and {Brugaletta, E.} and {Bucciarelli, B.} and {Burlacu, A.} and {Busonero, D.} and {Butkevich, A. G.} and {Buzzi, R.} and {Caffau, E.} and {Cancelliere, R.} and {Cannizzaro, G.} and {Cantat-Gaudin, T.} and {Carballo, R.} and {Carlucci, T.} and {Carrasco, J. M.} and {Casamiquela, L.} and {Castellani, M.} and {Castro-Ginard, A.} and {Charlot, P.} and {Chemin, L.} and {Chiavassa, A.} and {Cocozza, G.} and {Costigan, G.} and {Cowell, S.} and {Crifo, F.} and {Crosta, M.} and {Crowley, C.} and {Cuypers†, J.} and {Dafonte, C.} and {Damerdji, Y.} and {Dapergolas, A.} and {David, P.} and {David, M.} and {de Laverny, P.} and {De Luise, F.} and {De March, R.} and {de Martino, D.} and {de Souza, R.} and {de Torres, A.} and {Debosscher, J.} and {del Pozo, E.} and {Delbo, M.} and {Delgado, A.} and {Delgado, H. E.} and {Di Matteo, P.} and {Diakite, S.} and {Diener, C.} and {Distefano, E.} and {Dolding, C.} and {Drazinos, P.} and {Durán, J.} and {Edvardsson, B.} and {Enke, H.} and {Eriksson, K.} and {Esquej, P.} and {Eynard Bontemps, G.} and {Fabre, C.} and {Fabrizio, M.} and {Faigler, S.} and {Falcão, A. J.} and {Farràs Casas, M.} and {Federici, L.} and {Fedorets, G.} and {Fernique, P.} and {Figueras, F.} and {Filippi, F.} and {Findeisen, K.} and {Fonti, A.} and {Fraile, E.} and {Fraser, M.} and {Frézouls, B.} and {Gai, M.} and {Galleti, S.} and {Garabato, D.} and {García-Sedano, F.} and {Garofalo, A.} and {Garralda, N.} and {Gavel, A.} and {Gavras, P.} and {Gerssen, J.} and {Geyer, R.} and {Giacobbe, P.} and {Gilmore, G.} and {Girona, S.} and {Giuffrida, G.} and {Glass, F.} and {Gomes, M.} and {Granvik, M.} and {Gueguen, A.} and {Guerrier, A.} and {Guiraud, J.} and {Gutiérrez-Sánchez, R.} and {Haigron, R.} and {Hatzidimitriou, D.} and {Hauser, M.} and {Haywood, M.} and {Heiter, U.} and {Helmi, A.} and {Heu, J.} and {Hilger, T.} and {Hobbs, D.} and {Hofmann, W.} and {Holland, G.} and {Huckle, H. E.} and {Hypki, A.} and {Icardi, V.} and {Janßen, K.} and {Jevardat de Fombelle, G.} and {Jonker, P. G.} and {Juhász, Á. L.} and {Julbe, F.} and {Karampelas, A.} and {Kewley, A.} and {Klar, J.} and {Kochoska, A.} and {Kohley, R.} and {Kolenberg, K.} and {Kontizas, M.} and {Kontizas, E.} and {Koposov, S. E.} and {Kordopatis, G.} and {Kostrzewa-Rutkowska, Z.} and {Koubsky, P.} and {Lambert, S.} and {Lanza, A. F.} and {Lasne, Y.} and {Lavigne, J.-B.} and {Le Fustec, Y.} and {Le Poncin-Lafitte, C.} and {Lebreton, Y.} and {Leccia, S.} and {Leclerc, N.} and {Lecoeur-Taibi, I.} and {Lenhardt, H.} and {Leroux, F.} and {Liao, S.} and {Licata, E.} and {Lindstrøm, H. E. P.} and {Lister, T. A.} and {Livanou, E.} and {Lobel, A.} and {López, M.} and {Managau, S.} and {Mann, R. G.} and {Mantelet, G.} and {Marchal, O.} and {Marchant, J. M.} and {Marconi, M.} and {Marinoni, S.} and {Marschalkó, G.} and {Marshall, D. J.} and {Martino, M.} and {Marton, G.} and {Mary, N.} and {Massari, D.} and {Matijevič, G.} and {Mazeh, T.} and {McMillan, P. J.} and {Messina, S.} and {Michalik, D.} and {Millar, N. R.} and {Molina, D.} and {Molinaro, R.} and {Molnár, L.} and {Montegriffo, P.} and {Mor, R.} and {Morbidelli, R.} and {Morel, T.} and {Morris, D.} and {Mulone, A. F.} and {Muraveva, T.} and {Musella, I.} and {Nelemans, G.} and {Nicastro, L.} and {Noval, L.} and {O’Mullane, W.} and {Ordénovic, C.} and {Ordóñez-Blanco, D.} and {Osborne, P.} and {Pagani, C.} and {Pagano, I.} and {Pailler, F.} and {Palacin, H.} and {Palaversa, L.} and {Panahi, A.} and {Pawlak, M.} and {Piersimoni, A. M.} and {Pineau, F.-X.} and {Plachy, E.} and {Plum, G.} and {Poggio, E.} and {Poujoulet, E.} and {Prša, A.} and {Pulone, L.} and {Racero, E.} and {Ragaini, S.} and {Rambaux, N.} and {Ramos-Lerate, M.} and {Regibo, S.} and {Reylé, C.} and {Riclet, F.} and {Ripepi, V.} and {Riva, A.} and {Rivard, A.} and {Rixon, G.} and {Roegiers, T.} and {Roelens, M.} and {Romero-Gómez, M.} and {Rowell, N.} and {Royer, F.} and {Ruiz-Dern, L.} and {Sadowski, G.} and {Sagristà Sellés, T.} and {Sahlmann, J.} and {Salgado, J.} and {Salguero, E.} and {Sanna, N.} and {Santana-Ros, T.} and {Sarasso, M.} and {Savietto, H.} and {Schultheis, M.} and {Sciacca, E.} and {Segol, M.} and {Segovia, J. C.} and {Ségransan, D.} and {Shih, I-C.} and {Siltala, L.} and {Silva, A. F.} and {Smart, R. L.} and {Smith, K. W.} and {Solano, E.} and {Solitro, F.} and {Sordo, R.} and {Soria Nieto, S.} and {Souchay, J.} and {Spagna, A.} and {Spoto, F.} and {Stampa, U.} and {Steele, I. A.} and {Steidelmüller, H.} and {Stephenson, C. A.} and {Stoev, H.} and {Suess, F. F.} and {Surdej, J.} and {Szabados, L.} and {Szegedi-Elek, E.} and {Tapiador, D.} and {Taris, F.} and {Tauran, G.} and {Taylor, M. B.} and {Teixeira, R.} and {Terrett, D.} and {Teyssandier, P.} and {Thuillot, W.} and {Titarenko, A.} and {Torra Clotet, F.} and {Turon, C.} and {Ulla, A.} and {Utrilla, E.} and {Uzzi, S.} and {Vaillant, M.} and {Valentini, G.} and {Valette, V.} and {van Elteren, A.} and {Van Hemelryck, E.} and {van Leeuwen, M.} and {Vaschetto, M.} and {Vecchiato, A.} and {Veljanoski, J.} and {Viala, Y.} and {Vicente, D.} and {Vogt, S.} and {von Essen, C.} and {Voss, H.} and {Votruba, V.} and {Voutsinas, S.} and {Walmsley, G.} and {Weiler, M.} and {Wertz, O.} and {Wevers, T.} and {Wyrzykowski, Ł.} and {Yoldas, A.} and {Žerjal, M.} and {Ziaeepour, H.} and {Zorec, J.} and {Zschocke, S.} and {Zucker, S.} and {Zurbach, C.} and {Zwitter, T.}},
	title = {Gaia Data Release 2 - Summary of the contents and survey properties},
	DOI= "10.1051/0004-6361/201833051",
	url= "https://doi.org/10.1051/0004-6361/201833051",
	journal = {\aap},
	year = 2018,
	volume = 616,
	pages = "A1",
}

@article{DR3-DPACP-185,
	author = {{Gaia Collaboration} and {Vallenari, A.} and {Brown, A. G. A.} and {Prusti, T.} and {de Bruijne, J. H. J.} and {Arenou, F.} and {Babusiaux, C.} and {Biermann, M.} and {Creevey, O. L.} and {Ducourant, C.} and {Evans, D. W.} and {Eyer, L.} and {Guerra, R.} and {Hutton, A.} and {Jordi, C.} and {Klioner, S. A.} and {Lammers, U. L.} and {Lindegren, L.} and {Luri, X.} and {Mignard, F.} and {Panem, C.} and {Pourbaix, D.} and {Randich, S.} and {Sartoretti, P.} and {Soubiran, C.} and {Tanga, P.} and {Walton, N. A.} and {Bailer-Jones, C. A. L.} and {Bastian, U.} and {Drimmel, R.} and {Jansen, F.} and {Katz, D.} and {Lattanzi, M. G.} and {van Leeuwen, F.} and {Bakker, J.} and {Cacciari, C.} and {Castañeda, J.} and {De Angeli, F.} and {Fabricius, C.} and {Fouesneau, M.} and {Frémat, Y.} and {Galluccio, L.} and {Guerrier, A.} and {Heiter, U.} and {Masana, E.} and {Messineo, R.} and {Mowlavi, N.} and {Nicolas, C.} and {Nienartowicz, K.} and {Pailler, F.} and {Panuzzo, P.} and {Riclet, F.} and {Roux, W.} and {Seabroke, G. M.} and {Sordo, R.} and {Thévenin, F.} and {Gracia-Abril, G.} and {Portell, J.} and {Teyssier, D.} and {Altmann, M.} and {Andrae, R.} and {Audard, M.} and {Bellas-Velidis, I.} and {Benson, K.} and {Berthier, J.} and {Blomme, R.} and {Burgess, P. W.} and {Busonero, D.} and {Busso, G.} and {Cánovas, H.} and {Carry, B.} and {Cellino, A.} and {Cheek, N.} and {Clementini, G.} and {Damerdji, Y.} and {Davidson, M.} and {de Teodoro, P.} and {Nuñez Campos, M.} and {Delchambre, L.} and {Dell’Oro, A.} and {Esquej, P.} and {Fernández-Hernández, J.} and {Fraile, E.} and {Garabato, D.} and {García-Lario, P.} and {Gosset, E.} and {Haigron, R.} and {Halbwachs, J.-L.} and {Hambly, N. C.} and {Harrison, D. L.} and {Hernández, J.} and {Hestroffer, D.} and {Hodgkin, S. T.} and {Holl, B.} and {Janßen, K.} and {Jevardat de Fombelle, G.} and {Jordan, S.} and {Krone-Martins, A.} and {Lanzafame, A. C.} and {Löffler, W.} and {Marchal, O.} and {Marrese, P. M.} and {Moitinho, A.} and {Muinonen, K.} and {Osborne, P.} and {Pancino, E.} and {Pauwels, T.} and {Recio-Blanco, A.} and {Reylé, C.} and {Riello, M.} and {Rimoldini, L.} and {Roegiers, T.} and {Rybizki, J.} and {Sarro, L. M.} and {Siopis, C.} and {Smith, M.} and {Sozzetti, A.} and {Utrilla, E.} and {van Leeuwen, M.} and {Abbas, U.} and {Ábrahám, P.} and {Abreu Aramburu, A.} and {Aerts, C.} and {Aguado, J. J.} and {Ajaj, M.} and {Aldea-Montero, F.} and {Altavilla, G.} and {Álvarez, M. A.} and {Alves, J.} and {Anders, F.} and {Anderson, R. I.} and {Anglada Varela, E.} and {Antoja, T.} and {Baines, D.} and {Baker, S. G.} and {Balaguer-Núñez, L.} and {Balbinot, E.} and {Balog, Z.} and {Barache, C.} and {Barbato, D.} and {Barros, M.} and {Barstow, M. A.} and {Bartolomé, S.} and {Bassilana, J.-L.} and {Bauchet, N.} and {Becciani, U.} and {Bellazzini, M.} and {Berihuete, A.} and {Bernet, M.} and {Bertone, S.} and {Bianchi, L.} and {Binnenfeld, A.} and {Blanco-Cuaresma, S.} and {Blazere, A.} and {Boch, T.} and {Bombrun, A.} and {Bossini, D.} and {Bouquillon, S.} and {Bragaglia, A.} and {Bramante, L.} and {Breedt, E.} and {Bressan, A.} and {Brouillet, N.} and {Brugaletta, E.} and {Bucciarelli, B.} and {Burlacu, A.} and {Butkevich, A. G.} and {Buzzi, R.} and {Caffau, E.} and {Cancelliere, R.} and {Cantat-Gaudin, T.} and {Carballo, R.} and {Carlucci, T.} and {Carnerero, M. I.} and {Carrasco, J. M.} and {Casamiquela, L.} and {Castellani, M.} and {Castro-Ginard, A.} and {Chaoul, L.} and {Charlot, P.} and {Chemin, L.} and {Chiaramida, V.} and {Chiavassa, A.} and {Chornay, N.} and {Comoretto, G.} and {Contursi, G.} and {Cooper, W. J.} and {Cornez, T.} and {Cowell, S.} and {Crifo, F.} and {Cropper, M.} and {Crosta, M.} and {Crowley, C.} and {Dafonte, C.} and {Dapergolas, A.} and {David, M.} and {David, P.} and {de Laverny, P.} and {De Luise, F.} and {De March, R.} and {De Ridder, J.} and {de Souza, R.} and {de Torres, A.} and {del Peloso, E. F.} and {del Pozo, E.} and {Delbo, M.} and {Delgado, A.} and {Delisle, J.-B.} and {Demouchy, C.} and {Dharmawardena, T. E.} and {Di Matteo, P.} and {Diakite, S.} and {Diener, C.} and {Distefano, E.} and {Dolding, C.} and {Edvardsson, B.} and {Enke, H.} and {Fabre, C.} and {Fabrizio, M.} and {Faigler, S.} and {Fedorets, G.} and {Fernique, P.} and {Fienga, A.} and {Figueras, F.} and {Fournier, Y.} and {Fouron, C.} and {Fragkoudi, F.} and {Gai, M.} and {Garcia-Gutierrez, A.} and {Garcia-Reinaldos, M.} and {García-Torres, M.} and {Garofalo, A.} and {Gavel, A.} and {Gavras, P.} and {Gerlach, E.} and {Geyer, R.} and {Giacobbe, P.} and {Gilmore, G.} and {Girona, S.} and {Giuffrida, G.} and {Gomel, R.} and {Gomez, A.} and {González-Núñez, J.} and {González-Santamaría, I.} and {González-Vidal, J. J.} and {Granvik, M.} and {Guillout, P.} and {Guiraud, J.} and {Gutiérrez-Sánchez, R.} and {Guy, L. P.} and {Hatzidimitriou, D.} and {Hauser, M.} and {Haywood, M.} and {Helmer, A.} and {Helmi, A.} and {Sarmiento, M. H.} and {Hidalgo, S. L.} and {Hilger, T.} and {Hładczuk, N.} and {Hobbs, D.} and {Holland, G.} and {Huckle, H. E.} and {Jardine, K.} and {Jasniewicz, G.} and {Jean-Antoine Piccolo, A.} and {Jiménez-Arranz, Ó.} and {Jorissen, A.} and {Juaristi Campillo, J.} and {Julbe, F.} and {Karbevska, L.} and {Kervella, P.} and {Khanna, S.} and {Kontizas, M.} and {Kordopatis, G.} and {Korn, A. J.} and {Kóspál, Á} and {Kostrzewa-Rutkowska, Z.} and {Kruszyńska, K.} and {Kun, M.} and {Laizeau, P.} and {Lambert, S.} and {Lanza, A. F.} and {Lasne, Y.} and {Le Campion, J.-F.} and {Lebreton, Y.} and {Lebzelter, T.} and {Leccia, S.} and {Leclerc, N.} and {Lecoeur-Taibi, I.} and {Liao, S.} and {Licata, E. L.} and {Lindstrøm, H. E. P.} and {Lister, T. A.} and {Livanou, E.} and {Lobel, A.} and {Lorca, A.} and {Loup, C.} and {Madrero Pardo, P.} and {Magdaleno Romeo, A.} and {Managau, S.} and {Mann, R. G.} and {Manteiga, M.} and {Marchant, J. M.} and {Marconi, M.} and {Marcos, J.} and {Marcos Santos, M. M. S.} and {Marín Pina, D.} and {Marinoni, S.} and {Marocco, F.} and {Marshall, D. J.} and {Martin Polo, L.} and {Martín-Fleitas, J. M.} and {Marton, G.} and {Mary, N.} and {Masip, A.} and {Massari, D.} and {Mastrobuono-Battisti, A.} and {Mazeh, T.} and {McMillan, P. J.} and {Messina, S.} and {Michalik, D.} and {Millar, N. R.} and {Mints, A.} and {Molina, D.} and {Molinaro, R.} and {Molnár, L.} and {Monari, G.} and {Monguió, M.} and {Montegriffo, P.} and {Montero, A.} and {Mor, R.} and {Mora, A.} and {Morbidelli, R.} and {Morel, T.} and {Morris, D.} and {Muraveva, T.} and {Murphy, C. P.} and {Musella, I.} and {Nagy, Z.} and {Noval, L.} and {Ocaña, F.} and {Ogden, A.} and {Ordenovic, C.} and {Osinde, J. O.} and {Pagani, C.} and {Pagano, I.} and {Palaversa, L.} and {Palicio, P. A.} and {Pallas-Quintela, L.} and {Panahi, A.} and {Payne-Wardenaar, S.} and {Peñalosa Esteller, X.} and {Penttilä, A.} and {Pichon, B.} and {Piersimoni, A. M.} and {Pineau, F.-X.} and {Plachy, E.} and {Plum, G.} and {Poggio, E.} and {Prša, A.} and {Pulone, L.} and {Racero, E.} and {Ragaini, S.} and {Rainer, M.} and {Raiteri, C. M.} and {Rambaux, N.} and {Ramos, P.} and {Ramos-Lerate, M.} and {Re Fiorentin, P.} and {Regibo, S.} and {Richards, P. J.} and {Rios Diaz, C.} and {Ripepi, V.} and {Riva, A.} and {Rix, H.-W.} and {Rixon, G.} and {Robichon, N.} and {Robin, A. C.} and {Robin, C.} and {Roelens, M.} and {Rogues, H. R. O.} and {Rohrbasser, L.} and {Romero-Gómez, M.} and {Rowell, N.} and {Royer, F.} and {Ruz Mieres, D.} and {Rybicki, K. A.} and {Sadowski, G.} and {Sáez Núñez, A.} and {Sagristà Sellés, A.} and {Sahlmann, J.} and {Salguero, E.} and {Samaras, N.} and {Sanchez Gimenez, V.} and {Sanna, N.} and {Santoveña, R.} and {Sarasso, M.} and {Schultheis, M.} and {Sciacca, E.} and {Segol, M.} and {Segovia, J. C.} and {Ségransan, D.} and {Semeux, D.} and {Shahaf, S.} and {Siddiqui, H. I.} and {Siebert, A.} and {Siltala, L.} and {Silvelo, A.} and {Slezak, E.} and {Slezak, I.} and {Smart, R. L.} and {Snaith, O. N.} and {Solano, E.} and {Solitro, F.} and {Souami, D.} and {Souchay, J.} and {Spagna, A.} and {Spina, L.} and {Spoto, F.} and {Steele, I. A.} and {Steidelmüller, H.} and {Stephenson, C. A.} and {Süveges, M.} and {Surdej, J.} and {Szabados, L.} and {Szegedi-Elek, E.} and {Taris, F.} and {Taylor, M. B.} and {Teixeira, R.} and {Tolomei, L.} and {Tonello, N.} and {Torra, F.} and {Torra, J.} and {Torralba Elipe, G.} and {Trabucchi, M.} and {Tsounis, A. T.} and {Turon, C.} and {Ulla, A.} and {Unger, N.} and {Vaillant, M. V.} and {van Dillen, E.} and {van Reeven, W.} and {Vanel, O.} and {Vecchiato, A.} and {Viala, Y.} and {Vicente, D.} and {Voutsinas, S.} and {Weiler, M.} and {Wevers, T.} and {Wyrzykowski, Ł.} and {Yoldas, A.} and {Yvard, P.} and {Zhao, H.} and {Zorec, J.} and {Zucker, S.} and {Zwitter, T.}},
	title = {Gaia Data Release 3 - Summary of the content and survey properties},
	DOI= "10.1051/0004-6361/202243940",
	url= "https://doi.org/10.1051/0004-6361/202243940",
	journal = {\aap},
	year = 2023,
	volume = 674,
	pages = "A1",
}

@article {DR1-DPACP-7,
       author = {{Fabricius}, C. and {Bastian}, U. and {Portell}, J. and {Casta{\~n}eda}, J. and {Davidson}, M. and {Hambly}, N.~C. and {Clotet}, M. and {Biermann}, M. and {Mora}, A. and {Busonero}, D. and {Riva}, A. and {Brown}, A.~G.~A. and {Smart}, R. and {Lammers}, U. and {Torra}, J. and {Drimmel}, R. and {Gracia}, G. and {L{\"o}ffler}, W. and {Spagna}, A. and {Lindegren}, L. and {Klioner}, S. and {Andrei}, A. and {Bach}, N. and {Bramante}, L. and {Br{\"u}semeister}, T. and {Busso}, G. and {Carrasco}, J.~M. and {Gai}, M. and {Garralda}, N. and {Gonz{\'a}lez-Vidal}, J.~J. and {Guerra}, R. and {Hauser}, M. and {Jordan}, S. and {Jordi}, C. and {Lenhardt}, H. and {Mignard}, F. and {Messineo}, R. and {Mulone}, A. and {Serraller}, I. and {Stampa}, U. and {Tanga}, P. and {van Elteren}, A. and {van Reeven}, W. and {Voss}, H. and {Abbas}, U. and {Allasia}, W. and {Altmann}, M. and {Anton}, S. and {Barache}, C. and {Becciani}, U. and {Berthier}, J. and {Bianchi}, L. and {Bombrun}, A. and {Bouquillon}, S. and {Bourda}, G. and {Bucciarelli}, B. and {Butkevich}, A. and {Buzzi}, R. and {Cancelliere}, R. and {Carlucci}, T. and {Charlot}, P. and {Collins}, R. and {Comoretto}, G. and {Cross}, N. and {Crosta}, M. and {de Felice}, F. and {Fienga}, A. and {Figueras}, F. and {Fraile}, E. and {Geyer}, R. and {Hernandez}, J. and {Hobbs}, D. and {Hofmann}, W. and {Liao}, S. and {Licata}, E. and {Martino}, M. and {McMillan}, P.~J. and {Michalik}, D. and {Morbidelli}, R. and {Parsons}, P. and {Pecoraro}, M. and {Ramos-Lerate}, M. and {Sarasso}, M. and {Siddiqui}, H. and {Steele}, I. and {Steidelm{\"u}ller}, H. and {Taris}, F. and {Vecchiato}, A. and {Abreu}, A. and {Anglada}, E. and {Boudreault}, S. and {Cropper}, M. and {Holl}, B. and {Cheek}, N. and {Crowley}, C. and {Fleitas}, J.~M. and {Hutton}, A. and {Osinde}, J. and {Rowell}, N. and {Salguero}, E. and {Utrilla}, E. and {Blagorodnova}, N. and {Soffel}, M. and {Osorio}, J. and {Vicente}, D. and {Cambras}, J. and {Bernstein}, H. -H.},
        title = "{Gaia Data Release 1. Pre-processing and source list creation}",
      journal = {\aap},
         year = 2016,
        month = nov,
       volume = {595},
          eid = {A3},
        pages = {A3},
          doi = {10.1051/0004-6361/201628643},
archivePrefix = {arXiv},
       eprint = {1609.04273},
 primaryClass = {astro-ph.IM},
       adsurl = {https://ui.adsabs.harvard.edu/abs/2016A&A...595A...3F}
}

@article{Lindegren2012,
	author = {{Lindegren}, L. and {Lammers}, U. and {Hobbs}, D. and {O’Mullane}, W. and {Bastian}, U. and {Hernández}, J.},
	title = {The astrometric core solution for the Gaia mission - Overview of models, algorithms, and software implementation},
	DOI= "10.1051/0004-6361/201117905",
	url= "https://doi.org/10.1051/0004-6361/201117905",
	journal = {\aap},
	year = 2012,
	volume = 538,
	pages = "A78",
	month = "",
}

@article{EDR3-DPACP-130,
	author = {{Gaia Collaboration} and {Brown, A. G. A.} and {Vallenari, A.} and {Prusti, T.} and {de Bruijne, J. H. J.} and {Babusiaux, C.} and {Biermann, M.} and {Creevey, O. L.} and {Evans, D. W.} and {Eyer, L.} and {Hutton, A.} and {Jansen, F.} and {Jordi, C.} and {Klioner, S. A.} and {Lammers, U.} and {Lindegren, L.} and {Luri, X.} and {Mignard, F.} and {Panem, C.} and {Pourbaix, D.} and {Randich, S.} and {Sartoretti, P.} and {Soubiran, C.} and {Walton, N. A.} and {Arenou, F.} and {Bailer-Jones, C. A. L.} and {Bastian, U.} and {Cropper, M.} and {Drimmel, R.} and {Katz, D.} and {Lattanzi, M. G.} and {van Leeuwen, F.} and {Bakker, J.} and {Cacciari, C.} and {Castañeda, J.} and {De Angeli, F.} and {Ducourant, C.} and {Fabricius, C.} and {Fouesneau, M.} and {Frémat, Y.} and {Guerra, R.} and {Guerrier, A.} and {Guiraud, J.} and {Jean-Antoine Piccolo, A.} and {Masana, E.} and {Messineo, R.} and {Mowlavi, N.} and {Nicolas, C.} and {Nienartowicz, K.} and {Pailler, F.} and {Panuzzo, P.} and {Riclet, F.} and {Roux, W.} and {Seabroke, G. M.} and {Sordo, R.} and {Tanga, P.} and {Thévenin, F.} and {Gracia-Abril, G.} and {Portell, J.} and {Teyssier, D.} and {Altmann, M.} and {Andrae, R.} and {Bellas-Velidis, I.} and {Benson, K.} and {Berthier, J.} and {Blomme, R.} and {Brugaletta, E.} and {Burgess, P. W.} and {Busso, G.} and {Carry, B.} and {Cellino, A.} and {Cheek, N.} and {Clementini, G.} and {Damerdji, Y.} and {Davidson, M.} and {Delchambre, L.} and {Dell’Oro, A.} and {Fernández-Hernández, J.} and {Galluccio, L.} and {García-Lario, P.} and {Garcia-Reinaldos, M.} and {González-Núñez, J.} and {Gosset, E.} and {Haigron, R.} and {Halbwachs, J.-L.} and {Hambly, N. C.} and {Harrison, D. L.} and {Hatzidimitriou, D.} and {Heiter, U.} and {Hernández, J.} and {Hestroffer, D.} and {Hodgkin, S. T.} and {Holl, B.} and {Janßen, K.} and {Jevardat de Fombelle, G.} and {Jordan, S.} and {Krone-Martins, A.} and {Lanzafame, A. C.} and {Löffler, W.} and {Lorca, A.} and {Manteiga, M.} and {Marchal, O.} and {Marrese, P. M.} and {Moitinho, A.} and {Mora, A.} and {Muinonen, K.} and {Osborne, P.} and {Pancino, E.} and {Pauwels, T.} and {Petit, J.-M.} and {Recio-Blanco, A.} and {Richards, P. J.} and {Riello, M.} and {Rimoldini, L.} and {Robin, A. C.} and {Roegiers, T.} and {Rybizki, J.} and {Sarro, L. M.} and {Siopis, C.} and {Smith, M.} and {Sozzetti, A.} and {Ulla, A.} and {Utrilla, E.} and {van Leeuwen, M.} and {van Reeven, W.} and {Abbas, U.} and {Abreu Aramburu, A.} and {Accart, S.} and {Aerts, C.} and {Aguado, J. J.} and {Ajaj, M.} and {Altavilla, G.} and {Álvarez, M. A.} and {Álvarez Cid-Fuentes, J.} and {Alves, J.} and {Anderson, R. I.} and {Anglada Varela, E.} and {Antoja, T.} and {Audard, M.} and {Baines, D.} and {Baker, S. G.} and {Balaguer-Núñez, L.} and {Balbinot, E.} and {Balog, Z.} and {Barache, C.} and {Barbato, D.} and {Barros, M.} and {Barstow, M. A.} and {Bartolomé, S.} and {Bassilana, J.-L.} and {Bauchet, N.} and {Baudesson-Stella, A.} and {Becciani, U.} and {Bellazzini, M.} and {Bernet, M.} and {Bertone, S.} and {Bianchi, L.} and {Blanco-Cuaresma, S.} and {Boch, T.} and {Bombrun, A.} and {Bossini, D.} and {Bouquillon, S.} and {Bragaglia, A.} and {Bramante, L.} and {Breedt, E.} and {Bressan, A.} and {Brouillet, N.} and {Bucciarelli, B.} and {Burlacu, A.} and {Busonero, D.} and {Butkevich, A. G.} and {Buzzi, R.} and {Caffau, E.} and {Cancelliere, R.} and {Cánovas, H.} and {Cantat-Gaudin, T.} and {Carballo, R.} and {Carlucci, T.} and {Carnerero, M. I} and {Carrasco, J. M.} and {Casamiquela, L.} and {Castellani, M.} and {Castro-Ginard, A.} and {Castro Sampol, P.} and {Chaoul, L.} and {Charlot, P.} and {Chemin, L.} and {Chiavassa, A.} and {Cioni, M.-R. L.} and {Comoretto, G.} and {Cooper, W. J.} and {Cornez, T.} and {Cowell, S.} and {Crifo, F.} and {Crosta, M.} and {Crowley, C.} and {Dafonte, C.} and {Dapergolas, A.} and {David, M.} and {David, P.} and {de Laverny, P.} and {De Luise, F.} and {De March, R.} and {De Ridder, J.} and {de Souza, R.} and {de Teodoro, P.} and {de Torres, A.} and {del Peloso, E. F.} and {del Pozo, E.} and {Delbo, M.} and {Delgado, A.} and {Delgado, H. E.} and {Delisle, J.-B.} and {Di Matteo, P.} and {Diakite, S.} and {Diener, C.} and {Distefano, E.} and {Dolding, C.} and {Eappachen, D.} and {Edvardsson, B.} and {Enke, H.} and {Esquej, P.} and {Fabre, C.} and {Fabrizio, M.} and {Faigler, S.} and {Fedorets, G.} and {Fernique, P.} and {Fienga, A.} and {Figueras, F.} and {Fouron, C.} and {Fragkoudi, F.} and {Fraile, E.} and {Franke, F.} and {Gai, M.} and {Garabato, D.} and {Garcia-Gutierrez, A.} and {García-Torres, M.} and {Garofalo, A.} and {Gavras, P.} and {Gerlach, E.} and {Geyer, R.} and {Giacobbe, P.} and {Gilmore, G.} and {Girona, S.} and {Giuffrida, G.} and {Gomel, R.} and {Gomez, A.} and {Gonzalez-Santamaria, I.} and {González-Vidal, J. J.} and {Granvik, M.} and {Gutiérrez-Sánchez, R.} and {Guy, L. P.} and {Hauser, M.} and {Haywood, M.} and {Helmi, A.} and {Hidalgo, S. L.} and {Hilger, T.} and {Hładczuk, N.} and {Hobbs, D.} and {Holland, G.} and {Huckle, H. E.} and {Jasniewicz, G.} and {Jonker, P. G.} and {Juaristi Campillo, J.} and {Julbe, F.} and {Karbevska, L.} and {Kervella, P.} and {Khanna, S.} and {Kochoska, A.} and {Kontizas, M.} and {Kordopatis, G.} and {Korn, A. J.} and {Kostrzewa-Rutkowska, Z.} and {Kruszyńska, K.} and {Lambert, S.} and {Lanza, A. F.} and {Lasne, Y.} and {Le Campion, J.-F.} and {Le Fustec, Y.} and {Lebreton, Y.} and {Lebzelter, T.} and {Leccia, S.} and {Leclerc, N.} and {Lecoeur-Taibi, I.} and {Liao, S.} and {Licata, E.} and {Lindstrøm, E. P.} and {Lister, T. A.} and {Livanou, E.} and {Lobel, A.} and {Madrero Pardo, P.} and {Managau, S.} and {Mann, R. G.} and {Marchant, J. M.} and {Marconi, M.} and {Marcos Santos, M. M. S.} and {Marinoni, S.} and {Marocco, F.} and {Marshall, D. J.} and {Martin Polo, L.} and {Martín-Fleitas, J. M.} and {Masip, A.} and {Massari, D.} and {Mastrobuono-Battisti, A.} and {Mazeh, T.} and {McMillan, P. J.} and {Messina, S.} and {Michalik, D.} and {Millar, N. R.} and {Mints, A.} and {Molina, D.} and {Molinaro, R.} and {Molnár, L.} and {Montegriffo, P.} and {Mor, R.} and {Morbidelli, R.} and {Morel, T.} and {Morris, D.} and {Mulone, A. F.} and {Munoz, D.} and {Muraveva, T.} and {Murphy, C. P.} and {Musella, I.} and {Noval, L.} and {Ordénovic, C.} and {Orrù, G.} and {Osinde, J.} and {Pagani, C.} and {Pagano, I.} and {Palaversa, L.} and {Palicio, P. A.} and {Panahi, A.} and {Pawlak, M.} and {Peñalosa Esteller, X.} and {Penttilä, A.} and {Piersimoni, A. M.} and {Pineau, F.-X.} and {Plachy, E.} and {Plum, G.} and {Poggio, E.} and {Poretti, E.} and {Poujoulet, E.} and {Prša, A.} and {Pulone, L.} and {Racero, E.} and {Ragaini, S.} and {Rainer, M.} and {Raiteri, C. M.} and {Rambaux, N.} and {Ramos, P.} and {Ramos-Lerate, M.} and {Re Fiorentin, P.} and {Regibo, S.} and {Reylé, C.} and {Ripepi, V.} and {Riva, A.} and {Rixon, G.} and {Robichon, N.} and {Robin, C.} and {Roelens, M.} and {Rohrbasser, L.} and {Romero-Gómez, M.} and {Rowell, N.} and {Royer, F.} and {Rybicki, K. A.} and {Sadowski, G.} and {Sagristà Sellés, A.} and {Sahlmann, J.} and {Salgado, J.} and {Salguero, E.} and {Samaras, N.} and {Sanchez Gimenez, V.} and {Sanna, N.} and {Santoveña, R.} and {Sarasso, M.} and {Schultheis, M.} and {Sciacca, E.} and {Segol, M.} and {Segovia, J. C.} and {Ségransan, D.} and {Semeux, D.} and {Shahaf, S.} and {Siddiqui, H. I.} and {Siebert, A.} and {Siltala, L.} and {Slezak, E.} and {Smart, R. L.} and {Solano, E.} and {Solitro, F.} and {Souami, D.} and {Souchay, J.} and {Spagna, A.} and {Spoto, F.} and {Steele, I. A.} and {Steidelmüller, H.} and {Stephenson, C. A.} and {Süveges, M.} and {Szabados, L.} and {Szegedi-Elek, E.} and {Taris, F.} and {Tauran, G.} and {Taylor, M. B.} and {Teixeira, R.} and {Thuillot, W.} and {Tonello, N.} and {Torra, F.} and {Torra, J.} and {Turon, C.} and {Unger, N.} and {Vaillant, M.} and {van Dillen, E.} and {Vanel, O.} and {Vecchiato, A.} and {Viala, Y.} and {Vicente, D.} and {Voutsinas, S.} and {Weiler, M.} and {Wevers, T.} and {Wyrzykowski, Ł.} and {Yoldas, A.} and {Yvard, P.} and {Zhao, H.} and {Zorec, J.} and {Zucker, S.} and {Zurbach, C.} and {Zwitter, T.}},
	title = {Gaia Early Data Release 3 - Summary of the contents and survey properties},
	DOI= "10.1051/0004-6361/202039657",
	url= "https://doi.org/10.1051/0004-6361/202039657",
	journal = {\aap},
	year = 2021,
	volume = 649,
	pages = "A1",
}

@article{EDR3-DPACP-73,
	author = {{Rowell}, N. and {Davidson}, M. and {Lindegren}, L. and {van Leeuwen}, F. and {Castañeda}, J. and {Fabricius}, C. and {Bastian}, U. and {Hambly}, N. C. and {Hernández}, J. and {Bombrun}, A. and {Evans}, D. W. and {De Angeli}, F. and {Riello}, M. and {Busonero}, D. and {Crowley}, C. and {Mora}, A. and {Lammers}, U. and {Gracia}, G. and {Portell}, J. and {Biermann}, M. and {Brown}, A. G. A.},
	title = {Gaia Early Data Release 3 - Modelling and calibration of Gaia’s point and line spread functions},
	DOI= "10.1051/0004-6361/202039448",
	url= "https://doi.org/10.1051/0004-6361/202039448",
	journal = {\aap},
	year = 2021,
	volume = 649,
	pages = "A11",
}

@ARTICLE{anderson2000,
       author = {{Anderson}, Jay and {King}, Ivan R.},
        title = "{Toward High-Precision Astrometry with WFPC2. I. Deriving an Accurate Point-Spread Function}",
      journal = {\pasp},
         year = "2000",
        month = "Oct",
       volume = {112},
       number = {776},
        pages = {1360--1382},
          doi = {10.1086/316632},
archivePrefix = {arXiv},
       eprint = {astro-ph/0006325},
 primaryClass = {astro-ph},
       adsurl = {https://ui.adsabs.harvard.edu/abs/2000PASP..112.1360A}
}

@article{hirise,
author = {McEwen, Alfred S. and Eliason, Eric M. and Bergstrom, James W. and Bridges, Nathan T. and Hansen, Candice J. and Delamere, W. Alan and Grant, John A. and Gulick, Virginia C. and Herkenhoff, Kenneth E. and Keszthelyi, Laszlo and Kirk, Randolph L. and Mellon, Michael T. and Squyres, Steven W. and Thomas, Nicolas and Weitz, Catherine M.},
title = {Mars Reconnaissance Orbiter's High Resolution Imaging Science Experiment (HiRISE)},
journal = {Journal of Geophysical Research: Planets},
volume = {112},
number = {E5},
pages = {},
doi = {https://doi.org/10.1029/2005JE002605},
year = {2007}
}

@Article{lroc,
author={Robinson, M. S.
and Brylow, S. M.
and Tschimmel, M.
and Humm, D.
and Lawrence, S. J.
and Thomas, P. C.
and Denevi, B. W.
and Bowman-Cisneros, E.
and Zerr, J.
and Ravine, M. A.
and Caplinger, M. A.
and Ghaemi, F. T.
and Schaffner, J. A.
and Malin, M. C.
and Mahanti, P.
and Bartels, A.
and Anderson, J.
and Tran, T. N.
and Eliason, E. M.
and McEwen, A. S.
and Turtle, E.
and Jolliff, B. L.
and Hiesinger, H.},
title={Lunar Reconnaissance Orbiter Camera (LROC) Instrument Overview},
journal={Space Science Reviews},
year={2010},
month={Jan},
day={01},
volume={150},
number={1},
pages={81-124},
issn={1572-9672},
doi={10.1007/s11214-010-9634-2},
url={https://doi.org/10.1007/s11214-010-9634-2}
}

@ARTICLE{1998AJ....116.3040G,
       author = {{Gunn}, J.~E. and {Carr}, M. and {Rockosi}, C. and {Sekiguchi}, M. and {Berry}, K. and {Elms}, B. and {de Haas}, E. and {Ivezi{\'c}}, {\v{Z}} . and {Knapp}, G. and {Lupton}, R. and {Pauls}, G. and {Simcoe}, R. and {Hirsch}, R. and {Sanford}, D. and {Wang}, S. and {York}, D. and {Harris}, F. and {Annis}, J. and {Bartozek}, L. and {Boroski}, W. and {Bakken}, J. and {Haldeman}, M. and {Kent}, S. and {Holm}, S. and {Holmgren}, D. and {Petravick}, D. and {Prosapio}, A. and {Rechenmacher}, R. and {Doi}, M. and {Fukugita}, M. and {Shimasaku}, K. and {Okada}, N. and {Hull}, C. and {Siegmund}, W. and {Mannery}, E. and {Blouke}, M. and {Heidtman}, D. and {Schneider}, D. and {Lucinio}, R. and {Brinkman}, J.},
        title = "{The Sloan Digital Sky Survey Photometric Camera}",
      journal = {\aj},
         year = 1998,
        month = dec,
       volume = {116},
       number = {6},
        pages = {3040-3081},
          doi = {10.1086/300645},
archivePrefix = {arXiv},
       eprint = {astro-ph/9809085},
 primaryClass = {astro-ph},
       adsurl = {https://ui.adsabs.harvard.edu/abs/1998AJ....116.3040G}
}

@BOOK{2007ASSL..350.....V,
   author = {{van Leeuwen}, Floor},
    title = "{Hipparcos, the New Reduction of the Raw Data}",
booktitle = {Hipparcos, the New Reduction of the Raw Data},
     year = 2007,
publisher = {Springer Dordrecht},
   series = {Astrophysics and Space Science Library},
   volume = {350},
      doi = {10.1007/978-1-4020-6342-8},
   adsurl = {https://ui.adsabs.harvard.edu/abs/2007ASSL..350.....V}
}

@ARTICLE{Refregier_2003,
       author = {{Refregier}, Alexandre},
        title = "{Shapelets - I. A method for image analysis}",
      journal = {\mnras},
         year = "2003",
        month = "Jan",
       volume = {338},
       number = {1},
        pages = {35-47},
          doi = {10.1046/j.1365-8711.2003.05901.x},
archivePrefix = {arXiv},
       eprint = {astro-ph/0105178},
 primaryClass = {astro-ph},
       adsurl = {https://ui.adsabs.harvard.edu/abs/2003MNRAS.338...35R}
}

@BOOK{Bierman_1977,
   author = {{Bierman}, G.~J.},
    title = "{Factorization Methods for Discrete Sequential Estimation}",
booktitle = {Factorization Methods for Discrete Sequential Estimation, by G.~J.~Bierman.~ Mathematics in Science and Engineering, volume 128, 241 pages Academic Press, New York, NY, 1977},
     year = 1977,
publisher = {Dover},
   adsurl = {https://ui.adsabs.harvard.edu/abs/1977fmds.book.....B}
}

@UNPUBLISHED{LL:LL-084,
author = {{Lindegren}, L.},
title={{M}inimum-dimension {L}{S}{F} modelling},
institution={Lund Observatory},
year={2009},
month={August},
url={http://www.rssd.esa.int/doc_fetch.php?id=2915742},
note={{GAIA}-C3-TN-LU-LL-084},
type={Technical note}
}

@UNPUBLISHED{LL:LL-078,
author = {{Lindegren}, L.},
title={{A} general {M}aximum-{L}ikelihood algorithm for model fitting to {C}{C}{D} sample data},
institution={Lund Observatory},
year={2008},
month={November},
url={https://dms.cosmos.esa.int/COSMOS/doc_fetch.php?id=2861864},
note={{GAIA}-C3-TN-LU-LL-078},
type={Technical note} }

@UNPUBLISHED{LL:LL-090,
author = {{Lindegren}, L.},
title={{A} posteriori reduction of the set of basis functions for {L}{S}{F} modelling},
institution={Lund Observatory},
year={2010},
month={October},
url={https://dms.cosmos.esa.int/COSMOS/doc_fetch.php?id=3049411},
note={{GAIA}-C3-TN-LU-LL-090},
type={Technical Note} }

@ARTICLE{2023arXiv231200488H,
       author = {{H{\o}g}, Erik},
        title = "{GaiaNIR: Note on processing and photometry}",
      journal = {arXiv e-prints},
         year = 2023,
        month = dec,
          eid = {arXiv:2312.00488},
        pages = {arXiv:2312.00488},
          doi = {10.48550/arXiv.2312.00488},
archivePrefix = {arXiv},
       eprint = {2312.00488},
 primaryClass = {astro-ph.IM},
       adsurl = {https://ui.adsabs.harvard.edu/abs/2023arXiv231200488H}
}

@ARTICLE{hobbs2021,
       author = {{Hobbs}, D. and {Brown}, A. and {H{\o}g}, E. and {Jordi}, C. and {Kawata}, D. and
       {Tanga}, P. and {Klioner}, S. and {Sozzetti}, A. and {Wyrzykowski}, {\L}. and {Walton}, N. and
       {Vallenari}, A. and {Makarov}, V. and {Rybizki}, J. and {Jiménez-Esteban}, F. and {Caballero}, J. and
       {McMillan}, P. and {Secrest}, N. and {Mor}, R. and {Andrews}, J. and {Zwitter}, T. and
       {Chiappini}, C. and {Fynbo}, J. and {Ting}, Y. and {Hestroffer}, D. and {Lindegren}, L. and
       {McArthur}, B. and {Gouda}, N. and {Moore}, A. and {Gonzalez}, O. and {Vaccari}, M.},
        title = "{All-sky visible and near infrared space astrometry}",
      journal = {Experimental Astronomy},
         year = "2021",
       volume = {51},
        pages = {783--843},
          doi = {10.1007/s10686-021-09705-z}
}

@misc{rixon2023,
  author = {{Rixon}, G. and {Walton}, N. and {Busso}, G. and {Baker}, I. and {Gonzalez}, O. and {Miller}, C. and {Kawata}, D. and {Deason}, A. and {Fattahi}, A. and {Helly}, J.},
  title = "{Detector options for GaiaNIR}",
  url = {https://www.astro.lu.se/sites/astro.lu.se/files/2023-07/Rixon-Lund-Jul-2023.pdf},
  year = {2023},
  month = jul,
  day = {20}
}

@ARTICLE{2021ExA....51..783H,
       author = {{Hobbs}, David and {Brown}, Anthony and {H{\o}g}, Erik and {Jordi}, Carme and {Kawata}, Daisuke and {Tanga}, Paolo and {Klioner}, Sergei and {Sozzetti}, Alessandro and {Wyrzykowski}, {\L}ukasz and {Walton}, Nicholas and {Vallenari}, Antonella and {Makarov}, Valeri and {Rybizki}, Jan and {Jim{\'e}nez-Esteban}, Fran and {Caballero}, Jos{\'e} A. and {McMillan}, Paul J. and {Secrest}, Nathan and {Mor}, Roger and {Andrews}, Jeff J. and {Zwitter}, Toma{\v{z}} and {Chiappini}, Cristina and {Fynbo}, Johan P.~U. and {Ting}, Yuan-Sen and {Hestroffer}, Daniel and {Lindegren}, Lennart and {McArthur}, Barbara and {Gouda}, Naoteru and {Moore}, Anna and {Gonzalez}, Oscar A. and {Vaccari}, Mattia},
        title = "{All-sky visible and near infrared space astrometry}",
      journal = {Experimental Astronomy},
         year = 2021,
        month = jun,
       volume = {51},
       number = {3},
        pages = {783-843},
          doi = {10.1007/s10686-021-09705-z},
archivePrefix = {arXiv},
       eprint = {1907.12535},
 primaryClass = {astro-ph.IM},
       adsurl = {https://ui.adsabs.harvard.edu/abs/2021ExA....51..783H}
}

@article {2014JInst...9C3048A,
	title = {The brighter-fatter effect and pixel correlations in CCD sensors},
	journal = {Journal of Instrumentation},
	volume = {9},
	year = {2014},
	month = {mar},
	pages = {C03048},
	doi = {10.1088/1748-0221/9/03/C03048},
	author = {P. Antilogus and Astier, P. and Doherty, P. and A. Guyonnet and Regnault, N.}
}

@article{EDR3-DPACP-117,
author = {{Riello}, M. and {De Angeli}, F. and {Evans}, D.~W. and {Montegriffo}, P. and {Carrasco}, J.~M. and others},
	title = "{Gaia Early Data Release 3: Photometric content and validation}",
	DOI= "10.1051/0004-6361/202039587",
	url= "https://doi.org/10.1051/0004-6361/202039587",
	journal = {\aap},
	year = 2020,
}

@ARTICLE{2015JInst..10C5032G,
       author = {{Gruen}, D. and {Bernstein}, G.~M. and {Jarvis}, M. and {Rowe}, B. and {Vikram}, V. and {Plazas}, A.~A. and {Seitz}, S.},
        title = "{Characterization and correction of charge-induced pixel shifts in DECam}",
      journal = {Journal of Instrumentation},
         year = 2015,
        month = may,
       volume = {10},
       number = {5},
          eid = {C05032},
        pages = {C05032},
          doi = {10.1088/1748-0221/10/05/C05032},
archivePrefix = {arXiv},
       eprint = {1501.02802},
 primaryClass = {astro-ph.IM},
       adsurl = {https://ui.adsabs.harvard.edu/abs/2015JInst..10C5032G}
}

@ARTICLE{2023A&A...670A.118A,
       author = {{Astier}, Pierre and {Regnault}, Nicolas},
        title = "{Correction of the brighter-fatter effect on the CCDs of Hyper Suprime-Cam}",
      journal = {\aap},
         year = 2023,
        month = feb,
       volume = {670},
          eid = {A118},
        pages = {A118},
          doi = {10.1051/0004-6361/202245407},
archivePrefix = {arXiv},
       eprint = {2301.03274},
 primaryClass = {astro-ph.IM},
       adsurl = {https://ui.adsabs.harvard.edu/abs/2023A&A...670A.118A}
}

@ARTICLE{2011MNRAS.414.2215P,
   author = {{Prod'homme}, T. and {Brown}, A.~G.~A. and {Lindegren}, L. and 
	{Short}, A.~D.~T. and {Brown}, S.~W.},
    title = "{Electrode level Monte Carlo model of radiation damage effects on astronomical CCDs}",
  journal = {\mnras},
archivePrefix = "arXiv",
   eprint = {1103.3630},
 primaryClass = "astro-ph.IM",
     year = 2011,
    month = jul,
   volume = 414,
    pages = {2215-2228},
      doi = {10.1111/j.1365-2966.2011.18537.x},
   adsurl = {http://adsabs.harvard.edu/abs/2011MNRAS.414.2215P}
}

@ARTICLE{2013MNRAS.430.3078S,
       author = {{Short}, A. and {Crowley}, C. and {de Bruijne}, J.~H.~J. and {Prod'homme}, T.},
        title = "{An analytical model of radiation-induced Charge Transfer Inefficiency for CCD detectors}",
      journal = {\mnras},
         year = 2013,
        month = apr,
       volume = {430},
       number = {4},
        pages = {3078-3085},
          doi = {10.1093/mnras/stt114},
archivePrefix = {arXiv},
       eprint = {1302.1416},
 primaryClass = {astro-ph.IM},
       adsurl = {https://ui.adsabs.harvard.edu/abs/2013MNRAS.430.3078S}
}

@ARTICLE{2022JATIS...8a6003A,
       author = {{Ahmed}, Saad and {Hall}, David and {Crowley}, Cian and {Skottfelt}, Jesper and {Dryer}, Ben and {Seabroke}, George and {Hernandez}, Jose and {Holland}, Andrew},
        title = "{Understanding the evolution of radiation damage on the Gaia CCDs after 72 months at L2}",
      journal = {Journal of Astronomical Telescopes, Instruments, and Systems},
         year = 2022,
        month = jan,
       volume = {8},
          eid = {016003},
        pages = {016003},
          doi = {10.1117/1.JATIS.8.1.016003},
       adsurl = {https://ui.adsabs.harvard.edu/abs/2022JATIS...8a6003A}
}

@article{pemnu,
	author = {{Hambly}, N.~C. and {Cropper}, M. and {Boudreault}, S. and {Crowley}, C. and {Kohley}, R. and {de Bruijne}, J.~H.~J. and {Dolding}, C. and {Fabricius}, C. and {Seabroke}, G. and {Davidson}, M. and {Rowell}, N. and {Collins}, R. and {Cross}, N. and {Mart\'{\i}n-Fleitas}, J. and {Baker}, S. and {Smith}, M. and {Sartoretti}, P. and {Marchal}, O. and {Katz}, D. and {De Angeli}, F. and {Busso}, G. and {Riello}, M. and {Allende Prieto}, C. and {Els}, S. and {Corcione}, L. and {Masana}, E. and {Luri}, X. and {Chassat}, F. and {Fusero}, F. and {Pasquier}, J.~F. and {V\'etel}, C. and {Sarri}, G. and {Gare}, P.},
	title = {Gaia Data Release 2 - Calibration and mitigation of electronic offset effects in the data},
	DOI= "10.1051/0004-6361/201832716",
	url= "https://doi.org/10.1051/0004-6361/201832716",
	journal = {\aap},
	year = 2018,
	volume = 616,
	pages = "A15",
}

@ARTICLE{2013MNRAS.430.3155S,
       author = {{Seabroke}, G.~M. and {Prod'homme}, T. and {Murray}, N.~J. and {Crowley}, C. and {Hopkinson}, G. and {Brown}, A.~G.~A. and {Kohley}, R. and {Holland}, A.},
        title = "{Digging supplementary buried channels: investigating the notch architecture within the CCD pixels on ESA's Gaia satellite}",
      journal = {\mnras},
         year = 2013,
        month = apr,
       volume = {430},
       number = {4},
        pages = {3155-3170},
          doi = {10.1093/mnras/stt121},
archivePrefix = {arXiv},
       eprint = {1302.1873},
 primaryClass = {astro-ph.IM},
       adsurl = {https://ui.adsabs.harvard.edu/abs/2013MNRAS.430.3155S}
}

@article{EDR3-DPACP-128,
	author = {{Lindegren}, Lennart and {Klioner}, S. A. and {Hern\'andez}, J. and {Bombrun}, A. and {Ramos-Lerate}, M. and {Steidelm\"uller}, H. and {Bastian}, U. and {Biermann}, M. and {de Torres}, A. and {Gerlach}, E. and {Geyer}, R. and {Hilger}, T. and {Hobbs}, D. and {Lammers}, U. and {McMillan}, P. J. and {Stephenson}, C. A. and {Casta\~neda}, J. and {Davidson}, M.},
	title = "{Gaia Early Data Release 3. The astrometric solution}",
	journal = {\aap},
    year = 2021,
    month = may,
    volume = {649},
    eid = {A2},
    pages = {A2},
    doi = {10.1051/0004-6361/202039709},
	archivePrefix = {arXiv},
    eprint = {2012.03380},
    primaryClass = {astro-ph.IM},
    adsurl = {https://ui.adsabs.harvard.edu/abs/2021A&A...649A...2L}}

@ARTICLE{2012MNRAS.419.2995P,
       author = {{Prod'homme}, T. and {Holl}, B. and {Lindegren}, L. and {Brown}, A.~G.~A.},
        title = "{The impact of CCD radiation damage on Gaia astrometry - I. Image location estimation in the presence of radiation damage}",
      journal = {\mnras},
         year = 2012,
        month = feb,
       volume = {419},
       number = {4},
        pages = {2995-3017},
          doi = {10.1111/j.1365-2966.2011.19934.x},
archivePrefix = {arXiv},
       eprint = {1110.1547},
 primaryClass = {astro-ph.IM},
       adsurl = {https://ui.adsabs.harvard.edu/abs/2012MNRAS.419.2995P}
}

@ARTICLE{2012MNRAS.422.2786H,
       author = {{Holl}, B. and {Prod'homme}, T. and {Lindegren}, L. and {Brown}, A.~G.~A.},
        title = "{The impact of CCD radiation damage on Gaia astrometry - II. Effect of image location errors on the astrometric solution}",
      journal = {\mnras},
         year = 2012,
        month = jun,
       volume = {422},
       number = {4},
        pages = {2786-2807},
          doi = {10.1111/j.1365-2966.2012.20429.x},
       adsurl = {https://ui.adsabs.harvard.edu/abs/2012MNRAS.422.2786H}
}

@ARTICLE{2026MNRAS.546f2186M,
       author = {{Massey}, Richard and {Kegerreis}, Jacob A. and {Barrios}, Juan Paolo Lorenzo Gerardo and {Nightingale}, James W. and {Hayes}, Richard G. and {Lagattuta}, David and {Lentz}, Zane D. and {Leroy}, Gavin and {Skottfelt}, Jesper and {Vecchi}, Felix and {von Wietersheim-Kramsta}, Maximilian},
        title = "{Radiation damage to the Hubble Space Telescope during two solar cycles, and correction of charge transfer inefficiency using ArCTIc}",
      journal = {\mnras},
         year = 2026,
        month = feb,
       volume = {546},
       number = {2},
          eid = {staf2186},
        pages = {staf2186},
          doi = {10.1093/mnras/staf2186},
archivePrefix = {arXiv},
       eprint = {2509.05057},
 primaryClass = {astro-ph.IM},
       adsurl = {https://ui.adsabs.harvard.edu/abs/2026MNRAS.546f2186M}
}

@ARTICLE{2011BASI...39..289G,
       author = {{Green}, D.~A.},
        title = "{A colour scheme for the display of astronomical intensity images}",
      journal = {Bulletin of the Astronomical Society of India},
         year = 2011,
        month = jun,
       volume = {39},
        pages = {289-295},
archivePrefix = {arXiv},
       eprint = {1108.5083},
 primaryClass = {astro-ph.IM},
       adsurl = {https://ui.adsabs.harvard.edu/abs/2011BASI...39..289G}
}
\begin{acknowledgements}
This work has made use of data from the European Space Agency (ESA) mission
\gaia (\url{https://www.cosmos.esa.int/gaia}), processed by the \gaia
Data Processing and Analysis Consortium (DPAC,
\url{https://www.cosmos.esa.int/web/gaia/dpac/consortium}). Funding for the DPAC
has been provided by national institutions, in particular the institutions
participating in the \gaia Multilateral Agreement. The \gaia\
mission website is:
\url{http://www.cosmos.esa.int/gaia}.

The work described in this paper has been financially supported by
the United Kingdom Science and Technology Facilities Council (STFC) and the
United Kingdom Space Agency (UKSA) through the following grants:
ST/N000595/1,
ST/S000976/1,
ST/W002493/1, ST/X001601/1,
ST/N000595/1, and UKRI9452:APP68395;
by the Spanish MICIN/AEI/10.13039/501100011033 and 
``ERDF - A way of making Europe" by the European Union through grants
PID2021-122842OB-C21 and PID2024-157964OB-C21, and the Institute of Cosmos
Sciences University of Barcelona (ICCUB, Unidad de Excelencia Mar\'{i}a de
Maeztu) through grant CEX2024-001451-M and the project 2021-SGR-00679 GRC de
l'Ag\`encia de Gesti\'o d'Ajuts Universitaris i de Recerca (Generalitat de
Catalunya); by ESA PRODEX 4000145062;
and by the Swedish National Space Board (SNSB/Rymdstyrelsen).
Finally, the authors also acknowledge the computer resources from MareNostrum,
and the technical expertise and assistance provided by the Red Espa\~{n}ola de
Supercomputaci\'{o}n at the Barcelona Supercomputing Center, Centro Nacional de
Supercomputaci\'{o}n.
\end{acknowledgements}
\begin{appendix}
\nolinenumbers
\section{Along-scan pixel scale estimation \label{app:al_pix_scale}}
As explained in Sect.~\ref{sec:terms} the AL angular pixel scale $\alPixScale$
is required in order to convert the stellar image drift rate from angular units
(mas\,sec$^{-1}$) to spatial units (pix\,sec$^{-1}$), for input to the PLSF modelling. During
development of the PSF model described in this paper we found that the nominal
value of the AL angular pixel scale (a single, fixed value of
$\alPixScale=58.933$ mas\,pix$^{-1}$) is not sufficiently accurate for this. In reality,
the pixel scale varies between the two telescopes due to differences in the
focal length. It also varies in time due to evolution in the focus, and across
the focal plane with several possible causes, such as optical aberrations,
variations in pixel pitch and CCD vertical mounting error.
As such, a calibrated value of $\alPixScale$ is required in order to account
for these effects. 

While the geometric part of the AGIS instrument model \citep[see][section
3.4]{Lindegren2012} includes no explicit calibration of $\alPixScale$, we
found that it can be estimated to sufficient accuracy by using the calibrated
angular positions of the CCD fiducial lines.
Specifically, the calibrated $\eta$ field angle for a particular CCD gate
provides an estimate of the angular position of the gate fiducial line, which
itself lies at a fixed position in the along scan direction on the CCD (e.g. see
Table~\ref{tab:ccdgates}). By comparing the $\eta$ value for several different
CCD gates within the same device, we can estimate the dependence of $\eta$ on
the TDI line number, the gradient of which corresponds to $\alPixScale$,
i.e.~$\alPixScale = \partial \eta(\tau) / \partial \tau$.

This is depicted in Fig.~\ref{fig:alPixScaleExample}, where in the upper panel
we plot $\eta$ against TDI line number for the five longest CCD gates. The
positions of the gate fiducial lines are indicated with the vertical grey lines.
Long gates provide better constraint as the fiducial lines are further apart,
so gates shorter than 9 are not used.
The purple line is a linear fit to the points, with the lower panel depicting
the residuals. The (constant) gradient of the fitted line gives the calibrated
value of $\alPixScale$, which in this example is $58.927$ mas\,pix$^{-1}$.
In principle a higher order fit could be performed, which would allow the
estimated pixel scale to vary in the AL direction. However, a constant value was
found to be sufficient, which is fortunate since a significant dependence of
$\alPixScale$ on TDI line number would greatly complicate the PSF modelling.
Finally, this algorithm cannot be applied to SM devices because only GATE12 is
present. Instead, we use the equivalent AF1 value and apply a small correction
factor that was determined by a separate optimisation procedure.
\begin{figure}
\resizebox{\hsize}{!}{\includegraphics{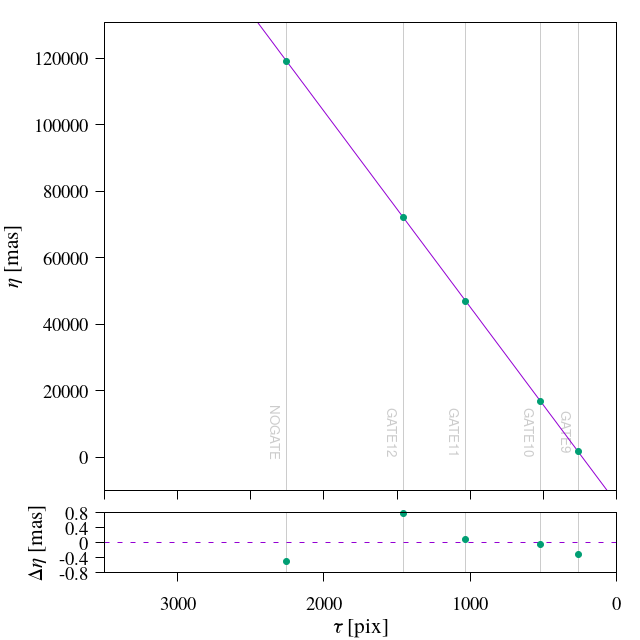}}
\caption{Estimation of the FOV1 AL pixel scale in device ROW1 AF7, at
AC coordinate 1000 and at revolution 4500. The vertical grey lines indicate the
positions of the fiducial lines for the 5 longest CCD gates, which are the ones
used operationally to estimate the pixel scale.}
\label{fig:alPixScaleExample}
\end{figure}
In Fig.~\ref{fig:alPixScaleFpaFov1} we depict the variation in AL pixel scale
across the focal plane, for FOV1 at revolution 4500. The variation is generally
quite smooth.
\begin{figure}
\resizebox{\hsize}{!}{\includegraphics{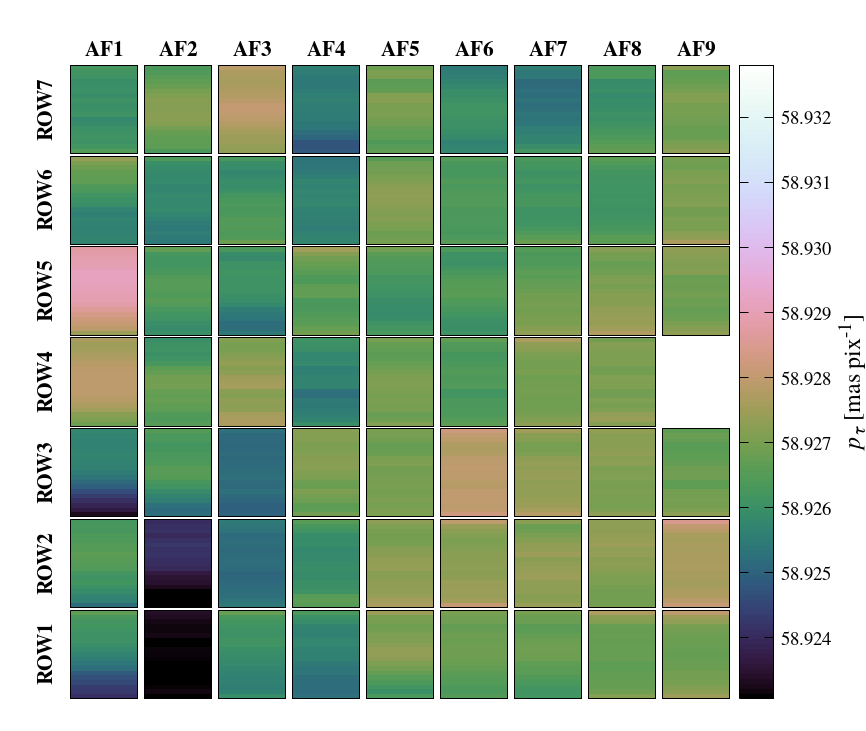}}
\caption{Variation in AL pixel scale across the focal plane as
determined by the algorithm described in this section, for FOV1 at revolution
4500.}
\label{fig:alPixScaleFpaFov1}
\end{figure}
In Fig.~\ref{fig:alPixScaleTime} we depict the time evolution in the AL pixel
scale for one particular calibration. The discontinuities are associated with
thermal decontaminations, refocuses and other major mission events.
\begin{figure}
\resizebox{\hsize}{!}{\includegraphics{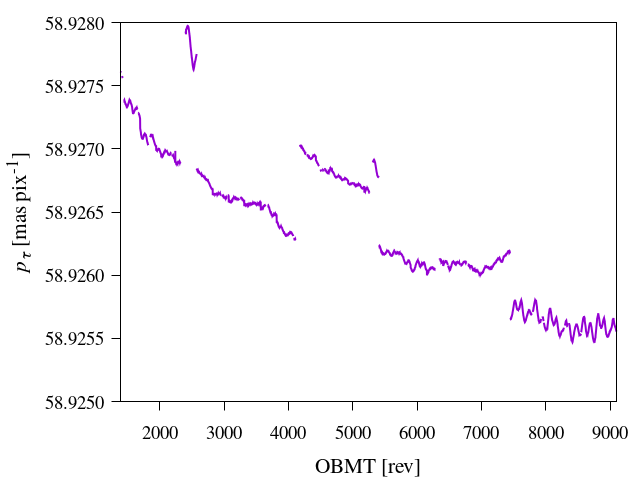}}
\caption{Evolution in AL pixel scale over most of the DR4 time
range, for one particular calibration (FOV1 ROW1 AF7 at AC coordinate 1000).}
\label{fig:alPixScaleTime}
\end{figure}
\section{Native AC rate estimation \label{app:native_ac_rate}}
The native AC rate $\dot{\mu}_0$ (see Sect.~\ref{sec:terms}) is the AC rate at
which the AC smearing is minimal.
It is offset from zero due to a combination of optical effects and small
orientation errors in the placement of the CCDs in the focal plane. The value of
$\dot{\mu}_0$ is required in order to convert the source AC motion to the
corresponding AC smearing width, and it therefore needs to be measured and
incorporated into the calibration pipeline.

Fortunately, $\dot{\mu}_0$ can be measured directly from the observations in a
relatively simple manner, using the principle that sources moving at
$\dot{\mu}_0$ are minimally smeared AC. We first select a large number of
ungated 2D observations, which are then debiased, background-subtracted,
normalised, and finally marginalised to form the 1D LSF in the AC direction.
These are aligned according to the predicted AC location of the source, and
binned into narrow ranges of AC rate. Within each bin we estimate the AC
full-width-half-maximum (FWHM) of the observations, by fitting a polynomial to
the samples then finding the maximum and the two adjacent half-maximum
locations. This is performed independently for each FOV and CCD combination;
in Fig.~\ref{fig:acFwhmVersusAcRate} we present two plots depicting the
polynomial fits for two different AC rate bins in FOV1 ROW3 AF5. The three
vertical lines indicate the maximum and half-maximum locations from which the AC
FWHM is derived, with the resulting values indicated in the figures.
\begin{figure}
\resizebox{\hsize}{!}{\includegraphics{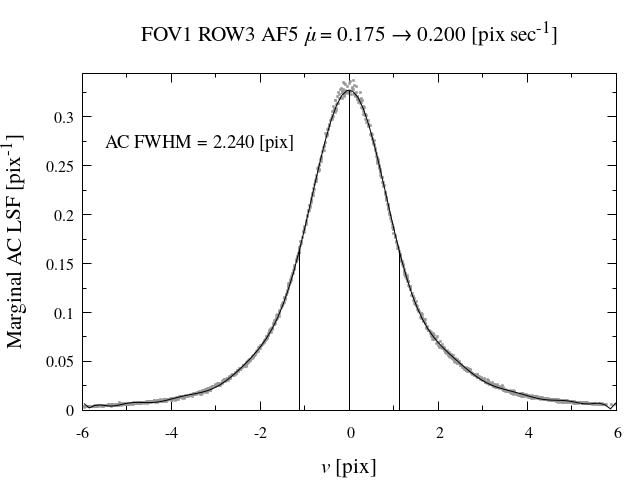}}
\resizebox{\hsize}{!}{\includegraphics{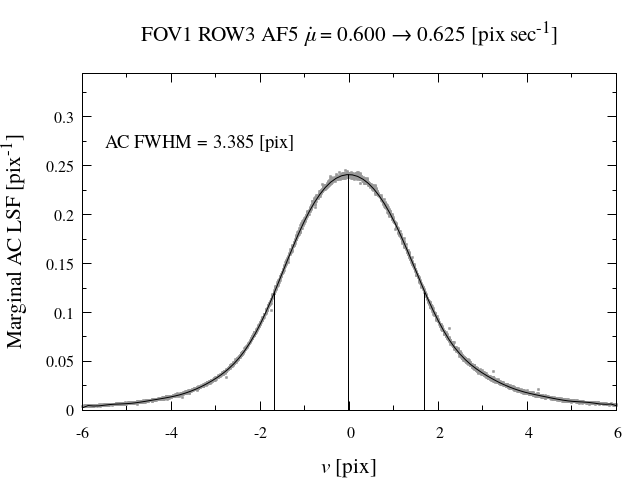}}
\caption{Example estimation of the AC FWHM as a function of AC rate,
for one particular FOV (FOV1) and CCD (ROW3 AF5). The upper and lower panels
correspond to ungated 2D observations in the AC rate ranges 0.175-0.200 and
0.600-0.625 pix\,sec$^{-1}$, respectively. The grey points are samples from the
marginalised and normalised AC LSFs of a large number of observations, aligned
according to the AC location of the source. The black curve is a polynomial fit
to the samples, with the vertical lines indicating the maximum and half-maximum
locations from which the AC FWHM (quoted in the figures) is derived.}
\label{fig:acFwhmVersusAcRate}
\end{figure}

This procedure results in a set of discrete samples of the AC FWHM as a function
of AC rate, with observations at high positive and negative AC rate having the
largest AC FWHM and the minimum falling close to zero. The precise location of the
minimum gives the value for $\dot{\mu}_0$; this needs to be estimated to an
accuracy finer than the bin size in AC rate, so we must fit an interpolating
function to the samples.
Empirically, the resulting profile has a roughly hyperbolic form, and we found
that it can be well modelled in all devices and FOVs by the
pseudo-hyperbolic function
\begin{equation*}
f(x) = \left(A + B(x-C)^2\right)^{\frac{1}{D}} + \left(E +
F(x-C)^4\right)^{\frac{1}{G}} \,,
\end{equation*}
where $A,B,C,D,E,F,G$ are free parameters that are fitted using the
Levenberg-Marquardt algorithm. The native AC rate corresponds to the translation
from the origin, which is given by the parameter $C$.
In Fig.~\ref{fig:native_ac_rate_row3_af5} we depict this fitting procedure for
the two FOVs in a single CCD (ROW3 AF5).
\begin{figure}
\resizebox{\hsize}{!}{\includegraphics{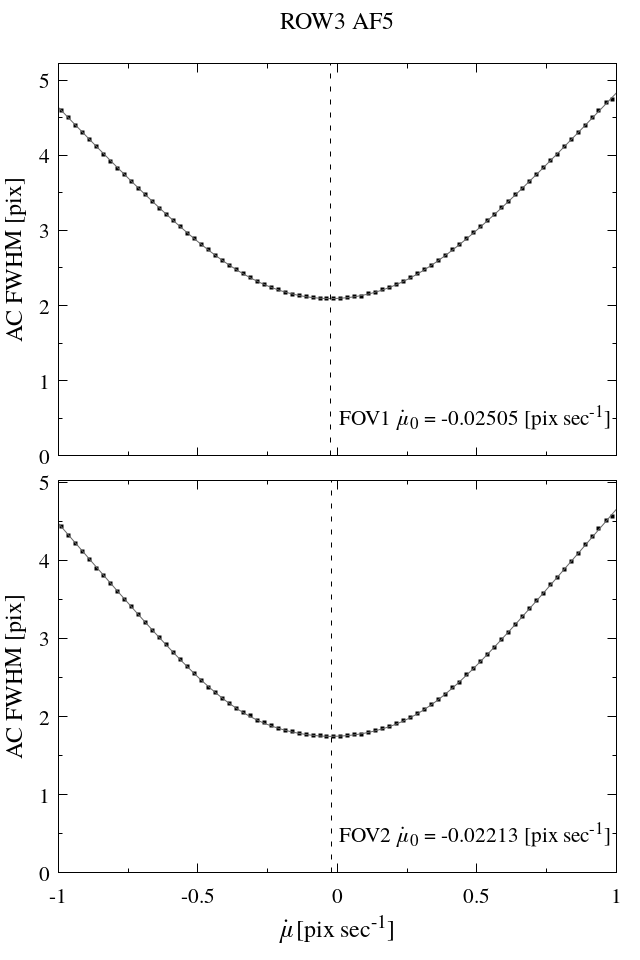}}
\caption{Example estimation of the native AC rate $\dot{\mu}_0$ for
FOV1 (upper panel) and FOV2 (lower) in the same device (ROW3 AF5). The value of
$\dot{\mu}_0$ coincides with the AC rate for which the AC FWHM is minimised, as
determined by fitting an appropriate pseudo-hyperbolic function (grey line)
through the data (black dots). The vertical dashed lines indicate the minimum of
the interpolating function, which provides the estimate of $\dot{\mu}_0$.}
\label{fig:native_ac_rate_row3_af5}
\end{figure}

This procedure estimates an independent, constant value of $\dot{\mu}_0$ for
each FOV and CCD. Within the same device the two FOVs have similar values of
$\dot{\mu}_0$ due partly to the common CCD rotation error and partly to similar
projection effects in the same region of the focal plane.
This is evident in Fig.~\ref{fig:native_ac_rate_correlation}, which shows the
correlation of $\dot{\mu}_0$ between FOV1 and FOV2, for all AF devices. SM
devices are excluded because each is used by only one of the two FOVs.
\begin{figure}
\resizebox{\hsize}{!}{\includegraphics{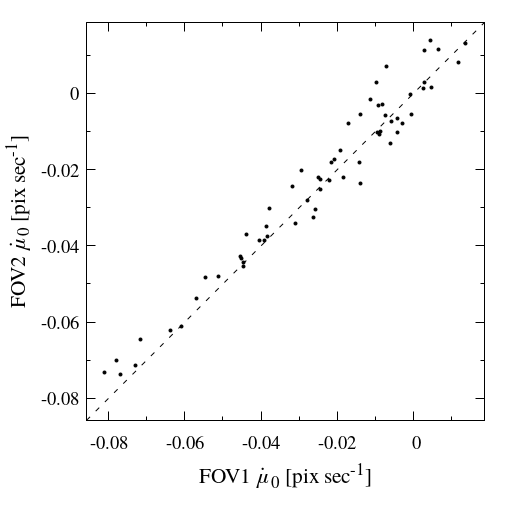}}
\caption{Native AC rate $\dot{\mu}_0$ in FOV1 and FOV2 for all AF devices.}
\label{fig:native_ac_rate_correlation}
\end{figure}
Note that the use of a single, constant value of $\dot{\mu}_0$ for each FOV and
CCD ignores any possible time dependence or spatial variation within each
device. We checked for time dependence in $\dot{\mu}_0$ and found it to be
negligible. Spatial variation in the AC direction within each device is likely
to be more significant, albeit small, and in future data processing we may allow
$\dot{\mu}_0$ to additionally depend on AC position in the CCD.
\section{CCD device types and twins \label{app:ccd_types}}
In Fig.~\ref{fig:deviceTypes} we present the CCD type information for all SM
and AF devices, where the type refers to whether the CCD was manufactured
from the left half (TYPE-01) or the right half (TYPE-02) of the circular silicon
wafer.
As explained in Sect.~\ref{sec:corner}, the type has an impact on spatial
variations in the CCD response and the behaviour of different gates, due to the
corner effect.
\begin{figure}
\resizebox{\hsize}{!}{\includegraphics[trim={1cm 0 5cm 0},clip]{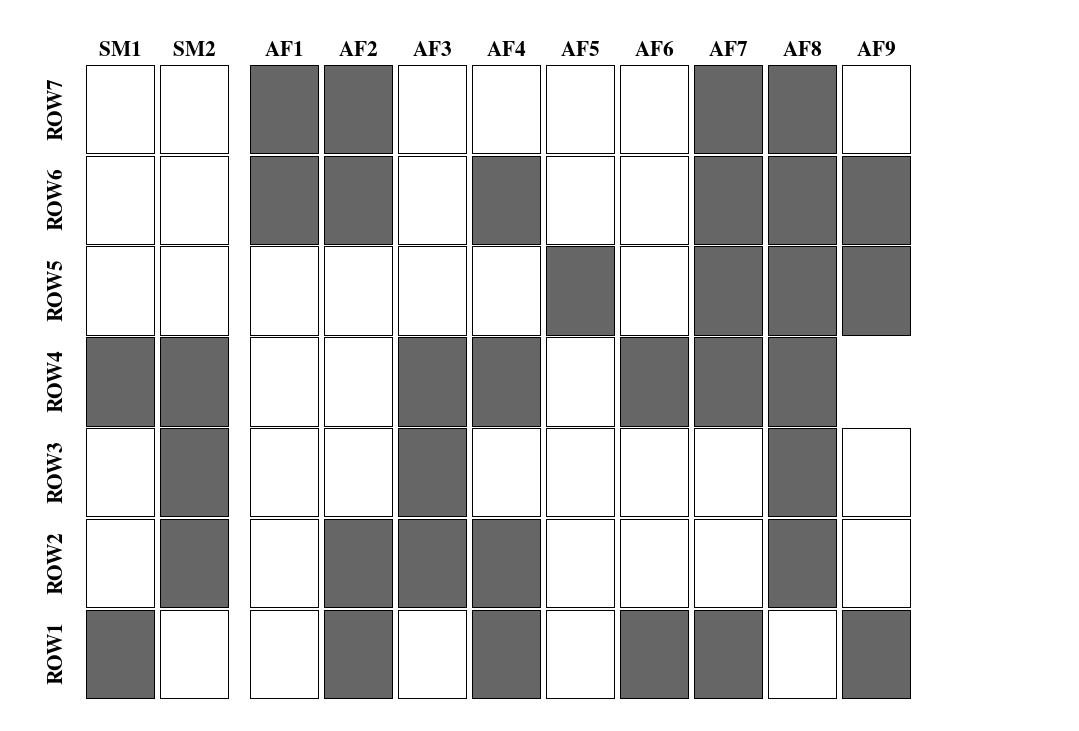}}
\caption{Layout of the SM and AF part of the focal plane and the
distribution of CCD types among the devices; TYPE-01 devices are indicated in
grey, and TYPE-02 devices are indicated in white.}
\label{fig:deviceTypes}
\end{figure}
In Table~\ref{tab:devicePairs} we list all twin CCDs in SM and AF, where a
`twin' refers to two devices that have been manufactured from the same
wafer. Twin devices can sometimes exhibit similar behaviour. Most devices are
not twinned with another.
\begin{table}
\caption{\label{tab:devicePairs}Twin CCDs among all SM and AF devices.}
\centering
\begin{tabular}{cc}
\toprule
TYPE-01 & TYPE-02 \\
\midrule
ROW1  AF4 & ROW5  SM1 \\
ROW1  AF9 & ROW7  AF3 \\
ROW2  AF2 & ROW3  AF5 \\
ROW2  AF4 & ROW3  AF9 \\
ROW2  AF8 & ROW2  AF5 \\
ROW4  SM1 & ROW2  AF7 \\
ROW4  AF4 & ROW6  AF6 \\
ROW4  AF7 & ROW2  SM1 \\
ROW6  AF8 & ROW7  AF6 \\
ROW6  AF9 & ROW3  AF4 \\
\bottomrule
\end{tabular}
\tablefoot{Each row corresponds to a pair of TYPE-01 and TYPE-02 devices created
from the left (column 1) and right (column 2) half of the same silicon wafer.}
\end{table}
\section{Numerical integration by Gauss-Legendre quadrature
\label{app:num_int}}
The PSF model presented in Eq.~\ref{eqn:analyticIntEffPsfExcTdi} and its
associated first derivatives include three integrals over $\tau$ that must be
evaluated numerically. For convenience, the integrands will be denoted
$P(\tau)$, $Q(\tau)$ and $R(\tau)$ where
\begin{multline} \label{eqn:num_int}
P(\tau) = \al_i(u + \Delta u(\tau,\dot{\eta})) \, \ac_j(v + \Delta v(\tau,\dot{\zeta}))\\
Q(\tau) = \al_i^{\prime}(u + \Delta u(\tau,\dot{\eta})) \, \ac_j(v + \Delta v(\tau,\dot{\zeta}))\\
R(\tau) = \al_i(u + \Delta u(\tau,\dot{\eta})) \, \ac_j^{\prime}(v + \Delta v(\tau,\dot{\zeta}))\\
\end{multline}
A few selected examples of $P(\tau)$, $Q(\tau)$ and $R(\tau)$ are depicted in
Fig.~\ref{fig:pqr}.
\begin{figure}
\resizebox{\hsize}{!}{\includegraphics{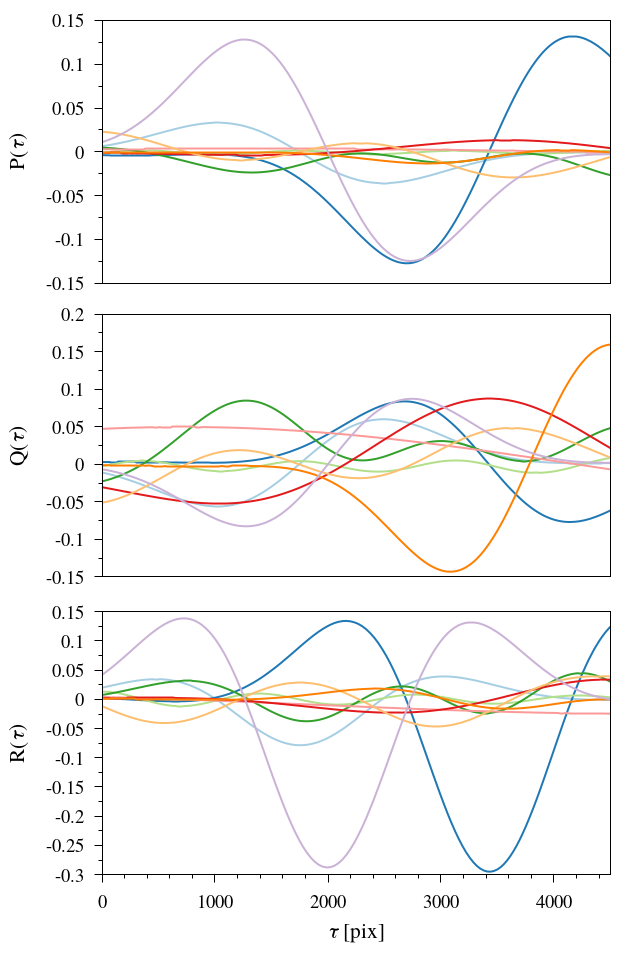}}
\caption{Selected examples of the $P(\tau)$, $Q(\tau)$ and $R(\tau)$
functions from Eq.~\ref{eqn:num_int} that must be integrated numerically.
These have been generated for a range of values of the
$u,v,\dot{\eta},\dot{\zeta},i$ and $j$ parameters encountered in routine
processing, and correspond to NOGATE for which the $\tau$ coordinate spans 1 to
4500.}
\label{fig:pqr}
\end{figure}
Many different numerical integration methods exist, and while the accuracy of
any method generally depends on the kind of functions that are being integrated,
this must also be balanced against our implementation requirements of low
execution time and memory footprint. We tested both Romberg's method and
Gauss-Legendre quadrature,
and found the latter to have significantly lower numerical integration error for
the same number of function evaluations. In Fig.~\ref{fig:gaussSteps} we
present the relative error in the Gauss-Legendre algorithm as a function of the
number of steps used, for each of the functions in Fig.~\ref{fig:pqr}
\begin{figure}
\resizebox{\hsize}{!}{\includegraphics{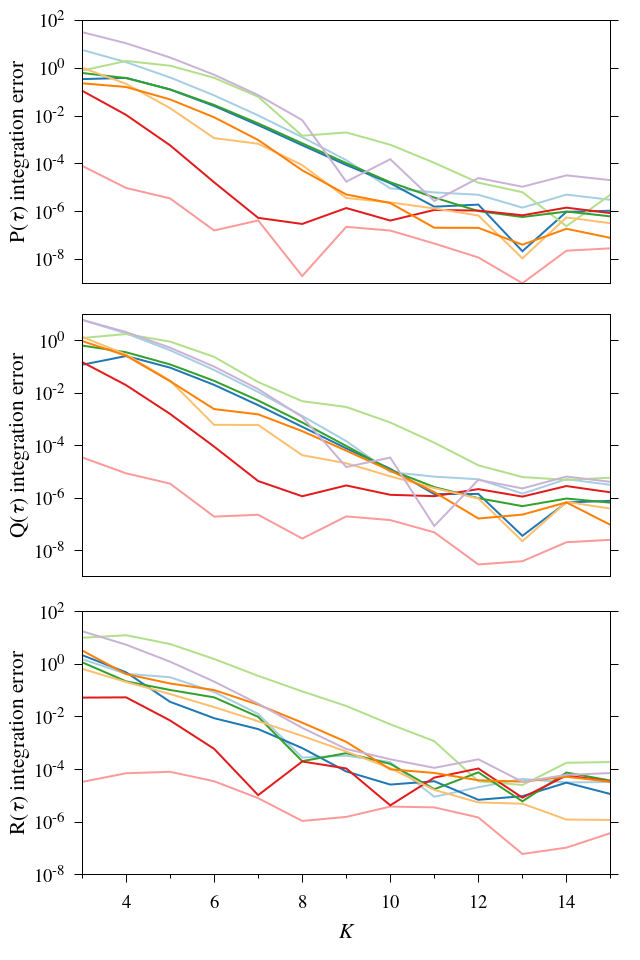}}
\caption{Relative error in the Gauss-Legendre numerical integration of
the functions in Fig.~\ref{fig:pqr} for different numbers of steps $K$.
The relative integration error is defined as $|(I_K - I_{200})/I_{200}|$,
where $I_K$ is the result using $K$ steps.}
\label{fig:gaussSteps}
\end{figure}
We selected the $K=9$ step algorithm as an appropriate balance of the
integration error and number of function evaluations that optimises the
available execution time. The corresponding function evaluation points $\tau_k$
and associated weight factors $w_k$ for $k=0$ to $K-1$ are listed in
Table~\ref{tab:gaussLegendreWeights}; these are the values used in
Eq.~\ref{eqn:numIntEffPsfExcTdi} and throughout this work.
\begin{table}
\caption{\label{tab:gaussLegendreWeights}Gauss-Legendre quadrature
configuration.}
\centering
\begin{tabular}{cccc}
\toprule
$k$ &  \multicolumn{2}{c}{$\tau_k$ [pix]} & $w_k$ \\
    & NOGATE            & GATE12          &       \\
\midrule 
0 &	4428.4 & 2859.8	& 0.08127 \\
1 &	4131.2 & 2667.8	& 0.18065 \\
2 &	3630.3 & 2344.4	& 0.26061 \\
3 &	2979.9 & 1924.5	& 0.31235 \\
4 &	2250.5 & 1453.5	& 0.33024 \\
5 &	1521.1 & 982.5	& 0.31235 \\
6 &	870.7  & 562.6	& 0.26061 \\
7 &	369.8  & 239.2	& 0.18065 \\
8 &	72.6   & 47.2	& 0.08127 \\
\bottomrule
\end{tabular}
\tablefoot{
$\tau_k$ depends on the CCD gate, and values are listed
for NOGATE and GATE12 only. The corresponding values for other gates can be
obtained by scaling according to the $\tau$ range of the gate.
}
\end{table}

\end{appendix}

\end{document}